\documentclass[twocolumn,aps,pre,amsmath,showpacs]{revtex4}

\usepackage{latexsym}
\usepackage{graphicx}
\usepackage{amsfonts}

\usepackage{amssymb}                                                          
\begin{document}

\title{Nonequilibrium mode-coupling theory for uniformly sheared systems}
\author{Song-Ho Chong}
\affiliation{Institute for Molecular Science,
Okazaki 444-8585, Japan}
\author{Bongsoo Kim}
\affiliation{Department of Physics, Changwon National University,
Changwon 641-773, Korea}

\date{\today}

\begin{abstract}

We develop a nonequilibrium mode-coupling theory for uniformly sheared systems
starting from microscopic, thermostatted SLLOD equations of motion.
Our theory aims at describing stationary-state properties including rheological ones
of sheared systems, and this is accomplished via two steps. 
Firstly, a set of self-consistent equations is formulated based on the
projection-operator formalism and on the mode-coupling approach
for the transient density correlators
which measure the correlations between the density fluctuations 
in the initial equilibrium state and the ones
at later times after the shearing force is turned on. 
The transient time-correlation function formalism is then used
which, combined with the mode-coupling approximation, expresses
stationary-state properties in terms of the transient density correlators.
A detailed comparison of our theory is also presented with the related
mode-coupling theory which is based on the Smoluchowski equation
for Brownian particles under stationary shearing. 

\end{abstract}


\maketitle

\section{Introduction}
\label{sec:introduction}

Nonlinear rheological behavior of glassy materials under stationary shearing 
has attracted considerable attention in recent years 
since it provides additional insight into the physics of glass 
transition~\cite{Larson99,Liu01,Yamamoto98b,Berthier02,Petekidis04,Besseling07,Schall07,Fuchs08}.
For such systems driven far from equilibrium, the shear rate should be
regarded as a relevant control parameter rather than as a small perturbation~\cite{Liu98}.
In this paper, we develop a nonequilibrium statistical mechanical theory
for glass-forming systems in which the shear rate as well as 
temperature and density can be handled as external control parameters.
This will be done by extending the projection-operator formalism~\cite{Hansen86} and 
the mode-coupling theory (MCT)~\cite{Goetze91b} to nonequilibrium systems.

MCT has been known as the most successful microscopic theory
for the glass transition.
Indeed, extensive tests of the theoretical predictions carried out so far
against experimental data and computer-simulation results
suggest that the theory deals properly with some essential features of 
glass-forming systems~\cite{Goetze92,Goetze99}.
It is therefore natural that extensions of MCT have been attempted 
to stationary sheared systems. 

At present there exist two different approaches in such nonequilibrium extensions
of MCT: one based on steady-state 
fluctuations~\cite{Miyazaki-sheared-MCT-all,Hayakawa08}, and the other
based on the transient time-correlation function (TTCF) 
formalism~\cite{Fuchs-Cates-sheared-MCT,Fuchs05}.
In the former approach, basic objects are the steady-state density correlators 
defined with fluctuations around the stationary state. 
Rheological properties like the shear stress are then expressed 
within the mode-coupling approximation 
in terms of these steady-state correlators. 
With the same spirit as MCT for quiescent systems~\cite{Goetze91b},
the structure factor $S_{\bf q}^{\rm ss}$ of the stationary state, 
which now depends on the shear rate as well as on the wave ``vector'' ${\bf q}$,
enters as input into the equations describing the dynamics.
At first sight, such an approach looks quite reasonable, but in fact
it possesses a conceptual problem.
For example, the following exact relation holds between the
interaction part of the steady-state shear stress $\sigma_{\rm ss}^{\rm int}$ 
and the steady-state pair-correlation function $g_{\rm ss}({\bf r})$
\begin{equation}
\sigma_{\rm ss}^{\rm int} = 
\frac{\rho^{2}}{2} \int d{\bf r} \, g_{\rm ss}({\bf r}) 
\frac{xy}{r} \frac{d \phi(r)}{dr},
\label{eq:ss-shear-Kirkwood}
\end{equation}
for a uniform shear with velocity along the $x$-axis and its gradient
along the $y$-axis~\cite{Kirkwood49}.
Here $\rho$ denotes the average number density, and $\phi(r)$ the
pair-interaction potential.
Since $S_{\bf q}^{\rm ss}$ is related to the Fourier transform of $g_{\rm ss}({\bf r})$,
Eq.~(\ref{eq:ss-shear-Kirkwood}) states that 
$S_{\bf q}^{\rm ss}$ and $\sigma_{\rm ss}^{\rm int}$ should be
handled on an equal footing, but this aspect is missing in the
steady-state-fluctuations approach of Refs.~\cite{Miyazaki-sheared-MCT-all,Hayakawa08}
where $S_{\bf q}^{\rm ss}$ is treated as the input while
$\sigma_{\rm ss}^{\rm int}$ is the output. 
In addition, it is assumed in Ref.~\cite{Miyazaki-sheared-MCT-all} that
the fluctuation-dissipation theorem (FDT) holds also
in the nonequilibrium stationary state.
The use of such an assumption is unjustified since the
violations of the FDT have been reported in the computer-simulation 
study of sheared systems~\cite{Berthier02}.

On the other hand, no such problems arise in 
the theory developed by Fuchs and Cates (FC)~\cite{Fuchs-Cates-sheared-MCT,Fuchs05}
which is based on the TTCF formalism~\cite{Evans90}. 
Starting from the Smoluchowski equation for interacting Brownian particles
under stationary shearing,
the FC theory aims at describing steady-state properties via two steps:
firstly, the MCT equations for the {\em transient}
density correlators --
the correlators between the density fluctuations 
in the equilibrium starting state and the ones
at later times after the shearing force is turned on --
are formulated, and then the TTCF formalism is used which, combined with the
mode-coupling approximation, expresses
stationary-state properties in terms of these transient correlators.
In this approach, only {\em equilibrium} static structure factor
is required as input, whereas the steady-state structure factor $S_{\bf q}^{\rm ss}$ as well as 
the shear stress $\sigma_{\rm ss}^{\rm int}$ are the output
of the theory.
Thus, the aforementioned conceptual problem in 
Refs.~\cite{Miyazaki-sheared-MCT-all,Hayakawa08} does not apply here. 
Furthermore, it is in principle possible using the FC theory to investigate 
the violations of the FDT, although this issue has not yet been addressed.

It is expected on physical grounds that the microscopic dynamics
does not matter as far as the long-time glassy dynamics is concerned.
Indeed, it was argued that the equilibrium MCT leads to the
same glass transition scenario for both Newtonian and Brownian microscopic 
dynamics~\cite{Goetze91b,Goetze92,Szamel91},
and this was confirmed by computer-simulation studies~\cite{Gleim98,Szamel04b}.
However, it is not {\em a priori} obvious whether such an equivalent
long-time dynamics holds true also for nonequilibrium sheared systems.

In this paper, we develop a nonequilibrium MCT starting from the 
SLLOD equations~\cite{Evans90} -- Newtonian equations of motion under stationary shearing --
which have been widely adopted in simulation studies
of homogeneously sheared systems (see, e.g., Refs.~\cite{Berthier02,Varnik06}).
Our theory follows the FC formulation in that
the MCT equations for the transient density correlators
are derived first, and then the TTCF formalism
is used for describing stationary-state properties.
However, we found that, although it is not
difficult to adapt the FC formulation in Ref.~\cite{Fuchs-Cates-sheared-MCT,Fuchs05}
to the SLLOD equations at the formal level,
the resulting equations are too cumbersome 
to be useful in practice.
We therefore developed an alternative formulation
to be presented in the following.
It is found that a new memory kernel enters into our
nonequilibrium MCT equations reflecting the 
non-Hermitian nature of the relevant Liouville operator, which is absent in the FC theory
formulated with the Brownian microscopic dynamics. 
In what circumstances 
this additional memory kernel from our theory matters is an open question.
We shall elaborate on this at the end of the paper.

The paper is organized as follows.
In Sec.~\ref{sec:microscopic-starting-point},
we derive exact microscopic equations 
and relations for systems subjected to stationary shearing.
These exact results serve a basis 
for the development of our nonequilibrium MCT.
We will then derive a set of self-consistent
equations for the transient density correlators
based on the projection-operator formalism (Sec.~\ref{sec:GLE})
and on the mode-coupling approach (Sec.~\ref{sec:MCT}).
It is then argued in Sec.~\ref{sec:steady-state-properties}
how the steady-state properties can be
evaluated within the mode-coupling approximation
based on the knowledge of the transient density correlators.
The paper is summarized in Sec.~\ref{sec:summary}, where
a detailed comparison of our theory is also presented with the FC theory.
Appendix~\ref{appen:derivation} is devoted to 
a summary of miscellaneous materials which are necessary
in the main text, and to 
various technical manipulations in the
derivations of some equations.
Appendix~\ref{appen:isotropic-app} describes details
of the isotropic approximation which is useful in practical
applications of our theory to systems where anisotropy
in the density fluctuations is small.

\section{Microscopic starting points}
\label{sec:microscopic-starting-point}

In this section, we derive exact microscopic equations 
and relations subjected to stationary shearing along with thermostat.
These exact results serve a basis for developing
a nonequilibrium MCT for sheared systems to be presented in later sections. 

\subsection{SLLOD equations of motion}
\label{subsec:SLLOD}

We shall consider a system of $N$ atoms of mass $m$ 
in a volume $V$ subjected to stationary shearing characterized by the
shear-rate tensor $\mbox{\boldmath $\kappa$}$.
For a simple uniform shear with velocity
along the $x$-axis and its gradient along the $y$-axis, which we consider
throughout this paper, the shear-rate tensor is 
$\kappa_{\lambda \mu} = \dot{\gamma} \delta_{\lambda x} \delta_{\mu y}$
with $\dot{\gamma}$ denoting the strain rate.
It is postulated that the applied shear induces a homogeneous 
streaming-velocity profile 
${\bf u}({\bf r}) = \mbox{\boldmath $\kappa$}\cdot {\bf r}$
at position ${\bf r}$, assuming that no spontaneous symmetry breaking
takes place. 
Newtonian equations of motion describing such a homogeneously 
sheared system are the thermostatted SLLOD equations~\cite{Evans90}, 
\begin{subequations}
\label{eq:SLLOD}
\begin{eqnarray}
\dot{\bf r}_{i} &=& \frac{{\bf p}_{i}}{m} +
\mbox{\boldmath $\kappa$} \cdot {\bf r}_{i},
\label{eq:SLLOD-a}
\\
\dot{\bf p}_{i} &=& {\bf F}_{i} - 
\mbox{\boldmath $\kappa$} \cdot {\bf p}_{i} - \alpha {\bf p}_{i}. 
\label{eq:SLLOD-b}
\end{eqnarray}
\end{subequations}
Here ${\bf r}_{i}$ and ${\bf p}_{i}$ refer to the position 
and momentum of the $i$th particle, 
${\bf F}_{i} = - \partial U / \partial {\bf r}_{i}$ with the total
interaction potential $U$ 
is the conservative force exerted on the $i$th particle by 
other particles, and
$\alpha {\bf p}_{i}$ is the thermostatting term
which prevents the system from heating up 
due to the work done on it by the shearing force. 
The momenta $\{ {\bf p}_{i} \}$, referred to as
the SLLOD momenta, are peculiar with respect to the
streaming velocity 
${\bf u}({\bf r}_{i}) = \mbox{\boldmath $\kappa$} \cdot {\bf r}_{i}$
at the particle position ${\bf r}_{i}$, 
and satisfy $\sum_{i} {\bf p}_{i} = 0$.

The thermostatting multiplier $\alpha$ controls 
the kinetic temperature or some other quantity such as the
internal energy.
There exist various types of thermostats --
stochastic or deterministic, and reversible or irreversible ones -- 
considered in the literature.
Among them, the Gaussian isokinetic thermostat
has acquired a respected status and a special importance~\cite{Evans90}.
However, from a fundamental point of view, there is no privileged thermostat,
and one should not attribute a fundamental role
to special assumptions about such models since 
they simply describe various ways to take out energy from the system.
Indeed, it has been conjectured that different thermostats may lead to the same
steady-state properties, in the usual sense of the macroscopic
equivalence of equilibrium ensembles~\cite{Gallavotti98}.
Although no proof is yet available,
it is at least reasonable to expect that 
steady-state properties do not significantly depend on the types of the thermostats, 
and this has been tacitly assumed in simulation studies where
various models have been used as  practical means to control the temperature.

In the present work, we shall adopt a constant-$\alpha$ thermostat
in which the multiplier $\alpha$ can be regarded as a ``friction'' constant.
As we will see later, this thermostat greatly simplifies
the equations to be handled compared, e.g., to the corresponding
equations under the Gaussian isokinetic thermostat whose
multiplier $\alpha_{\rm G}$ reads~\cite{Evans90}
\begin{equation}
\alpha_{\rm G} = 
\sum_{i} {\bf p}_{i} \cdot
( {\bf F}_{i} - \mbox{\boldmath $\kappa$} \cdot {\bf p}_{i}) \, / \,
\sum_{i} {\bf p}_{i}^{2}.
\label{eq:alpha-Gaussian}
\end{equation}
How the steady-state temperature can be controlled
with the constant-$\alpha$ model will be discussed in Sec.~\ref{subsec:TTCF}.

\subsection{The Liouville equation}

For nonequilibrium systems described by the SLLOD equations, 
the form of the Liouville equation commonly used 
for Hamiltonian systems should be properly generalized to take into account
the effect of phase-space compression~\cite{Evans90}.
The Liouville equation for the nonequilibrium 
phase-space distribution function $f({\bf \Gamma},t)$,
where ${\bf \Gamma} = ({\bf r}^{N}, {\bf p}^{N})$
stands for a phase-space point, 
is given by
\begin{equation}
\frac{\partial f({\bf \Gamma},t)}{\partial t} =
- \Bigl[ \, \dot{\bf \Gamma} \cdot 
       \frac{\partial}{\partial {\bf \Gamma}} +
       \Lambda({\bf \Gamma}) \,
\Bigr] f({\bf \Gamma},t) \equiv
- i {\cal L}^{\dagger} f({\bf \Gamma},t).
\label{eq:Lf}
\end{equation}
The operator $i{\cal L}^{\dagger}$ is called the
$f$-Liouvillean, and $\Lambda({\bf \Gamma})$ 
defined by
\begin{equation}
\Lambda({\bf \Gamma}) \equiv
\frac{\partial}{\partial {\bf \Gamma}} \cdot
\dot{\bf \Gamma},
\label{eq:Lambda}
\end{equation}
is referred to as the phase-space compression factor.
For the SLLOD equations (\ref{eq:SLLOD})
with constant $\alpha$, one obtains 
\begin{equation}
\Lambda({\bf \Gamma}) =
\sum_{i}
\Bigl( \,
  \frac{\partial}{\partial {\bf r}_{i}} \cdot \dot{\bf r}_{i} +
  \frac{\partial}{\partial {\bf p}_{i}} \cdot \dot{\bf p}_{i} \,
\Bigr)  = - 3 N \alpha.
\label{eq:SLLOD-Lambda}
\end{equation}
The formal solution to the Liouville equation (\ref{eq:Lf}) reads
\begin{equation}
f({\bf \Gamma},t) = \exp( - i {\cal L}^{\dagger} t) \,
f({\bf \Gamma},0),
\label{eq:f-propagator}
\end{equation}
where $\exp( - i {\cal L}^{\dagger} t)$ is called the $f$-propagator.

The time evolution of phase variables, 
which by definition do not depend on time explicitly
and whose time dependence comes solely from that of the phase ${\bf \Gamma}$,
is determined by
\begin{equation}
\frac{d}{dt} A({\bf \Gamma}) =
\dot{\bf \Gamma} \cdot 
\frac{\partial}{\partial {\bf \Gamma}}
A({\bf \Gamma}) \equiv
i {\cal L} A({\bf \Gamma}).
\label{eq:Lp}
\end{equation}
The operator $i{\cal L}$ is referred to as the
$p$-Liouvillean.
The formal solution to this equation can be written 
in terms of the
$p$-propagator $\exp( i {\cal L} t)$ as
\begin{equation}
A({\bf \Gamma},t) =
\exp(i {\cal L} t)
A({\bf \Gamma}).
\label{eq:p-propagator}
\end{equation}

Let us summarize here for later use 
relations between $f$- and $p$-Liouvilleans and 
corresponding propagators. 
It follows from Eqs.~(\ref{eq:Lf}) and (\ref{eq:Lp}) that
\begin{equation}
i{\cal L}^{\dagger}({\bf \Gamma}) = 
i{\cal L}({\bf \Gamma}) + \Lambda({\bf \Gamma}).
\label{eq:relation-Liouville-operators}
\end{equation}
One can show that
$i{\cal L}$ and $i{\cal L}^{\dagger}$
are adjoint operators, and this is why
the notation $i {\cal L}^{\dagger}$ is used for the
$f$-Liouvillean:
\begin{equation}
\int d{\bf \Gamma} \,
[ i {\cal L} A({\bf \Gamma}) ] \, B({\bf \Gamma}) =
- \int d{\bf \Gamma} \,
A({\bf \Gamma}) \,
[ i {\cal L}^{\dagger} B({\bf \Gamma})].
\label{eq:adjoint-property}
\end{equation}
This property can be proved from the integration by parts. 
By a repeated use of this property, the following
relation for the propagators can be derived:
\begin{equation}
\int d{\bf \Gamma} \,
[ e^{i {\cal L} t} A({\bf \Gamma})] \, B({\bf \Gamma}) =
\int d{\bf \Gamma} \,
A({\bf \Gamma}) \, [ e^{- i {\cal L}^{\dagger} t} B({\bf \Gamma})].
\label{eq:unrolling}
\end{equation} 
If the phase-space compression factor $\Lambda({\bf \Gamma})$
is identically zero, then $i{\cal L}^{\dagger} = i{\cal L}$ holds, 
and the Liouvillean becomes self-adjoint, or Hermitian.
In general, this is not the case for nonequilibrium systems.

\subsection{Nonequilibrium distribution function}

Let us consider an equilibrium system of temperature $T$ 
to which a constant shear rate $\dot{\gamma}$ is applied
at time $t=0$, and thereafter the system evolves according to
the SLLOD equations (\ref{eq:SLLOD}).
The $p$-Liouvillean is given by
\begin{subequations}
\label{eq:iL-SLLOD}
\begin{equation}
i {\cal L} = 
\left\{
\begin{array}{lc}
i {\cal L}_{0} & (t \le 0), \\
i {\cal L}_{0} + i {\cal L}_{\dot{\gamma}} + i {\cal L}_{\alpha} & (t > 0).
\end{array}
\right.
\label{eq:iL-SLLOD-a}
\end{equation}
Here, an unperturbed adiabatic or quiescent part ($i {\cal L}_{0}$),
a shear part $(i {\cal L}_{\dot{\gamma}})$,
and a thermostat part $(i {\cal L}_{\alpha})$ 
are respectively given by
\begin{eqnarray}
i {\cal L}_{0} &=& \sum_{i}
\Bigl[ \,
  \frac{{\bf p}_{i}}{m} \cdot \frac{\partial}{\partial {\bf r}_{i}} +
  {\bf F}_{i} \cdot \frac{\partial}{\partial {\bf p}_{i}} \,
\Bigr],
\label{eq:iL0}
\\
i {\cal L}_{\dot{\gamma}} &=& \sum_{i}
\Bigl[ \,
  (\mbox{\boldmath $\kappa$} \cdot {\bf r}_{i}) \cdot  
  \frac{\partial}{\partial {\bf r}_{i}} -
  (\mbox{\boldmath $\kappa$} \cdot {\bf p}_{i}) \cdot  
  \frac{\partial}{\partial {\bf p}_{i}} \,
\Bigr],
\label{eq:iL-dot-gamma}
\\
i {\cal L}_{\alpha} &=& \sum_{i}
  (- \alpha {\bf p}_{i}) \cdot 
\frac{\partial}{\partial {\bf p}_{i}}.
\label{eq:iL-alpha}
\end{eqnarray}
\end{subequations}
Since the phase-space distribution function at $t=0$ coincides with the 
equilibrium one, which we choose to be the 
canonical distribution, 
\begin{equation}
f_{\rm eq}({\bf \Gamma}) \equiv f({\bf \Gamma},0) =
\frac{1}{\cal Z}
e^{- \beta H_{0}({\bf \Gamma})}, \,\,\,
{\cal Z} = \int d{\bf \Gamma} \,
e^{- \beta H_{0}({\bf \Gamma})},
\end{equation}
where 
$\beta \equiv 1/k_{\rm B}T$ with $k_{\rm B}$ denoting Boltzmann's constant and
$H_{0} \equiv \sum_{i} {\bf p}_{i}^{2}/2m + U$, 
a formal solution to the Liouville equation (\ref{eq:Lf})
for $t > 0$ is given by
\begin{equation}
f({\bf \Gamma},t) = e^{- i {\cal L}^{\dagger} t}
f_{\rm eq}({\bf \Gamma}).
\label{eq:dist-dum-11}
\end{equation}
With the identity
\begin{equation}
e^{- i {\cal L}^{\dagger} t} =
1 + \int_{0}^{t} ds \,
e^{- i {\cal L}^{\dagger} s} 
(- i {\cal L}^{\dagger}),
\end{equation}
whose validity can easily be verified by differentiation with respect to $t$, 
Eq.~(\ref{eq:dist-dum-11}) can be expressed as
\begin{equation}
f({\bf \Gamma},t) = f_{\rm eq}({\bf \Gamma}) +
\int_{0}^{t} ds \, 
e^{- i {\cal L}^{\dagger} s}
(- i {\cal L}^{\dagger})
f_{\rm eq}({\bf \Gamma}).
\label{eq:dist-dum-12}
\end{equation}
Since 
$i {\cal L}_{0} f_{\rm eq}({\bf \Gamma}) = 0$,
we get from
Eqs.~(\ref{eq:SLLOD-Lambda}), (\ref{eq:relation-Liouville-operators}), and (\ref{eq:iL-SLLOD})
\begin{equation}
i {\cal L}^{\dagger} f_{\rm eq}({\bf \Gamma}) =
i {\cal L}_{\dot{\gamma}} f_{\rm eq}({\bf \Gamma}) +
i {\cal L}_{\alpha} f_{\rm eq}({\bf \Gamma}) -
3N\alpha f_{\rm eq}({\bf \Gamma}).
\label{eq:dist-dum-13}
\end{equation}
The first term in this expression is given by
\begin{eqnarray}
i {\cal L}_{\dot{\gamma}} f_{\rm eq}({\bf \Gamma}) &=&
\beta 
\sum_{i}
\Bigl[ 
  (\mbox{\boldmath $\kappa$} \cdot {\bf r}_{i}) \cdot {\bf F}_{i} +  
  (\mbox{\boldmath $\kappa$} \cdot {\bf p}_{i}) \cdot 
  \frac{{\bf p}_{i}}{m} 
\Bigr] \, f_{\rm eq}({\bf \Gamma}) 
\nonumber \\
&=&
\beta \mbox{\boldmath $\kappa$} \cdot \mbox{\boldmath $\sigma$}
f_{\rm eq}({\bf \Gamma}) =
\frac{\dot{\gamma}}{k_{\rm B}T} f_{\rm eq}({\bf \Gamma}) \sigma_{xy}.
\label{eq:dist-dum-14}
\end{eqnarray}
Here 
$\mbox{\boldmath $\sigma$}$ denotes the stress tensor whose
elements are 
\begin{equation}
\sigma_{\lambda \mu} =
\sum_{i}
[ \, p_{i}^{\lambda} p_{i}^{\mu} / m + r_{i}^{\lambda} F_{i}^{\mu} \, ]
\quad
(\lambda, \mu = x, y, z),
\label{eq:sigma-def}
\end{equation}
and in the final equality of Eq.~(\ref{eq:dist-dum-14}) we used the
specific form $\kappa_{\lambda \mu} = 
\dot{\gamma} \delta_{\lambda x} \delta_{\mu y}$
for the shear-rate tensor and the symmetry
$\sigma_{\lambda \mu} = \sigma_{\mu \lambda}$ of
the stress tensor
(see Appendix~\ref{appen:shear-stress}). 
The second term in Eq.~(\ref{eq:dist-dum-13}) is given by
\begin{equation}
i {\cal L}_{\alpha} f_{\rm eq}({\bf \Gamma}) =
\frac{2 \alpha}{k_{\rm B}T}
f_{\rm eq}({\bf \Gamma}) \, K({\bf \Gamma}),
\end{equation}
in terms of the kinetic energy 
$K({\bf \Gamma}) \equiv \sum_{i} {\bf p}_{i}^{2}/2m$.
We therefore obtain
\begin{equation}
i {\cal L}^{\dagger} f_{\rm eq}({\bf \Gamma}) =
\frac{\dot{\gamma}}{k_{\rm B}T} 
f_{\rm eq}({\bf \Gamma}) \sigma_{xy}({\bf \Gamma}) +
\frac{2 \alpha}{k_{\rm B}T} 
f_{\rm eq}({\bf \Gamma}) \delta K({\bf \Gamma}),
\label{eq:dist-dum-15}
\end{equation}
where we have introduced the kinetic energy fluctuation $\delta K({\bf \Gamma})$ defined as
\begin{equation}
\delta K({\bf \Gamma}) \equiv
K({\bf \Gamma}) - \frac{3}{2} N k_{\rm B}T.
\label{eq:deltaK-def}
\end{equation}
Substitution of Eq.~(\ref{eq:dist-dum-15})
into Eq.~(\ref{eq:dist-dum-12}) then yields
\begin{eqnarray}
f({\bf \Gamma},t) &=& f_{\rm eq}({\bf \Gamma}) 
- \frac{\dot{\gamma}}{k_{\rm B}T} 
\int_{0}^{t} ds \,
e^{- i {\cal L}^{\dagger} s}
[ f_{\rm eq}({\bf \Gamma}) \sigma_{xy}({\bf \Gamma}) ]
\nonumber \\
& & \quad
- \, 
\frac{2 \alpha}{k_{\rm B}T} 
\int_{0}^{t} ds \,
e^{- i {\cal L}^{\dagger} s}
[ f_{\rm eq}({\bf \Gamma}) \delta K({\bf \Gamma}) ].
\label{eq:neq-distribution}
\end{eqnarray}
This expression for the nonequilibrium phase-space distribution function
plays a fundamental role in the following. 
Notice that the last term in this expression vanishes
if the Gaussian isokinetic thermostat [see Eq.~(\ref{eq:alpha-Gaussian})]
is used for constraining $\delta K({\bf \Gamma})$ to zero. 
In this case, Eq.~(\ref{eq:neq-distribution}) reduces to 
the so-called Kawasaki distribution function for sheared systems~\cite{Evans90,Yamada67}.

\subsection{Transient time-correlation function formalism}

In contrast to equilibrium quantities,
the nonequilibrium ensemble average 
$\langle A(t) \rangle$ 
of a phase variable $A$
depends explicitly on the time $t$ 
past since the start of shearing.
Similarly, the time-correlation function
$\langle A(t+\tau) B(t)^{*} \rangle$
depends not only
on the time difference $\tau$ but also on $t$.
Using the nonequilibrium phase-space distribution
function $f({\bf \Gamma},t)$, 
$\langle A(t) \rangle$ and $\langle A(t+\tau) B(t)^{*} \rangle$
can be expressed as
\begin{eqnarray}
& &
\hspace{-0.5cm}
\langle A(t) \rangle =
\int d{\bf \Gamma} \,
f({\bf \Gamma},0) \, A(t) = 
\int d{\bf \Gamma} \,
f({\bf \Gamma},t) \, A(0),
\label{eq:neq-average}
\\
& &
\hspace{-0.5cm}
\langle A(t+\tau) B(t)^{*} \rangle =
\int d{\bf \Gamma} \,
f({\bf \Gamma},0) \,
A(t + \tau) B(t)^{*} 
\nonumber \\
& & \qquad \qquad \qquad 
=
\int d{\bf \Gamma} \,
f({\bf \Gamma},t) \,
A(\tau) B(0)^{*}.
\label{eq:neq-tcf}
\end{eqnarray}
The two representations in terms of
$f({\bf \Gamma},0)$ or 
$f({\bf \Gamma},t)$ are equivalent
because of the relation~(\ref{eq:unrolling}).
Hereafter, we shall reserve the notation
$\langle \cdots \rangle$
for representing the averaging over the
initial canonical distribution function
$f({\bf \Gamma},0) = f_{\rm eq}({\bf \Gamma})$:
\begin{equation}
\langle \cdots \rangle \equiv
\int d{\bf \Gamma} \, f_{\rm eq}({\bf \Gamma}) \cdots.
\label{eq:def-averaging}
\end{equation}
It should be remembered, however, that
the dynamics inside the brackets $\langle \cdots \rangle$ 
is governed by the
thermostatted SLLOD equations, 
and only averages like 
$\langle A(0) \rangle$ and
$\langle A(0) B(0)^{*} \rangle$
coincide with equilibrium quantities. 

Substituting Eq.~(\ref{eq:neq-distribution}) into 
Eqs.~(\ref{eq:neq-average}) and (\ref{eq:neq-tcf})
and then using Eq.~(\ref{eq:unrolling}), 
one obtains
\begin{widetext}
\begin{eqnarray}
& &
\langle A(t) \rangle = 
\langle A(0) \rangle 
- \frac{\dot{\gamma}}{k_{\rm B}T}
\int_{0}^{t} ds \,
\langle A(s) \sigma_{xy}(0) \rangle 
- \frac{2 \alpha}{k_{\rm B}T}
\int_{0}^{t} ds \,
\langle A(s) \delta K(0) \rangle,
\label{eq:neq-average-TTCF}
\\
& &
\langle A(t+\tau) B(t)^{*} \rangle = 
\langle A(\tau) B(0)^{*} \rangle 
- \frac{\dot{\gamma}}{k_{\rm B}T}
\int_{0}^{t} ds \,
\langle A(s+\tau) B(s)^{*} \sigma_{xy}(0) \rangle -
\frac{2 \alpha}{k_{\rm B}T}
\int_{0}^{t} ds \,
\langle A(s+\tau) B(s)^{*} \delta K(0) \rangle.
\label{eq:neq-tcf-TTCF}
\end{eqnarray}
\end{widetext}
The expression (\ref{eq:neq-average-TTCF})
relates the nonequilibrium value of a phase variable $A$ at
time $t$ to the integral of 
{\em transient} time-correlation function (TTCF)
$\langle A(s) \sigma_{xy}(0) \rangle$ 
-- the correlation between $\sigma_{xy}$ 
in the initial equilibrium state, 
$\sigma_{xy}(0)$,
and $A$ at time $s$ after the shearing force is 
turned on --
and another integral of TTCF
$\langle A(s) \delta K(0) \rangle$
formed with $\delta K(0)$. 
Equation (\ref{eq:neq-tcf-TTCF}) is a generalization 
of this TTCF expression
to the time-correlation function.

The system is said to be in a nonequilibrium steady state
if the ensemble averages of all phase variables become 
time-independent. 
Let us notice that 
the long-time limit of Eq.~(\ref{eq:neq-average-TTCF})
becomes constant if the system 
displays {\em mixing}~\cite{Evans90}.
This feature can be shown by taking a time
derivative of Eq.~(\ref{eq:neq-average-TTCF}):
\begin{equation}
\frac{d}{dt} \langle A(t) \rangle =
- \frac{\dot{\gamma}}{k_{\rm B}T}
\langle A(t) \sigma_{xy}(0) \rangle 
- \frac{2 \alpha}{k_{\rm B}T}
\langle A(t) \delta K(0) \rangle.
\label{eq:dA-over-dt}
\end{equation}
If the system displays mixing~\cite{Evans90},
then all the long-time correlations between
phase variables vanish.
We therefore obtain for $t \to \infty$
\begin{eqnarray}
\frac{d}{dt} \langle A(t) \rangle &\to&
- \frac{\dot{\gamma}}{k_{\rm B}T}
\langle A(t) \rangle
\langle \sigma_{xy}(0) \rangle 
\nonumber \\
& & \quad
- \,
\frac{2 \alpha}{k_{\rm B}T}
\langle A(t) \rangle
\langle \delta K(0) \rangle = 0,
\end{eqnarray}
since the equilibrium ensemble averages 
$\langle \sigma_{xy}(0) \rangle$ and
$\langle \delta K(0) \rangle$ are zero
[see Eqs.~(\ref{eq:virial-theorem}) and (\ref{eq:deltaK-def})]. 
This indicates that the long-time steady state average
of an arbitrary phase variable becomes constant,
i.e.,
\begin{equation}
\lim_{t \to \infty} \langle A(t) \rangle =
\langle A \rangle_{\rm ss},
\end{equation}
where the steady-state average, denoted by
$\langle \cdots \rangle_{\rm ss}$ hereafter,
is obtained from
the $t \to \infty$ limit of Eq.~(\ref{eq:neq-average-TTCF}):
\begin{widetext}
\begin{eqnarray}
\langle A \rangle_{\rm ss} =
\langle A(0) \rangle 
- \frac{\dot{\gamma}}{k_{\rm B}T}
\int_{0}^{\infty} ds \,
\langle A(s) \sigma_{xy}(0) \rangle -
\frac{2 \alpha}{k_{\rm B}T}
\int_{0}^{\infty} ds \,
\langle A(s) \delta K(0) \rangle.
\label{eq:ss-average}
\end{eqnarray}
Similarly, the $t \to \infty$ limit of 
$\langle A(t+\tau) B(t)^{*} \rangle$
becomes independent of $t$ 
since the time derivative of Eq.~(\ref{eq:neq-tcf-TTCF}),
\begin{eqnarray}
\frac{d}{dt}
\langle A(t+\tau) B(t)^{*} \rangle =
- \frac{\dot{\gamma}}{k_{\rm B}T}
\langle A(t+\tau) B(t)^{*} \sigma_{xy}(0) \rangle -
\frac{2 \alpha}{k_{\rm B}T}
\langle A(t+\tau) B(t)^{*} \delta K(0) \rangle,
\label{eq:dAB-over-dt}
\end{eqnarray}
becomes zero for $t \to \infty$
if the system exhibits mixing.
The steady-state time-correlation function
defined as
\begin{equation}
\langle A(\tau) B^{*} \rangle_{\rm ss} \equiv
\lim_{t \to \infty} 
\langle A(t+\tau) B(t)^{*} \rangle,
\label{eq:ss-tcf-def}
\end{equation}
is then given by
\begin{eqnarray}
\langle A(\tau) B^{*} \rangle_{\rm ss} =
\langle A(\tau) B(0)^{*} \rangle 
- \frac{\dot{\gamma}}{k_{\rm B}T}
\int_{0}^{\infty} ds \,
\langle A(s+\tau) B(s)^{*} \sigma_{xy}(0) \rangle -
\frac{2 \alpha}{k_{\rm B}T}
\int_{0}^{\infty} ds \,
\langle A(s+\tau) B(s)^{*} \delta K(0) \rangle.
\label{eq:ss-tcf}
\end{eqnarray}
The TTCF expressions (\ref{eq:ss-average}) and (\ref{eq:ss-tcf}), 
relating the steady-state quantities to the integrals
of TTCFs, can be considered as the
generalized Green-Kubo relations~\cite{Evans90}.

In deriving the nonequilibrium
Zwanzig-Mori-type equation of motion to be presented
in Sec.~\ref{sec:GLE}, 
it is necessary to know how the $p$-Liouvillean $i{\cal L}$
behaves inside the time-correlation function.
To this end, we first notice from Eq.~(\ref{eq:p-propagator}) 
\begin{equation}
\frac{d}{dt}
\langle A(t+\tau) B(t)^{*} \rangle =
\langle [ i {\cal L} A(t+\tau)] \,  B(t)^{*} \rangle +
\langle  A(t+\tau) \, [i {\cal L} B(t)]^{*} \rangle.
\label{eq:dAB-over-dt-2}
\end{equation}
Combined with Eq.~(\ref{eq:dAB-over-dt}), this yields
the desired result
\begin{eqnarray}
\langle i {\cal L} A(t+\tau) B(t)^{*} \rangle =
- \langle A(t+\tau) [i {\cal L} B(t)]^{*} \rangle
- \frac{\dot{\gamma}}{k_{\rm B}T}
\langle A(t+\tau) B(t)^{*} \sigma_{xy}(0) \rangle -
\frac{2 \alpha}{k_{\rm B}T}
\langle A(t+\tau) B(t)^{*} \delta K(0) \rangle.
\label{eq:Lp-inside}
\end{eqnarray}
\end{widetext}
For systems exhibiting mixing, there holds for $t \to \infty$
\begin{equation}
\langle [ i {\cal L} A(\tau) ] \, B^{*} \rangle_{\rm ss} =
- \langle A(\tau) \, [i {\cal L} B]^{*} \rangle_{\rm ss},
\label{eq:Lp-inside-steady-state}
\end{equation}
i.e., the $p$-Liouvillean becomes Hermitian
in the steady state.
This is expected since 
the time-translation symmetry is recovered 
in the stationary state. 

\subsection{Implication of translational invariance}

Since we are dealing with amorphous systems, 
the equilibrium distribution function 
$f_{\rm eq}({\bf \Gamma})$ is assumed to be translationally
invariant and isotropic.
In this subsection, it is shown that 
the nonequilibrium distribution function $f({\bf \Gamma},t)$ under shear
becomes anisotropic, but 
remains translationally invariant.
We then discuss an implication of this property.
Our treatment here follows the one presented in Ref.~\cite{Fuchs05}.

To this end, we shall consider global translation of all particle positions,
\begin{equation}
{\bf \Gamma} \to {\bf \Gamma}^{\prime}
\,\, \mbox{ where } \,\,
{\bf r}_{i}^{\prime} = {\bf r}_{i} + {\bf a}
\,\, \mbox{ for all } i,
\label{eq:shift-Gamma}
\end{equation}
which amounts to the shift ${\bf a}$ of the coordinate origin.
Under this shift, the nonequilibrium distribution 
$f({\bf \Gamma},t)$ given in Eq.~(\ref{eq:neq-distribution})
transforms to
\begin{eqnarray}
f({\bf \Gamma}^{\prime},t) &=& 
f_{\rm eq}({\bf \Gamma}) 
- \frac{\dot{\gamma}}{k_{\rm B}T} 
\int_{0}^{t} ds \,
e^{- i {\cal L}^{\dagger}({\bf \Gamma}^{\prime}) s}
[ f_{\rm eq}({\bf \Gamma}) \sigma_{xy}({\bf \Gamma}) ] 
\nonumber \\
& & \quad
- \, 
\frac{2 \alpha}{k_{\rm B}T} 
\int_{0}^{t} ds \,
e^{- i {\cal L}^{\dagger}({\bf \Gamma}^{\prime}) s}
[ f_{\rm eq}({\bf \Gamma}) \delta K({\bf \Gamma}) ].
\end{eqnarray}
Here we used 
\begin{equation}
f_{\rm eq}({\bf \Gamma}^{\prime}) = 
f_{\rm eq}({\bf \Gamma}), \,\,
\sigma_{xy}({\bf \Gamma}^{\prime}) = 
\sigma_{xy}({\bf \Gamma}), \,\,
\delta K({\bf \Gamma}^{\prime}) = \delta K({\bf \Gamma}).
\label{eq:trans-dum-11}
\end{equation}
These hold since $f_{\rm eq}$, 
$\sigma_{xy}$ [see Eq.~(\ref{eq:appen-sigma-def})], and
$\delta K$ depend on
momenta and particle separations only.
How the $f$-propagator transforms under 
${\bf \Gamma} \to {\bf \Gamma}^{\prime}$ is discussed
in Appendix~\ref{appen:translational-invariance} with the result
[see Eq.~(\ref{eq:appen-shifted-f})]
\begin{equation}
e^{- i {\cal L}^{\dagger}({\bf \Gamma}^{\prime}) t} =
e^{- i {\cal L}^{\dagger}({\bf \Gamma}) t} \, 
e^{- {\bf a} \cdot \mbox{\boldmath $\kappa$}^{\rm T} \cdot {\bf P} \, t}
\,\, \mbox{with} \,\,
{\bf P} \equiv \sum_{i} \frac{\partial}{\partial {\bf r}_{i}}.
\end{equation}
Here $\mbox{\boldmath $\kappa$}^{\rm T}$ denotes the
transposed matrix of $\mbox{\boldmath $\kappa$}$. 
Because of Eq.~(\ref{eq:trans-dum-11}), we have
${\bf P} f_{\rm eq}({\bf \Gamma}) = 0$, 
${\bf P} \sigma_{xy}({\bf \Gamma}) = 0$, and
${\bf P} \delta K({\bf \Gamma}) = 0$,
so that
\begin{equation}
e^{- i {\cal L}^{\dagger}({\bf \Gamma}^{\prime}) s}
[ f_{\rm eq}({\bf \Gamma}) 
\sigma_{xy}({\bf \Gamma}) ] =
e^{-i {\cal L}^{\dagger}({\bf \Gamma}) s}
[ f_{\rm eq}({\bf \Gamma}) \sigma_{xy}({\bf \Gamma}) ],
\label{eq:trans}
\end{equation}
and  a similar equation holds in which $\sigma_{xy}$ is
replaced by $\delta K$. 
Therefore, the nonequilibrium distribution 
function $f({\bf \Gamma},t)$ remains translationally invariant:
\begin{equation}
f({\bf \Gamma}',t) = f({\bf \Gamma},t).
\label{eq:shifted-distribution}
\end{equation}

We next consider how the wave-vector-dependent 
phase variable of the form
\begin{equation}
A_{\bf q}({\bf \Gamma},t) =
e^{ i {\cal L}({\bf \Gamma}) t }
\sum_{i} X_{i}^{A_{\bf q}}({\bf \Gamma}) \,
e^{ i {\bf q} \cdot {\bf r}_{i}},
\label{eq:Aq-def}
\end{equation}
transforms under the shift of the coordinate origin.
It is assumed that $X_{i}^{A_{\bf q}}({\bf \Gamma})$ is a
function of momenta and particle separations only, 
so that $X_{i}^{A_{\bf q}}({\bf \Gamma}^{\prime}) = 
X_{i}^{A_{\bf q}}({\bf \Gamma})$. 
For example, 
$X_{i}^{\rho_{\bf q}} = 1$ for density fluctuations,
$X_{i}^{j_{\bf q}^{\lambda}} = p_{i}^{\lambda}/m$
for current density fluctuations to be introduced below,
and $X_{i}^{\sigma_{\bf q}^{\lambda \mu}} =
p_{i}^{\lambda} p_{i}^{\mu} / m -
(1/2) \sum_{j \ne i}
(r_{ij}^{\lambda} r_{ij}^{\mu} / r_{ij}^{2})
P_{\bf q}({\bf r}_{ij})$
for the wave-vector dependent 
stress tensor [see Eq.~(\ref{eq:q-denendent-sigma})].
Using the result 
\begin{equation}
e^{i {\cal L}({\bf \Gamma}^{\prime}) t} =
e^{i {\cal L}({\bf \Gamma}) t} \, 
e^{{\bf a} \cdot \mbox{\boldmath $\kappa$}^{\rm T} \cdot {\bf P} \, t},
\end{equation}
for the $p$-propagator
which is also derived in Appendix~\ref{appen:translational-invariance}
[see Eq.~(\ref{eq:appen-shifted-p})],
one obtains
\begin{eqnarray}
A_{\bf q}({\bf \Gamma}^{\prime},t) &=&
e^{ i {\cal L}({\bf \Gamma}^{\prime}) t }
\sum_{i} X_{i}^{A_{\bf q}}({\bf \Gamma}^{\prime}) \,
e^{ i {\bf q} \cdot ({\bf r}_{i} + {\bf a})}
\nonumber \\
&=&
e^{i {\cal L}({\bf \Gamma}) t} \, 
e^{{\bf a} \cdot \mbox{\boldmath $\kappa$}^{\rm T} \cdot {\bf P} \, t}
\sum_{i} X_{i}^{A_{\bf q}}({\bf \Gamma}) \,
e^{ i {\bf q} \cdot ({\bf r}_{i} + {\bf a})} 
\nonumber \\
&=&
e^{i ({\bf q} + {\bf q} \cdot \mbox{\boldmath $\kappa$} t) \cdot {\bf a}}
A_{\bf q}({\bf \Gamma},t),
\label{eq:shifted-Aq}
\end{eqnarray}
where we used
$X_{i}^{A_{\bf q}}({\bf \Gamma}^{\prime}) =
X_{i}^{A_{\bf q}}({\bf \Gamma})$,
${\bf P} X_{i}^{A_{\bf q}}({\bf \Gamma}) = 0$,
and
$e^{{\bf a} \cdot \mbox{\boldmath $\kappa$}^{\rm T} \cdot {\bf P} \, t}
e^{i {\bf q} \cdot ({\bf r}_{i} + {\bf a})} =
e^{i {\bf q} \cdot \mbox{\boldmath $\kappa$} \cdot {\bf a} \, t}
e^{i {\bf q} \cdot ({\bf r}_{i} + {\bf a})}$. 

Since the integral over the phase space must agree for 
either integration variables ${\bf \Gamma}$ or ${\bf \Gamma}'$,
there holds
\begin{equation}
\langle A_{\bf q}(t) \rangle =
\int d{\bf \Gamma} \, f_{\rm eq}({\bf \Gamma}) 
A_{\bf q}({\bf \Gamma},t) =
\int d{\bf \Gamma}^{\prime} \, f_{\rm eq}({\bf \Gamma}^{\prime}) 
A_{\bf q}({\bf \Gamma}^{\prime},t).
\end{equation}
Using $f_{\rm eq}({\bf \Gamma}') = f_{\rm eq}({\bf \Gamma})$
and Eq.~(\ref{eq:shifted-Aq}),
one obtains
\begin{equation}
\langle A_{\bf q}(t) \rangle =
e^{i ({\bf q} + {\bf q} \cdot \mbox{\boldmath $\kappa$} \, t) 
   \cdot {\bf a}}
\langle A_{\bf q}(t) \rangle.
\end{equation}
This means that the nonequilibrium ensemble averages of
phase variables, including steady-state averages,
are non-vanishing for zero wave-vector only:
\begin{equation}
\langle A_{\bf q}(t) \rangle =
\delta_{{\bf q}, {\bf 0}} \langle A_{\bf q=0}(t) \rangle.
\end{equation}
Similarly, there must hold 
for nonequilibrium time-correlation functions
\begin{eqnarray}
& &
\langle A_{\bf q}(t+\tau) B_{\bf k}(t)^{*} \rangle =
\int d{\bf \Gamma} \, f({\bf \Gamma},t) 
A_{\bf q}({\bf \Gamma},\tau) B_{\bf k}({\bf \Gamma},0)^{*} 
\nonumber \\
& & \qquad \qquad \quad
= \int d{\bf \Gamma}^{\prime} \, f({\bf \Gamma}^{\prime},t) 
A_{\bf q}({\bf \Gamma}^{\prime},\tau) 
B_{\bf k}({\bf \Gamma}^{\prime},0)^{*}.
\end{eqnarray}
Using Eqs.~(\ref{eq:shifted-distribution}) and
(\ref{eq:shifted-Aq}), one finds
\begin{equation}
\langle A_{\bf q}(t+\tau) B_{\bf k}(t)^{*} \rangle =
e^{ i ({\bf q} 
    + {\bf q} \cdot \mbox{\boldmath $\kappa$} \, \tau
    - {\bf k}) \cdot {\bf a}} \,
\langle A_{\bf q}(t+\tau) B_{\bf k}(t)^{*} \rangle.
\label{eq:shifted-tcf}
\end{equation}
This means that 
$A_{\bf q}(t+\tau)$ is statistically
correlated with $B_{\bf k}(t)^{*}$ only if
${\bf k} = {\bf q}(\tau)$ 
with the {\em advected} wave vector 
${\bf q}(\tau) \equiv {\bf q} + {\bf q} \cdot \mbox{\boldmath $\kappa$} \, \tau$
during the time $\tau$, i.e.,
\begin{equation}
\langle A_{\bf q}(t+\tau) B_{\bf k}(t)^{*} \rangle =
\delta_{{\bf k}, {\bf q}(\tau)} \,
\langle A_{\bf q}(t+\tau) B_{{\bf q}(\tau)}(t)^{*} \rangle.
\label{eq:shifted-tcf-2}
\end{equation}
Thus, as in equilibrium systems,
a time-correlation function 
characterized by a single wave vector can be
defined: 
\begin{equation}
C_{\bf q}^{AB}(t+\tau,t) \equiv
\langle A_{\bf q}(t+\tau) B_{{\bf q}(\tau)}(t)^{*} \rangle.
\label{eq:shifted-tcf-3}
\end{equation}
For the shear-rate tensor
$\kappa_{\lambda \mu} = 
\dot{\gamma} \delta_{\lambda x} \delta_{\mu y}$,
the explicit expression for the advected wave vector reads
\begin{equation}
{\bf q}(\tau) = {\bf q} + {\bf q} \cdot \mbox{\boldmath $\kappa$} \tau =
(q_{x}, q_{y} + \dot{\gamma} \tau q_{x}, q_{z}).
\label{eq:advected-q}
\end{equation}
Equivalently, one can introduce a time-correlation
of the following form 
\begin{equation}
\tilde{C}_{\bf q}^{AB}(t+\tau,t) \equiv
\langle A_{{\bf q}(-\tau)}(t+\tau) B_{\bf q}(t)^{*} \rangle.
\label{eq:shifted-tcf-4}
\end{equation}
This also follows from Eq.~(\ref{eq:shifted-tcf})
by noting that
\begin{equation}
{\bf q} \cdot 
( \, {\bf I} + \mbox{\boldmath $\kappa$} t \, ) = {\bf k} 
\,\, \to \,\,
{\bf q} = {\bf k} \cdot 
( \, {\bf I} + \mbox{\boldmath $\kappa$} t \, )^{-1} =
{\bf k} \cdot
( \, {\bf I} - \mbox{\boldmath $\kappa$} t \, ),
\end{equation}
since the shear-rate tensor satisfies 
$\mbox{\boldmath $\kappa$} \cdot \mbox{\boldmath $\kappa$} = 0$.
In this paper, we shall mainly use the 
convention (\ref{eq:shifted-tcf-3}) for time-correlation
functions, and the convention (\ref{eq:shifted-tcf-4}) 
will be used only for the discussion in Sec.~\ref{subsec:GLE-1}.
 
Finally, we notice for later use
the following relation for time-correlation functions
involving three phase variables 
\begin{eqnarray}
& &
\langle A_{\bf q}(t+\tau) B_{{\bf q}(\tau)}(t)^{*} D_{\bf k}(t)^{*} 
\rangle 
\nonumber \\
& & \qquad \quad
=
\delta_{{\bf k}, {\bf 0}} \,
\langle A_{\bf q}(t+\tau) B_{{\bf q}(\tau)}(t)^{*} D_{{\bf k} = {\bf 0}}(t)^{*} \rangle,
\label{eq:shifted-tcf-11}
\end{eqnarray}
which can be derived in the same manner as Eq.~(\ref{eq:shifted-tcf-2}). 

\subsection{Implication of spatial inversion symmetry}

Let us notice that the SLLOD equations (\ref{eq:SLLOD})
are also invariant under spatial inversion ${\bf \Gamma} \to - {\bf \Gamma}$,
and hence, the $f$- and $p$-Liouvilleans have even parity,
$i {\cal L}^{\dagger}(-{\bf \Gamma}) = i {\cal L}^{\dagger}({\bf \Gamma})$ and
$i {\cal L}(-{\bf \Gamma}) = i {\cal L}({\bf \Gamma})$.
Since $f_{\rm eq}({\bf \Gamma})$, $\sigma_{xy}({\bf \Gamma})$, and
$\delta K({\bf \Gamma})$ also have even parity, so does the 
the nonequilibrium distribution function according to
Eq.~(\ref{eq:neq-distribution}): 
\begin{equation}
f(-{\bf \Gamma},t) = f({\bf \Gamma},t).
\label{eq:neq-distribution-spatial-inversion}
\end{equation}

We next consider how the wave-vector-dependent phase variable 
$A_{\bf q}({\bf \Gamma},t)$ of the form
given in Eq.~(\ref{eq:Aq-def}) transforms under spatial inversion.
It is assumed that $X_{i}^{A_{\bf q}}({\bf \Gamma})$ satisfies
\begin{equation}
X_{i}^{A_{\bf q}}(-{\bf \Gamma}) =
p_{A} X_{i}^{A_{-{\bf q}}}({\bf \Gamma}) =
p_{A} X_{i}^{A_{\bf q}}({\bf \Gamma})^{*},
\end{equation}
where $p_{A}$ denotes the parity of the variable $A$.
Three examples introduced below Eq.~(\ref{eq:Aq-def}) satisfy these relations
with $p_{\rho} = +1$, $p_{j^{\lambda}} = -1$, and $p_{\sigma^{\lambda \mu}} = +1$.
Then, it follows from Eq.~(\ref{eq:Aq-def}) and 
$i {\cal L}(-{\bf \Gamma}) = i {\cal L}({\bf \Gamma})$ that
\begin{equation}
A_{\bf q}(-{\bf \Gamma},t) = p_{A} A_{-{\bf q}}({\bf \Gamma},t) = 
p_{A} A_{\bf q}({\bf \Gamma},t)^{*}.
\label{eq:Aq-spatial-inversion}
\end{equation}

Let us consider an implication of Eqs.~(\ref{eq:neq-distribution-spatial-inversion}) and
(\ref{eq:Aq-spatial-inversion}) for the 
time correlation function $C_{\bf q}^{AB}(t+\tau,t)$ defined in Eq.~(\ref{eq:shifted-tcf-3}).
Since the integral over the phase space must agree for either 
${\bf \Gamma}$ or $-{\bf \Gamma}$, there holds
\begin{eqnarray}
& &
C_{\bf q}^{AB}(t+\tau,t) =
\int d{\bf \Gamma} \,
f({\bf \Gamma},t) A_{\bf q}({\bf \Gamma},\tau) B_{{\bf q}(\tau)}({\bf \Gamma},0)^{*} 
\nonumber \\
& & \,\,\,
=
\int d(-{\bf \Gamma}) \,
f(-{\bf \Gamma},t) A_{\bf q}(-{\bf \Gamma},\tau) B_{{\bf q}(\tau)}(-{\bf \Gamma},0)^{*}.
\end{eqnarray}
Using Eqs.~(\ref{eq:neq-distribution-spatial-inversion}) and
(\ref{eq:Aq-spatial-inversion}) and noting that 
$\int d{\bf \Gamma} \cdots = \int d(-{\bf \Gamma}) \cdots$
[e.g., $\int_{-\infty}^{\infty} dx_{i} \cdots \to \int_{\infty}^{-\infty} d(-x_{i}) \cdots =
\int_{-\infty}^{\infty} dx_{i} \cdots$ under $x_{i} \to - x_{i}$], one finds
\begin{equation}
C_{\bf q}^{AB}(t+\tau,t) =
p_{A} p_{B} C_{\bf q}^{AB}(t+\tau,t)^{*}.
\end{equation}
In particular, {\em autocorrelation} function is real:
\begin{equation}
C_{\bf q}^{AA}(t+\tau,t) =
C_{\bf q}^{AA}(t+\tau,t)^{*}.
\label{eq:reality-neq-tcf}
\end{equation}

\subsection{Steady-state properties}
\label{subsec:TTCF}

Among various stationary-state properties, 
we shall specifically be interested in this paper
in the steady-state shear stress, kinetic 
temperature, and density fluctuations.
Here we summarize the TTCF expressions for these
quantities. 

The steady-state shear stress shall be defined via
\begin{equation}
\sigma_{\rm ss} \equiv
- \langle \sigma_{xy} \rangle_{\rm ss} / V.
\label{eq:ss-shear-def}
\end{equation}
Since the equilibrium ensemble average of $\sigma_{xy}$
is zero, $\langle \sigma_{xy}(0) \rangle = 0$
[see Eq.~(\ref{eq:virial-theorem})],
one obtains from Eq.~(\ref{eq:ss-average})
the following TTCF expression for $\sigma_{\rm ss}$:
\begin{eqnarray}
\sigma_{\rm ss} &=&
\frac{\dot{\gamma}}{k_{\rm B}T V}
\int_{0}^{\infty} ds \,
\langle \sigma_{xy}(s) \sigma_{xy}(0) \rangle 
\nonumber \\
& & \quad
+ \,
\frac{2 \alpha}{k_{\rm B}T V}
\int_{0}^{\infty} ds \,
\langle \sigma_{xy}(s) \delta K(0) \rangle.
\label{eq:ss-shear-TTCF}
\end{eqnarray}

The steady-state temperature shall be defined as
\begin{equation}
T_{\rm ss} \equiv
\frac{2}{3N k_{\rm B}} \langle K \rangle_{\rm ss},
\label{eq:ss-temperature-def}
\end{equation}
in terms of the kinetic energy.
Let us show that $T_{\rm ss}$ is
connected to $\sigma_{\rm ss}$ via a
simple relation.
To this end, we notice that the rate of the 
change of the internal energy,
$H_{0} = K + U$, is given by 
\begin{equation}
\dot{H}_{0} = \sum_{i}
\Bigl[ \,
  \frac{\dot{\bf p}_{i} \cdot {\bf p}_{i}}{m} - 
  \dot{\bf r}_{i} \cdot {\bf F}_{i} \,
\Bigr] =
- \dot{\gamma} \sigma_{xy} - 2 \alpha K,
\label{eq:SLLOD-dot-H0-0}
\end{equation}
where the specific form 
$\kappa_{\lambda \mu} = 
\dot{\gamma} \delta_{\lambda x} \delta_{\mu y}$
for the shear-rate tensor 
and Eq.~(\ref{eq:sigma-def}) for the shear
stress have been used in the final equality.
Since there is no internal-energy change in the
steady state, i.e.,
$\langle \dot{H}_{0} \rangle_{\rm ss} = 0$, 
one finds from 
Eqs.~(\ref{eq:ss-shear-def}), (\ref{eq:ss-temperature-def}),
and (\ref{eq:SLLOD-dot-H0-0}) that
\begin{equation}
T_{\rm ss} =
\frac{\dot{\gamma}}{3 k_{\rm B} \rho \alpha} \sigma_{\rm ss},
\label{eq:ss-temperature-TCF}
\end{equation}
where $\rho = N/V$ denotes the average number density. 
Thus, it suffices to know $\sigma_{\rm ss}$ to obtain 
$T_{\rm ss}$.
This relation can also be used to control $T_{\rm ss}$
by varying the thermostatting multiplier $\alpha$.
However, a self-consistent treatment is necessary in order to set $T_{\rm ss}$
to a desired value, e.g., 
$T_{\rm ss} = T$ which mimics the Gaussian isokinetic
thermostat, since $\sigma_{\rm ss}$
also depends on $\alpha$.

In view of Eqs.~(\ref{eq:ss-tcf-def}) and (\ref{eq:shifted-tcf-3}), 
the steady-state correlator 
$F_{\bf q}^{\rm ss}(t)$ of the density fluctuations
\begin{equation}
\rho_{\bf q} (t)\equiv
\sum_{i} e^{i {\bf q} \cdot {\bf r}_{i}(t)} - N \delta_{{\bf q},{\bf 0}},
\label{eq:rho-def}
\end{equation}
shall be defined via
\begin{equation}
F_{\bf q}^{\rm ss}(t) \equiv
\lim_{s \to \infty}
\frac{1}{N} 
\langle \rho_{\bf q}(s+t) \, 
\rho_{{\bf q}(t)}(s)^{*} \rangle.
\end{equation}
One understands from Eq.~(\ref{eq:reality-neq-tcf}) 
that $F_{\bf q}^{\rm ss}(t)$ is a real function of time. 
From Eq.~(\ref{eq:ss-tcf}), one obtains the
following TTCF expression for $F_{\bf q}^{\rm ss}(t)$ 
\begin{widetext}
\begin{equation}
F_{\bf q}^{\rm ss}(t) = F_{\bf q}(t) - 
\frac{\dot{\gamma}}{N k_{\rm B}T}
\int_{0}^{\infty} ds \,
\langle 
  \rho_{\bf q}(s + t) \rho_{{\bf q}(t)}(s)^{*} \sigma_{xy}(0) 
\rangle -
\frac{2 \alpha}{N k_{\rm B}T}
\int_{0}^{\infty} ds \,
\langle 
  \rho_{\bf q}(s + t) \rho_{{\bf q}(t)}(s)^{*} \delta K(0)
\rangle,
\label{eq:ss-F-TTCF}
\end{equation}
in terms of the transient density correlator
defined by
\begin{equation}
F_{\bf q}(t) \equiv
\frac{1}{N} 
\langle \rho_{\bf q}(t) \, \rho_{{\bf q}(t)}(0)^{*} \rangle,
\label{eq:transient-F-def-0}
\end{equation}
and other transient cross correlators formed with 
$\sigma_{xy}(0)$ and $\delta K(0)$.
As a corollary, one gets
for the steady-state ``static'' or equal-time
structure factor $S_{\bf q}^{\rm ss} \equiv F_{\bf q}^{\rm ss}(t=0)$
\begin{equation}
S_{\bf q}^{\rm ss} = S_{q} -
\frac{\dot{\gamma}}{N k_{\rm B}T}
\int_{0}^{\infty} ds \,
\langle 
  \rho_{\bf q}(s) \rho_{\bf q}(s)^{*} \sigma_{xy}(0) 
\rangle -
\frac{2 \alpha}{N k_{\rm B}T}
\int_{0}^{\infty} ds \,
\langle 
  \rho_{\bf q}(s) \rho_{\bf q}(s)^{*} \delta K(0)
\rangle,
\label{eq:ss-S-TTCF}
\end{equation}
\end{widetext}
where $S_{q} = F_{\bf q}(t=0)$ denotes the equilibrium
static structure factor.
While $S_{q}$ depends only on the wave-vector modulus,
$q = | {\bf q} |$, reflecting the isotropy of the 
initial equilibrium state,
the steady-state structure factor $S_{\bf q}^{\rm ss}$ 
depends also on the direction of the wave vector 
due to the anisotropy of the sheared stationary state. 
It is also clear from Eq.~(\ref{eq:ss-S-TTCF}) that
$S_{\bf q}^{\rm ss}$ should be considered as a {\em dynamic} object
in the sense it is given by the time integrals of the transient time-correlation functions.

Finally, let us show the connection between the TTCF expressions for 
the steady-state shear stress $\sigma_{\rm ss}$ and the structure factor
 $S_{\bf q}^{\rm ss}$ given by 
Eqs.~(\ref{eq:ss-shear-TTCF}) and (\ref{eq:ss-S-TTCF}), respectively.
For this purpose, we first write $\sigma_{xy}(s)$ appearing 
in the integrand of Eq.~(\ref{eq:ss-shear-TTCF})
as [see Eq.~(\ref{eq:appen-sigma-def})]
\begin{widetext}
\begin{eqnarray}
\sigma_{xy}(s) &=& \sum_{i} \frac{p_{i}^{x}(s) p_{i}^{y}(s)}{m} -
\frac{1}{2} \sum_{i \ne j} \frac{r_{ij}^{x}(s) r_{ij}^{y}(s)}{r_{ij}(s)} \phi^{\prime}(r_{ij}(s))
\nonumber \\
&=& 
\sum_{i} \frac{p_{i}^{x}(s) p_{i}^{y}(s)}{m} -
\frac{N}{2} \int d{\bf r} \, \frac{xy}{r} \phi^{\prime}(r)
\int \frac{d{\bf q}}{(2 \pi)^{3}} \,
e^{- i {\bf q} \cdot {\bf r} }
\Bigl[ 
  \frac{1}{N} \rho_{\bf q}(s) \rho_{\bf q}(s)^{*} - 1 
\Bigr],
\end{eqnarray}
where we have used 
$f({\bf r}_{ij}) = \int d{\bf r} \, f({\bf r}) \delta({\bf r} - {\bf r}_{ij})$ and
$\delta( {\bf r} - {\bf r}_{ij}) =
(1/2\pi)^{3} \int d{\bf q} \,
e^{- i {\bf q} \cdot ({\bf r} - {\bf r}_{ij})}$
with ${\bf r}_{ij} = {\bf r}_{i}-{\bf r}_{j}$ 
in the final equality.
Substituting this into Eq.~(\ref{eq:ss-shear-TTCF}), one obtains
\begin{equation}
\sigma_{\rm ss} = - \frac{1}{V} 
\Bigl\langle  
\sum_{i} \frac{p_{i}^{x} p_{i}^{y}}{m}
\Bigr\rangle_{\rm ss} +
\frac{\rho}{2} \int d{\bf r} \, \frac{xy}{r} \phi^{\prime}(r)
\int \frac{d{\bf q}}{(2 \pi)^{3}} \,
e^{- i {\bf q} \cdot {\bf r} }
( S_{\bf q}^{\rm ss} - S_{q} ).
\label{eq:ss-shear-S-TTCF}
\end{equation}
\end{widetext}
Here, the definition (\ref{eq:ss-average}) for the steady-state average has been
exploited for the first term, and Eq.~(\ref{eq:ss-S-TTCF}) for the steady-state
structure factor $S_{\bf q}^{\rm ss}$ has been used for the second term.
Notice that $(S_{\bf q}^{\rm ss} - S_{q})$ in the second term can be
replaced, e.g., by $(S_{\bf q}^{\rm ss} - 1)$, since isotropic terms do not
survive after spatial integral involving the anisotropic term $xy$.
Equation~(\ref{eq:ss-shear-S-TTCF}) simply expresses that anisotropic
density fluctuations are responsible for the steady-state shear stress.
Since the steady-state pair correlation function, $g_{\rm ss}({\bf r})$, is 
related to $S_{\bf q}^{\rm ss}$ via
\begin{equation}
\rho [ g_{\rm ss}({\bf r}) - 1 ] =
\int \frac{d{\bf q}}{(2\pi)^{3}} \,
e^{- i {\bf q} \cdot {\bf r}} (S_{\bf q}^{\rm ss} - 1 ),
\end{equation}
one understands that the interaction part of 
Eq.~(\ref{eq:ss-shear-S-TTCF}) is equivalent to Eq.~(\ref{eq:ss-shear-Kirkwood}). 
Such an equal handling of $\sigma_{\rm ss}$ and $S_{\bf q}^{\rm ss}$
based on the TTCF formalism  
is expected since no approximation has yet been introduced. 

In the following sections, 
we will first derive a set of self-consistent
equations for the transient 
density correlators $F_{\bf q}(t)$
using the projection-operator formalism (Sec.~\ref{sec:GLE})
and the mode-coupling approach (Sec.~\ref{sec:MCT}).
We will then argue that the mentioned TTCF expressions
for the steady-state properties can be
evaluated within the mode-coupling approximation
based on the knowledge of $F_{\bf q}(t)$ 
(Sec.~\ref{sec:steady-state-properties}).
In this way, we construct the nonequilibrium MCT for stationary sheared systems.

\section{Zwanzig-Mori-type equations}
\label{sec:GLE}

In this section, we derive exact Zwanzig-Mori-type
equations of motion for the transient density correlator
$F_{\bf q}(t)$ for a system that is initially at 
equilibrium and subsequently subjected to
stationary shearing along with thermostat. 
A ``standard'' approach~\cite{Hansen86} for a quiescent system is that 
a Zwanzig-Mori equation for a correlator $\langle A(t) A(0)^{*} \rangle$ 
of some phase variable $A$ evolving with 
time-{\em independent} $p$-Liouvillean is derived based on the
{\em static} projection operator onto the subspace spanned by $A$. 
As we will see below, due to the presence of the time-dependent wave-vector advection
${\bf q}(t)$, this standard approach should be appropriately generalized for sheared systems.
We start our discussion by pointing this out. 

\subsection{Difficulties in applying previous formulations}
\label{subsec:GLE-1}

Recently, McPhie {\em et al.}~\cite{McPhie01}
developed a projection-operator formalism
which generalizes the standard approach
to nonequilibrium systems and allows one to derive a 
Zwanzig-Mori-type equation for a {\em transient} correlator
$\langle A(t) A(0)^{*} \rangle$.
Their formalism is also based on the time-independent 
$p$-Liouvillean and on the static projection operator.
Using the convention (\ref{eq:shifted-tcf-4}),
one can introduce the transient density correlator of the form
\begin{equation}
\tilde{F}_{\bf q}(t) =
\frac{1}{N}
\langle \rho_{{\bf q}(-t)}(t) \, \rho_{\bf q}(0)^{*} \rangle.
\end{equation}
Thus, apparently, 
there seems no problem to apply the formalism developed in
Ref.~\cite{McPhie01} by setting
$A(t) = \rho_{{\bf q}(-t)}(t)$.
However, $\rho_{{\bf q}(-t)}(t)$ is {\em not} a phase
variable since its time-evolution
is also affected by the wave-vector advection,
${\bf q}(-t)$,
and its equation of motion cannot be written
solely in terms of the time-independent $p$-Liouvillean as 
in Eq.~(\ref{eq:Lp}).
Their formalism, therefore, cannot be directly applied to 
derive the equation for $\tilde{F}_{\bf q}(t)$. 
Nevertheless, we mention here that 
our equations of motion derived below
resemble those presented in Ref.~\cite{McPhie01} in that
new memory kernels enter in addition to
the one familiar in the equilibrium Zwanzig-Mori equations.

More recently, Fuchs and Cates~\cite{Fuchs05} 
derived the Zwanzig-Mori-type
equation for $F_{\bf q}(t)$,
starting from the Smoluchowski equation
for interacting Brownian particles under stationary shearing. 
It is not difficult, at least formally, to adapt their 
formulation to the SLLOD equations, 
and we briefly summarize here its consequences. 

Because of the equivalence of the particles,
the transient density correlator $F_{\bf q}(t)$ 
defined in Eq.~(\ref{eq:transient-F-def-0})
can be written as
\begin{equation}
F_{\bf q}(t) = 
\frac{1}{N} 
\langle \rho_{{\bf q}(t)}(0)^{*} \rho_{\bf q}(t) \rangle =
\langle 
  \rho_{\bf q}^{s \, *} \,
  e^{- i {\bf q} \cdot \mbox{\boldmath $\kappa$} 
  \cdot {\bf r}_{s} t}
  e^{i {\cal L} t} \, \rho_{\bf q}
\rangle,
\label{eq:FC-dum-1}
\end{equation}
where $\rho_{\bf q}^{s} \equiv e^{i {\bf q} \cdot {\bf r}_{s}}$
denotes the density of a single tagged particle
(labeled $s$), which is identical to the others.
Hereafter, the absence of the argument $t$ implies that
the associated quantities are evaluated at $t=0$.
By this trick of singling out a particle, the motion of the
collective density fluctuations $\rho_{\bf q}$ 
can be described by one, but time-{\em dependent}, 
$p$-Liouvillean $i {\cal L}_{s}(t)$ 
defined via
\begin{equation}
\frac{\partial}{\partial t}
e^{ - i {\bf q} \cdot \mbox{\boldmath $\kappa$} 
    \cdot {\bf r}_{s} t}
e^{i {\cal L} t} \equiv
i {\cal L}_{s}(t) \,
e^{ - i {\bf q} \cdot \mbox{\boldmath $\kappa$} 
    \cdot {\bf r}_{s} t}
e^{i {\cal L} t}.
\label{eq:FC-dum-2}
\end{equation}
Based on the $p$-Liouvillean $i {\cal L}$ for the
SLLOD equations, 
the operator $i {\cal L}_{s}(t)$ 
can be worked out explicitly, 
\begin{equation}
i {\cal L}_{s} (t) =
i {\cal L} - i {\bf q} \cdot \mbox{\boldmath $\kappa$} 
\cdot {\bf r}_{s} +
i {\bf q} \cdot \mbox{\boldmath $\kappa$} 
\cdot ({\bf p}_{s}/m) t.
\label{eq:FC-dum-3}
\end{equation}
Integrating Eq.~(\ref{eq:FC-dum-2}) in time, one obtains
\begin{equation}
F_{\bf q}(t) = 
\Bigl\langle 
  \rho_{\bf q}^{s \, *}
  e_{+}^{\int_{0}^{t} d\tau \, i {\cal L}_{s}(\tau)} 
  \rho_{\bf q}
\Bigr\rangle.
\label{eq:FC-dum-4}
\end{equation}
Here $e_{+}$ denotes the time-ordered exponential,
where earlier times appear on the right.
This expression also explains why the formalism
developed in Ref.~\cite{McPhie01}, which is based on the
time-{\em independent} $p$-Liouvillean, cannot 
deal with $F_{\bf q}(t)$.

Equation~(\ref{eq:FC-dum-4}) can be handled by manipulations
based on the static projection operator
${\cal P}_{s} = \rho_{\bf q} \rangle
(1/S_{q}) \langle \rho_{\bf q}^{s \, *}$,
and one can derive the following exact Zwanzig-Mori-type
equation of motion for $F_{\bf q}(t)$
in the same manner as detailed in Ref.~\cite{Fuchs05}:
\begin{subequations}
\label{eq:FC-Zwanzig-Mori}
\begin{equation}
\frac{\partial}{\partial t} F_{\bf q}(t) -
\frac{1}{S_{q}}
\Bigl[
  {\bf q} \cdot \mbox{\boldmath $\kappa$} \cdot
  \frac{\partial}{\partial {\bf q}} S_{q}
\Bigr] 
F_{\bf q}(t) +
\int_{0}^{t} ds \,
K_{\bf q}(t,s) 
F_{\bf q}(s) = 0.
\label{eq:FC-dum-11}
\end{equation}
Here the memory kernel is given by
\begin{eqnarray}
K_{\bf q}(t,t^{\prime}) &=&
- \frac{1}{S_{q}}
\Bigl\langle 
\rho_{\bf q}^{s \, *}
i {\cal L}_{s}(t) {\cal Q}_{s} 
\nonumber \\
& & 
\times \,
e_{+}^{\int_{t^{\prime}}^{t} d\tau \, 
i {\cal Q}_{s} {\cal L}_{s}(\tau)  {\cal Q}_{s} } \,
{\cal Q}_{s} i {\cal L}_{s}(t^{\prime}) \rho_{\bf q}
\Bigr\rangle,
\label{eq:FC-dum-12}
\end{eqnarray}
\end{subequations}
in which ${\cal Q}_{s} \equiv I - {\cal P}_{s}$
with $I$ denoting the identity operator.
Equations~(\ref{eq:FC-Zwanzig-Mori})
serve as the starting equations for
Brownian particles exhibiting overdamped dynamics, 
since in this case
the velocity entering into
${\cal Q}_{s} i {\cal L}_{s}(t) \rho_{\bf q}$
is proportional to the force, and hence,
$K(t,t')$ essentially describes the 
fluctuating-force correlations.
Such an incorporation of the fluctuating-force correlations
is essential in developing self-consistent equations 
for $F_{\bf q}(t)$. 

On the other hand, 
we need an additional Zwanzig-Mori-type
equation for $K(t,t')$ in constructing a self-consistent theory 
since the time derivative of the (peculiar) momentum is 
proportional to the force in
the SLLOD equations (\ref{eq:SLLOD}).
For this purpose, one needs to introduce a 
time-{\em dependent} projection operator
onto the subspace spanned by
${\cal Q}_{s} i {\cal L}_{s}(t) \rho_{\bf q}$.
We found that the resulting equation of motion
for $K(t,t')$ is too cumbersome to be adopted 
as our starting equation.
(See, e.g., Ref.~\cite{Latz-all}
for the application of the time-dependent
projection-operator formalism.) 

Here we shall take an alternative route.
Adopting the original definition
\begin{equation}
F_{\bf q}(t) = 
\frac{1}{N} 
\langle 
\rho_{\bf q}(t) \,
\rho_{{\bf q}(t)}(0)^{*}
\rangle =
\frac{1}{N}
\langle 
[e^{i {\cal L} t} \rho_{\bf q}] \,
\rho_{{\bf q}(t)}^{*}
\rangle,
\label{eq:F-tr-def}
\end{equation}
of the transient density correlator, we will first derive
an exact continuity equation which relates $F_{\bf q}(t)$
to the transient cross correlator
\begin{equation}
H_{\bf q}^{\lambda}(t) = 
\frac{1}{N} 
\langle 
j_{\bf q}^{\lambda}(t) \,
\rho_{{\bf q}(t)}(0)^{*}
\rangle =
\frac{1}{N}
\langle 
[e^{i {\cal L} t} j_{\bf q}^{\lambda}] \,
\rho_{{\bf q}(t)}^{*}
\rangle,
\label{eq:H-tr-def}
\end{equation}
for $\lambda = x, y, z$, 
formed with the current density fluctuations $j_{\bf q}^{\lambda}$ defined by
\begin{equation}
j_{\bf q}^{\lambda} = \sum_{i} 
\frac{p_{i}^{\lambda}}{m} 
e^{i {\bf q} \cdot {\bf r}_{i}}.
\label{eq:j-def}
\end{equation}
Here it is necessary to take into account
all the $\lambda$ components of $j_{\bf q}^{\lambda}$ 
due to the anisotropic nature of the sheared system.
We will then derive 
a Zwanzig-Mori-type equation of motion for 
$H_{\bf q}^{\lambda}(t)$, which can be done via 
a partial use of the static projection operator
as we will see below. 

\subsection{Continuity equation}

We start with the time-evolution equation for the number density
fluctuations. 
Since [see Eqs.~(\ref{eq:iL-SLLOD})]
\begin{eqnarray}
i {\cal L} \rho_{\bf q} &=&
\big( i {\cal L}_{0} +
i {\cal L}_{\dot{\gamma}} +
i {\cal L}_{\alpha} \big) \rho_{\bf q}
\nonumber \\
&=&
i {\bf q} \cdot {\bf j}_{\bf q} +
{\bf q}\cdot \mbox{\boldmath $\kappa$} \cdot \frac{\partial}{\partial {\bf q}} 
\rho_{\bf q},
\label{eq:GLE-tr-dum-02}
\end{eqnarray}
one finds the following continuity equation for the sheared system
relating the partial time derivative of $\rho_{\bf q}(t) = e^{i {\cal L} t} \rho_{\bf q}$
to $j_{\bf q}^{\lambda}(t) = e^{i {\cal L} t} j_{\bf q}^{\lambda}$: 
\begin{eqnarray}
\Bigl[ 
  \frac{\partial}{\partial t} - 
  {\bf q} \cdot \mbox{\boldmath $\kappa$} \cdot \frac{\partial}{\partial {\bf q}} 
\Bigr]
\rho_{\bf q}(t) = i{\bf q}\cdot {\bf j}_{\bf q}(t).
\label{eq:BK-dum-12}
\end{eqnarray}
Likewise the partial time derivative of the density fluctuation at the advected
wave vector, $\rho_{{\bf q}(t)}^{*}$, is given by
\begin{equation}
\Bigl[ 
  \frac{\partial}{\partial t} - 
  {\bf q} \cdot \mbox{\boldmath $\kappa$} \cdot\frac{\partial}{\partial {\bf q}} 
\Bigr]
\rho_{{\bf q}(t)}^{*}  = 0,
\label{eq:GLE-tr-new-dum-11}
\end{equation}
in deriving which we have noticed that
the shear-rate tensor satisfies 
$\mbox{\boldmath $\kappa$} \cdot \mbox{\boldmath $\kappa$} = 0$.
One can readily obtain from the above two equations that the transient density correlator 
$F_{\bf q}(t) = \langle \rho_{\bf q}(t) \rho^*_{{\bf q}(t)}\rangle/N$ and
the transient cross correlator 
$H_{\bf q}^{\lambda}(t) = \langle j_{\bf q}^{\lambda}(t) \rho^*_{{\bf q}(t)}\rangle/N$ 
is connected via
\begin{equation}
\Bigl[ 
  \frac{\partial}{\partial t} - 
  {\bf q} \cdot \mbox{\boldmath $\kappa$} \cdot \frac{\partial}{\partial {\bf q}} 
\Bigr]
F_{\bf q}(t)  = i {\bf q} \cdot {\bf H}_{\bf q}(t).
\label{eq:continuity-equation-tr}
\end{equation}

\subsection{Exact equation for the transient cross correlator}

We next derive an exact equation of motion for the transient cross correlator
$H_{\bf q}^{\lambda}(t)$. 
We start from 
\begin{eqnarray}
i {\cal L} j_{\bf q}^{\lambda} &=&
\big( i {\cal L}_{0} +
i {\cal L}_{\dot{\gamma}} +
i {\cal L}_{\alpha} \big) j_{\bf q}^{\lambda} 
\nonumber \\
&=&
i {\cal L}_{0} j_{\bf q}^{\lambda} +
{\bf q}\cdot \mbox{\boldmath $\kappa$} \cdot \frac{\partial}{\partial {\bf q}} 
j_{\bf q}^{\lambda} -
( \mbox{\boldmath $\kappa$} \cdot {\bf j}_{\bf q} )^{\lambda} -
\alpha j_{\bf q}^{\lambda}.
\label{eq:GLE-tr-dum-12}
\end{eqnarray}
One therefore gets for $j_{\bf q}^{\lambda}(t) = e^{i {\cal L} t} j_{\bf q}$
\begin{equation}
\Bigl[ 
  \frac{\partial}{\partial t} - 
  {\bf q} \cdot \mbox{\boldmath $\kappa$} \cdot \frac{\partial}{\partial {\bf q}} 
\Bigr]
j_{\bf q}^{\lambda}(t) =
e^{i {\cal L}t} i {\cal L}_{0} j_{\bf q}^{\lambda} -
[ \mbox{\boldmath $\kappa$} \cdot {\bf j}_{\bf q}(t) ]^{\lambda} -
\alpha j_{\bf q}^{\lambda}(t).
\end{equation}
It is straightforward to obtain from this equation and Eq.~(\ref{eq:GLE-tr-new-dum-11})
for the correlator
$H_{\bf q}^{\lambda}(t) = \langle j_{\bf q}^{\lambda}(t) \rho^*_{{\bf q}(t)}\rangle/N$ 
\begin{eqnarray}
& &
\Bigl[ \,
  \frac{\partial}{\partial t} -
  {\bf q} \cdot \mbox{\boldmath $\kappa$} \cdot
  \frac{\partial}{\partial {\bf q}} \,
\Bigr] H_{\bf q}^{\lambda}(t) =
\frac{1}{N}
\langle [ e^{i {\cal L} t} i {\cal L}_{0} j_{\bf q}^{\lambda} ] \,
\rho_{{\bf q}(t)}^{*}
\rangle 
\nonumber \\
& & \qquad \qquad \qquad \qquad \quad 
- \,
[\mbox{\boldmath $\kappa$} \cdot {\bf H}_{\bf q}(t)]^{\lambda} -
\alpha H_{\bf q}^{\lambda}(t).
\label{eq:GLE-Hq-lambda-tr-1}
\end{eqnarray}
It is already clear at this point that
one cannot derive a closed equation for 
the ``longitudinal'' component 
${\bf q} \cdot {\bf H}_{\bf q}(t)$ alone 
due to the presence of the second term on the right-hand side of 
Eq.~(\ref{eq:GLE-Hq-lambda-tr-1}).
This reflects the anisotropic nature of the sheared system.
We also notice that the thermostatting multiplier
$\alpha$ can be taken outside of
the ensemble average in the last term of Eq.~(\ref{eq:GLE-Hq-lambda-tr-1})
since we have adopted
the constant-$\alpha$ model. 
If, for example, the Gaussian isokinetic multiplier $\alpha_{\rm G}$ 
were used [see Eq.~(\ref{eq:alpha-Gaussian})],
then one would have to consider an 
additional equation of motion for
$(1/N) \langle [ e^{i {\cal L} t} \alpha_{\rm G} j_{\bf q}^{\lambda} ] \,
\rho_{{\bf q}(t)}^{*} \rangle$.
Thus, a considerable simplification is achieved via the
adoption of the constant-$\alpha$ model.

\subsection{Projection-operator formalism}
\label{subsec:projection-operator-formalism}

In the following, we shall apply a 
projection-operator formalism, but {\em only} to 
the first term on the
right-hand side of Eq.~(\ref{eq:GLE-Hq-lambda-tr-1}).
As will be shown below, 
this can be done via a static projection operator, 
and thereby the aforementioned difficulty connected with
Eq.~(\ref{eq:FC-dum-12}) can be avoided.
In this way, we complete the derivation of the Zwanzig-Mori-type
equation of motion for $H_{\bf q}^{\lambda}(t)$, which
together with the continuity equation (\ref{eq:continuity-equation-tr})
provides our starting equations for developing a nonequilibrium
MCT for transient density correlators. 

For this purpose, 
let us introduce the static projection operator
${\cal P}$ onto the subspace spanned by
$\rho_{\bf k}$ and $j_{\bf k}^{\mu}$ $(\mu = x, y, z)$.
Since 
$\langle \rho_{\bf k} \, \rho_{{\bf k}'}^{*} \rangle = 
\delta_{{\bf k}, {\bf k}'} NS_{k}$,
$\langle j_{\bf k}^{\lambda} \, j_{{\bf k}'}^{\mu \, *} \rangle =
\delta_{{\bf k}, {\bf k}'} \delta_{\lambda \mu} N v^{2}$
with $v = \sqrt{k_{\rm B}T/m}$ denoting the thermal velocity, and
$\langle \rho_{\bf k} \, j_{{\bf k}'}^{\mu \, *} \rangle = 0$
(remember that the averaging is over the initial canonical
distribution), 
the projection operator ${\cal P}$ is given by
\begin{equation}
{\cal P} X =
\sum_{\bf k}
\langle X \rho_{\bf k}^{*} \rangle
\frac{1}{N S_{k}} \rho_{\bf k} +
\sum_{\bf k}
\sum_{\mu}
\langle X j_{\bf k}^{\mu \, *} \rangle
\frac{1}{N v^{2}} j_{\bf k}^{\mu}.
\label{eq:P-tr-def}
\end{equation}
The complementary projection operator is defined by
${\cal Q} \equiv I - {\cal P}$.
One can easily show that ${\cal P}$ and ${\cal Q}$ are 
idempotent and Hermitian.

The time evolution of $i {\cal L}_{0} j_{\bf q}^{\lambda}$
appearing in the first term on the right-hand side
of Eq.~(\ref{eq:GLE-Hq-lambda-tr-1})
shall then be separated into parts
correlated and uncorrelated with
$\{ \rho_{\bf k}, j_{\bf k}^{\mu} \}$:
\begin{equation}
e^{i {\cal L} t} i {\cal L}_{0} j_{\bf q}^{\lambda} =
e^{i {\cal L} t} {\cal P} i {\cal L}_{0} j_{\bf q}^{\lambda} +
e^{i {\cal L} t} {\cal Q} i {\cal L}_{0} j_{\bf q}^{\lambda}.
\label{eq:GLE-tr-dum-21}
\end{equation}
As derived in Appendix~\ref{appen:GLE-tr-dum-27},
one obtains
\begin{equation}
{\cal P} i {\cal L}_{0} j_{\bf q}^{\lambda} =
i q_{\lambda} \frac{v^{2}}{S_{q}} \rho_{\bf q},
\label{eq:GLE-tr-dum-27}
\end{equation}
and hence, 
the first term on the right-hand side of Eq.~(\ref{eq:GLE-tr-dum-21})
is given by
\begin{equation}
e^{i {\cal L} t} 
{\cal P} i {\cal L}_{0} j_{\bf q}^{\lambda} =
i q_{\lambda} \frac{v^{2}}{S_{q}} 
e^{i {\cal L} t} 
\rho_{\bf q}.
\label{eq:GLE-tr-dum-28}
\end{equation}

For the second term on the right-hand side of 
Eq.~(\ref{eq:GLE-tr-dum-21}), we use the identity
\begin{equation}
e^{i {\cal L} t} =
e^{{\cal Q} i {\cal L} t} +
\int_{0}^{t} ds \, e^{i {\cal L} (t-s)}
{\cal P} i {\cal L} e^{{\cal Q} i {\cal L} s}
\label{eq:operator-identity-1}
\end{equation}
to obtain
\begin{eqnarray}
& &
e^{i {\cal L} t} {\cal Q} i {\cal L}_{0} j_{\bf q}^{\lambda} =
e^{i {\cal QLQ} t} 
{\cal Q} i {\cal L}_{0} j_{\bf q}^{\lambda} 
\nonumber \\
& & \qquad 
+ \,
\int_{0}^{t} ds \, e^{i {\cal L} (t-s)}
{\cal P} i {\cal L} e^{i {\cal QLQ} s}
{\cal Q} i {\cal L}_{0} j_{\bf q}^{\lambda}.
\label{eq:GLE-tr-dum-31}
\end{eqnarray}
In the right-hand side of this equation, 
we have used 
$e^{{\cal Q} i {\cal L} t} {\cal Q} = e^{i {\cal QLQ} t} {\cal Q}$  
which holds due to the idempotency of the operator ${\cal Q}$. 
Let us introduce
\begin{equation}
R_{\bf q}^{\lambda}(t) \equiv 
e^{i {\cal QLQ} t} R_{\bf q}^{\lambda},
\label{eq:R-tr-def-1}
\end{equation}
with
\begin{equation}
R_{\bf q}^{\lambda} \equiv {\cal Q} i {\cal L}_{0} j_{\bf q}^{\lambda} =
i {\cal L}_{0} j_{\bf q}^{\lambda} - i q_{\lambda}
\frac{v^{2}}{S_{q}} \rho_{\bf q},
\label{eq:R-tr-def-2}
\end{equation}
where we have used Eq.~(\ref{eq:GLE-tr-dum-27}).
Since 
${\cal Q} \rho_{{\bf q}(t)}^{*} = {\cal Q} j_{{\bf q}(t)}^{\mu \, *} = 0$,
there holds
\begin{equation}
\langle R_{\bf q}^{\lambda}(t) \rho_{{\bf q}(t)}^{*} \rangle = 0
\quad \mbox{ and } \quad
\langle R_{\bf q}^{\lambda}(t) j_{{\bf q}(t)}^{\mu \, *} \rangle = 0.
\label{eq:GLE-tr-dum-32}
\end{equation}
Thus, $R_{\bf q}^{\lambda}(t)$ 
is always uncorrelated with
$\{ \rho_{\bf k}, j_{\bf k}^{\mu} \}$,
and we follow the usual convention to call this phase
variable the random or fluctuating force.

In terms of the fluctuating force $R_{\bf q}^{\lambda}(t)$,
the second term in Eq.~(\ref{eq:GLE-tr-dum-31})
can be expressed as
\begin{widetext}
\begin{eqnarray}
& &
\int_{0}^{t} ds \,
e^{i {\cal L} (t-s)} {\cal P} i {\cal L}
R_{\bf q}^{\lambda}(s) =
\int_{0}^{t} ds \,
\sum_{\bf k} 
\langle 
  [ i {\cal L} R_{\bf q}^{\lambda}(s) ] \, \rho_{\bf k}^{*}
\rangle
\frac{1}{N S_{k}}
e^{i {\cal L} (t-s)} \rho_{\bf k} +
\int_{0}^{t} ds \,
\sum_{\bf k} \sum_{\mu}
\langle 
  [ i {\cal L} R_{\bf q}^{\lambda}(s) ] \, j_{\bf k}^{\mu \, *}
\rangle
\frac{1}{N v^{2}}
e^{i {\cal L} (t-s)} j_{\bf k}^{\mu}
\nonumber \\
& & \qquad \qquad \qquad \quad 
=
\int_{0}^{t} ds \,
\langle 
  [ i {\cal L} R_{\bf q}^{\lambda}(s) ] \, \rho_{{\bf q}(s)}^{*} 
\rangle
\frac{1}{N S_{q(s)}} 
e^{i {\cal L} (t-s)} \rho_{{\bf q}(s)} +
\sum_{\mu} 
\int_{0}^{t} ds \,
\langle 
  [ i {\cal L} R_{\bf q}^{\lambda}(s) ] \, 
  j_{{\bf q}(s)}^{\mu \, *}
\rangle
\frac{1}{N v^{2}}
e^{i {\cal L} (t-s)} j_{{\bf q}(s)}^{\mu},
\label{eq:GLE-tr-dum-33}
\end{eqnarray}
where the last equality holds since
$\langle R_{\bf q}^{\lambda}(s) f_{\bf k}^{*} \rangle$ is nonzero
only for ${\bf k} = {\bf q}(s)$
[see Eq.~(\ref{eq:shifted-tcf-2})].
We also noticed that the 
ensemble averaged terms are independent of the phase
and are unaffected by the propagator.
The evaluation of the ensemble averaged terms 
in the integrands 
of Eq.~(\ref{eq:GLE-tr-dum-33}) is presented in 
Appendix~\ref{appen:GLE-tr-dum-45-and-GLE-tr-dum-57},
and the results are given by
\begin{eqnarray}
\langle
[ i {\cal L} R_{\bf q}^{\lambda}(s) ] \, \rho_{{\bf q}(s)}^{*} 
\rangle &=&
- \frac{\dot{\gamma}}{k_{\rm B}T}
\langle
R_{\bf q}^{\lambda}(s) \,
{\cal Q} [\rho_{{\bf q}(s)}^{*} \sigma_{xy}] 
\rangle 
- \frac{2 \alpha}{k_{\rm B}T}
\langle
R_{\bf q}^{\lambda}(s) \,
{\cal Q} [\rho_{{\bf q}(s)}^{*} \delta K]
\rangle,
\label{eq:GLE-tr-dum-45}
\\
\langle 
[ i {\cal L} R_{\bf q}^{\lambda}(s) ] \, j_{{\bf q}(s)}^{\mu \, *}
\rangle &=&
- \langle 
R_{\bf q}^{\lambda}(s) \, R_{{\bf q}(s)}^{\mu \, *} 
\rangle -
\frac{\dot{\gamma}}{k_{\rm B}T}
\langle 
R_{\bf q}^{\lambda}(s) \,
{\cal Q} [j_{{\bf q}(s)}^{\mu \, *} \sigma_{xy}] 
\rangle -
\frac{2 \alpha}{k_{\rm B}T}
\langle 
R_{\bf q}^{\lambda}(s) \,
{\cal Q} [j_{{\bf q}(s)}^{\mu \, *} \delta K]
\rangle.
\label{eq:GLE-tr-dum-57}
\end{eqnarray}
Let us notice that 
Eq.~(\ref{eq:GLE-tr-dum-57}) has been simplified
due to the adoption of the
constant-$\alpha$ model for the thermostat
[see the comment below Eq.~(\ref{eq:GLE-tr-dum-52})]:
otherwise, e.g., when the Gaussian isokinetic thermostat is used,
one has to add a term
$\langle R_{\bf q}^{\lambda}(s) \, 
{\cal Q} [ \alpha_{\rm G} j_{{\bf q}(s)}^{\mu \, *} ] \rangle$
to the right-hand side of Eq.~(\ref{eq:GLE-tr-dum-57}).

With Eqs.~(\ref{eq:GLE-tr-dum-31})--(\ref{eq:R-tr-def-2}) and
(\ref{eq:GLE-tr-dum-33})--(\ref{eq:GLE-tr-dum-57}), we now obtain
\begin{eqnarray}
& &
e^{i {\cal L} t} {\cal Q} i {\cal L}_{0} j_{\bf q}^{\lambda} =
R_{\bf q}^{\lambda}(t) -
\sum_{\mu}
\int_{0}^{t} ds \,
M_{\bf q}^{\lambda \mu}(s) \,
e^{i {\cal L} (t-s)} j_{{\bf q}(s)}^{\mu} +
\dot{\gamma}
\int_{0}^{t} ds \,
i L_{\bf q}^{\lambda}(s) \,
e^{i {\cal L} (t-s)} \rho_{{\bf q}(s)} 
\nonumber \\
& & \quad 
- \,
\dot{\gamma} \sum_{\mu}
\int_{0}^{t} ds \,
L_{\bf q}^{\prime \, \lambda \mu}(s) \,
e^{i {\cal L} (t-s)} j_{{\bf q}(s)}^{\mu} +
\alpha
\int_{0}^{t} ds \,
i N_{\bf q}^{\lambda}(s) \,
e^{i {\cal L} (t-s)} \rho_{{\bf q}(s)} -
\alpha \sum_{\mu}
\int_{0}^{t} ds \,
N_{\bf q}^{\prime \, \lambda \mu}(s) \,
e^{i {\cal L} (t-s)} j_{{\bf q}(s)}^{\mu}.
\label{eq:GLE-tr-dum-61}
\end{eqnarray}
Here we have introduced the following memory kernels: 
\begin{eqnarray}
& &
M_{\bf q}^{\lambda \mu}(t) \equiv
\frac{1}{N v^{2}}
\langle R_{\bf q}^{\lambda}(t) \, R_{{\bf q}(t)}^{\mu \, *} \rangle,
\label{eq:M-tr-def}
\\
& &
L_{\bf q}^{\lambda}(t) \equiv
i \frac{1}{N k_{\rm B} T S_{q(t)}}
\langle R_{\bf q}^{\lambda}(t) \, 
{\cal Q} [ \rho_{{\bf q}(t)}^{*} \sigma_{xy} ] \rangle,
\label{eq:L-tr-def}
\\
& &
L_{\bf q}^{\prime \, \lambda \mu}(t)  \equiv
\frac{m}{N(k_{\rm B}T)^{2}}
\langle R_{\bf q}^{\lambda}(t) \, 
{\cal Q} [ j_{{\bf q}(t)}^{\mu \, *} \sigma_{xy} ] \rangle.
\label{eq:L-prime-tr-def}
\\
& &
N_{\bf q}^{\lambda}(t)  \equiv
i \frac{2}{N k_{\rm B} T S_{q(t)}}
\langle R_{\bf q}^{\lambda}(t) \, 
{\cal Q} [ \rho_{{\bf q}(t)}^{*} \delta K ] \rangle,
\label{eq:N-tr-def}
\\
& &
N_{\bf q}^{\prime \, \lambda \mu}(t)  \equiv
\frac{2m}{N(k_{\rm B}T)^{2}}
\langle R_{\bf q}^{\lambda}(t) \, 
{\cal Q} [ j_{{\bf q}(t)}^{\mu \, *} \delta K ] \rangle.
\label{eq:N-prime-tr-def}
\end{eqnarray}
Substituting 
Eqs.~(\ref{eq:GLE-tr-dum-21}),
(\ref{eq:GLE-tr-dum-28}), and
(\ref{eq:GLE-tr-dum-61}) along with
Eq.~(\ref{eq:GLE-tr-dum-32}) 
into the first term on the right-hand side of 
Eq.~(\ref{eq:GLE-Hq-lambda-tr-1}), we finally obtain
the following Zwanzig-Mori-type equation for 
$H_{\bf q}^{\lambda}(t)$:
\begin{eqnarray}
\Bigl[ \,
  \frac{\partial}{\partial t} -
  {\bf q} \cdot \mbox{\boldmath $\kappa$} \cdot
  \frac{\partial}{\partial {\bf q}} \,
\Bigr] H_{\bf q}^{\lambda}(t) &=&
i q_{\lambda} \frac{v^{2}}{S_{q}} F_{\bf q}(t) -
[\mbox{\boldmath $\kappa$} \cdot {\bf H}_{\bf q}(t)]^{\lambda} -
\alpha H_{\bf q}^{\lambda}(t) -
\sum_{\mu}
\int_{0}^{t} ds \,
M_{\bf q}^{\lambda \mu}(s) \,
H_{{\bf q}(s)}^{\mu}(t-s)
\nonumber \\
& & \quad
+ \,
\dot{\gamma}
\int_{0}^{t} ds \,
i L_{\bf q}^{\lambda}(s) \,
F_{{\bf q}(s)}(t-s) -
\dot{\gamma} \sum_{\mu}
\int_{0}^{t} ds \,
L_{\bf q}^{\prime \, \lambda \mu}(s) \,
H_{{\bf q}(s)}^{\mu}(t-s)
\nonumber \\
& & \qquad
+ \,
\alpha
\int_{0}^{t} ds \,
i N_{\bf q}^{\lambda}(s) \,
F_{{\bf q}(s)}(t-s) -
\alpha \sum_{\mu}
\int_{0}^{t} ds \,
N_{\bf q}^{\prime \, \lambda \mu}(s) \,
H_{{\bf q}(s)}^{\mu}(t-s).
\label{eq:GLE-Hq-lambda-tr}
\end{eqnarray}
\end{widetext}
Here, we have noticed 
$(1/N)
\langle [ e^{i {\cal L} (t-s)} \rho_{{\bf q}(s)} ] \,
\rho_{{\bf q}(t)}^{*} \rangle =
F_{{\bf q}(s)}(t-s)$ and
$(1/N)
\langle [ e^{i {\cal L} (t-s)} j_{{\bf q}(s)}^{\mu} ] \,
\rho_{{\bf q}(t)}^{*} \rangle =
H_{{\bf q}(s)}^{\mu}(t-s)$.
One can easily confirm that these are
consistent with the 
definitions (\ref{eq:F-tr-def}) and (\ref{eq:H-tr-def}).

The memory kernel $M_{\bf q}^{\lambda \mu}(t)$
describing the fluctuating force correlations is already
familiar from the equilibrium 
Zwanzig-Mori equation of motion for the density correlator~\cite{Goetze91b}.
The additional memory kernels $L_{\bf q}^{\lambda}(t)$ and 
$L_{\bf q}^{\prime \, \lambda \mu}(t)$
are due to couplings between the fluctuating force and the shear stress, and
$N_{\bf q}^{\lambda}(t)$ and 
$N_{\bf q}^{\prime \, \lambda \mu}(t)$
are associated with couplings between
the fluctuating force and temperature fluctuations.
In the following section, we introduce mode-coupling
approximations for these memory kernels to obtain
a set of self-consistent equations of motion
for the transient density correlators.

\section{Mode-coupling approximation}
\label{sec:MCT}

We have encountered five memory kernels in the
Zwanzig-Mori-type exact equations of motion for the
transient correlators. 
We need to invoke approximations for these memory kernels 
in order to obtain closed equations for $F_{\bf q}(t)$. 
In this section, we apply the mode-coupling approximations~\cite{Goetze91b}
to these memory kernels.

The basic idea behind MCT is that the
fluctuation of a given dynamical variable decays,
at intermediate and long times, predominantly
into pairs of hydrodynamic modes associated with
quasi-conserved dynamical variables.
It is therefore reasonable to expect that the decay of the memory
kernels at intermediate and long times is dominated
by those mode correlations which have the longest
relaxation times.
The sluggishness of the structural relaxation processes
in glass-forming systems suggests that the slow decay
of the memory kernels is basically due to
couplings to pair-density modes.
The simplest way to extract such a 
slowly decaying part is to introduce another projection
operator ${\cal P}_{2}$ 
which projects any variable onto the subspace
spanned by $\rho_{\bf k} \rho_{\bf p}$, i.e.,
\begin{equation}
{\cal P}_{2} X =
\sum_{{\bf k} > {\bf p}}
\langle X \rho_{\bf k}^{*} \rho_{\bf p}^{*} \rangle
\frac{1}{N^{2} S_{k} S_{p}}
\rho_{\bf k} \rho_{\bf p}.
\label{eq:P2-def}
\end{equation}
Here we already used
the static version of the
factorization approximation introduced below
[see Eq.~(\ref{eq:factorization-app})].
It is readily verified that ${\cal P}_{2}$ is
idempotent and Hermitian.

The first approximation in the mode-coupling approach 
thus corresponds to replacing the propagator 
$e^{i {\cal QLQ} t}$ governing the time-evolution of the 
memory kernels by its projection on the
subspace spanned by the pair-density modes,
$e^{i {\cal QLQ} t} \approx 
{\cal P}_{2} e^{i {\cal QLQ} t} {\cal P}_{2}$.
Under this approximation, the memory kernel 
$M_{\bf q}^{\lambda \mu}(t)$ defined in Eq.~(\ref{eq:M-tr-def})
is given by
\begin{eqnarray}
M_{\bf q}^{\lambda \mu}(t) &\approx&
\frac{1}{Nv^{2}}
\langle 
  [ {\cal P}_{2} e^{i {\cal QLQ} t} {\cal P}_{2}
  R_{\bf q}^{\lambda} ] \, 
  R_{{\bf q}(t)}^{\mu \, *}
\rangle 
\nonumber \\
&=&
\frac{1}{Nv^{2}}
\langle 
  [ e^{i {\cal QLQ} t} {\cal P}_{2}
  R_{\bf q}^{\lambda} ] \, 
  {\cal P}_{2} R_{{\bf q}(t)}^{\mu \, *}
\label{eq:MCT-tr-dum-21}
\rangle.
\end{eqnarray}
The expression for the projected
random force ${\cal P}_{2} R_{\bf q}^{\lambda}$ 
is derived in Appendix~\ref{appen:MCT-tr-dum-22}
within the convolution approximation 
for triple correlations,
\begin{equation}
\langle
\rho_{\bf q} \rho_{\bf k}^{*} \rho_{\bf p}^{*}
\rangle \approx
\delta_{{\bf q}, {\bf k} + {\bf p}} \, 
N S_{q} S_{k} S_{p},
\label{eq:convolution-app}
\end{equation}
and is given by
\begin{equation}
{\cal P}_{2} R_{\bf q}^{\lambda} =
-i \frac{\rho v^{2}}{N}
\sum_{{\bf k} > {\bf p}}
\delta_{{\bf q}, {\bf k}+{\bf p}}
[ k_{\lambda} c_{k} + p_{\lambda} c_{p} ]
\rho_{\bf k} \rho_{\bf p}.
\label{eq:MCT-tr-dum-22}
\end{equation}
Here $c_{q}$ is the direct correlation function 
defined via
\begin{equation}
\rho c_{q} = 1 - \frac{1}{S_{q}}.
\label{eq:c-def}
\end{equation}
Substituting Eq.~(\ref{eq:MCT-tr-dum-22}) 
into Eq.~(\ref{eq:MCT-tr-dum-21}), we obtain
\begin{eqnarray}
M_{\bf q}^{\lambda \mu}(t) &=&
\frac{\rho^{2} v^{2}}{N^{3}}
\sum_{{\bf k} > {\bf p}} \sum_{{\bf k}^{\prime} > {\bf p}^{\prime}}
\delta_{{\bf q}, {\bf k}+{\bf p}} \,
\delta_{{\bf q}(t), {\bf k}^{\prime}+{\bf p}^{\prime}} 
[ k_{\lambda} c_{k} + p_{\lambda} c_{p} ]
\nonumber \\
& & \times \,
[ k^{\prime}_{\mu} c_{k^{\prime}} + 
  p^{\prime}_{\mu} c_{p^{\prime}} ]
\langle
  [ e^{i {\cal QLQ} t} \rho_{\bf k} \rho_{\bf p} ] 
  \rho_{{\bf k}^{\prime}}^{*} \rho_{{\bf p}^{\prime}}^{*}
\rangle.
\label{eq:MCT-tr-dum-23}
\end{eqnarray}
The final approximation in the mode-coupling approach 
is to factorize averages of products,
evolving in time with the propagator $e^{i {\cal QLQ} t}$, 
into products of averages formed with the
variables evolving with $e^{i {\cal L} t}$
(factorization approximation):
\begin{eqnarray}
& &
\langle
  [ e^{i {\cal QLQ} t} \rho_{\bf k} \rho_{\bf p} ] \,
  \rho_{{\bf k}^{\prime}}^{*} \rho_{{\bf p}^{\prime}}^{*}
\rangle \approx
\langle
  [ e^{i {\cal L} t} \rho_{\bf k} ] \, \rho_{{\bf k}^{\prime}}^*
\rangle \, 
\langle
  [ e^{i {\cal L} t} \rho_{\bf p} ] \, \rho_{{\bf p}^{\prime}}^*
\rangle 
\nonumber \\
& & \qquad \qquad \qquad 
=
\delta_{{\bf k}^{\prime}, {\bf k}(t)}
\delta_{{\bf p}^{\prime}, {\bf p}(t)}
N^{2} F_{\bf k}(t) F_{\bf p}(t).
\label{eq:factorization-app}
\end{eqnarray}
Here the translational invariance of the sheared system is taken into account
[see Eq.~(\ref{eq:shifted-tcf-2})].
Applying this approximation to Eq.~(\ref{eq:MCT-tr-dum-23}),
we obtain
\begin{eqnarray}
& &
M_{\bf q}^{\lambda \mu}(t) =
\frac{\rho v^{2}}{2 (2 \pi)^{3}}
\int d{\bf k} \,
[ k_{\lambda} c_{k} + p_{\lambda} c_{p} ]
\nonumber \\
& & \qquad 
\times \,
[ k_{\mu}(t) c_{k(t)} + p_{\mu}(t) c_{p(t)} ]
F_{\bf k}(t) F_{\bf p}(t),
\label{eq:MCT-tr-dum-24}
\end{eqnarray}
where the wave vector ${\bf p}$ in this and the following expressions for the memory kernels
abbreviates ${\bf p} \equiv {\bf q}-{\bf k}$, and should not be
confused with the momentum variable. 

In the absence of shear, the MCT expression (\ref{eq:MCT-tr-dum-24})
reduces to the one familiar from 
the equilibrium MCT~\cite{Goetze91b} describing nonlinear interactions
of density fluctuations, called the cage effect, relevant for structural slowing down.
The matrix structure as well as the wave-vector dependence in 
$M_{\bf q}^{\lambda \mu}(t)$ can be simplified, i.e., 
it can be decomposed into longitudinal and transversal components
which depend on the modulus $q = |{\bf q}|$ only, and this is possible 
because of the isotropic nature of the quiescent equilibrium system.
In the presence of shear, on the other hand, 
the ``dephasing'' of the vertex function in Eq.~(\ref{eq:MCT-tr-dum-24})
occurs, which reduces the nonlinear interactions, and hence,  
enhances the structural relaxation. 
In addition, the structure of $M_{\bf q}^{\lambda \mu}(t)$ cannot be
simplified in a mentioned way 
due to the wave-vector dependence of the vertex function and  
of the transient density correlators,
which are associated with the anisotropic nature of the sheared system.

The memory kernel $L_{\bf q}^{\lambda}(t)$ defined in Eq.~(\ref{eq:L-tr-def})
can be handled in a similar manner under the mode-coupling approximation,
and its detailed derivation is presented in 
Appendix~\ref{appen:MCT-tr-dum-41} with the result
\begin{widetext}
\begin{equation}
L_{\bf q}^{\lambda}(t) = -
\frac{v^{2}}{2 (2 \pi)^{3}}
\int d{\bf k} \,
[ \, k_{\lambda} c_{k} + p_{\lambda} c_{p} \, ] 
\Bigl[ \,
  \frac{k_{x} k_{y}(t)}{k(t)} \frac{S_{k(t)}^{\prime}}{S_{k(t)}} +
  \frac{p_{x} p_{y}(t)}{p(t)} \frac{S_{p(t)}^{\prime}}{S_{p(t)}} \,
\Bigr]
F_{\bf k}(t) F_{\bf p}(t).
\label{eq:MCT-tr-dum-41}
\end{equation}
\end{widetext}
Here $S_{q}^{\prime} \equiv \partial S_{q}/\partial q$. 
It is anticipated that this memory kernel becomes relevant
only if significant anisotropy is developed in the density fluctuations.
This is because the shear stress $\sigma_{xy}$ entering into 
its defining equation (\ref{eq:L-tr-def}),
which is reflected in the quantities in the second square brackets in
Eq.~(\ref{eq:MCT-tr-dum-41}),
is intrinsically an anisotropic quantity. 
For example, one finds from Eq.~(\ref{eq:MCT-tr-dum-41})
that $L_{\bf q}^{\lambda}(0) = 0$ reflecting the isotropy 
of the initial equilibrium state. 

The other memory kernels defined in 
Eqs.~(\ref{eq:L-prime-tr-def})--(\ref{eq:N-prime-tr-def}) are 
found to vanish under the mode-coupling approximation
as demonstrated in Appendixes~\ref{appen:MCT-tr-dum-81} and \ref{appen:MCT-tr-dum-91}:
\begin{eqnarray}
L_{\bf q}^{\prime \, \lambda \mu}(t) = 0.
\label{eq:MCT-tr-dum-81}
\\
N_{\bf q}^{\lambda}(t) = 0, \quad 
N_{\bf q}^{\prime \, \lambda \mu}(t) = 0.
\label{eq:MCT-tr-dum-91}
\end{eqnarray}
Thus, only the memory kernels
$M_{\bf q}^{\lambda \mu}(t)$ and 
$L_{\bf q}^{\lambda}(t)$ survive under the mode-coupling approximation
formulated with the projection operator ${\cal P}_{2}$.

\section{Steady-state properties}
\label{sec:steady-state-properties}

In this section, we provide the TTCF expressions
for the steady-state properties (see Sec.~\ref{subsec:TTCF})
under the mode-coupling approximation.
This enables one to obtain the stationary-state properties
based on the knowledge of the transient density correlators $F_{\bf q}(t)$. 

\subsection{Remarks on TTCF expressions}
\label{subsec:remarks-TTCF}

Let us first notice that the transient time-correlation functions
appearing in the TTCF expressions in Sec.~\ref{subsec:TTCF}
can be abbreviated as
\begin{equation}
G_{X}(t) \equiv
\langle [ e^{i {\cal L} t} X ] \, \sigma_{xy} \rangle, \,\,\,
H_{X}(t) \equiv
\langle [ e^{i {\cal L} t} X ] \, \delta K \rangle.
\label{eq:G-H-dum-1}
\end{equation}
For example, the TTCF formed with $\sigma_{xy}(0)$ 
in Eq.~(\ref{eq:ss-shear-TTCF}) is given by
$\langle [ e^{i {\cal L} s} \sigma_{xy} ] \, \sigma_{xy} \rangle$,
and the one in Eq.~(\ref{eq:ss-F-TTCF}) by
$\langle [ e^{i {\cal L} s} \{ \rho_{\bf q}(t) 
\rho_{{\bf q}(t)}^{*} \} ] \,
\sigma_{xy} \rangle$.

As discussed in Appendix~\ref{appen:remark-TTCF},
the functions $G_{X}(t)$ and $H_{X}(t)$ evolve in time
within the subspace orthogonal to 
$\{ \rho_{\bf k}, j_{\bf k}^{\mu} \}$,
i.e., there hold
\begin{subequations}
\label{eq:G-H-dum-2}
\begin{eqnarray}
G_{X}(t) &=& 
\langle [ e^{i {\cal QLQ} t} {\cal Q} X ] \, {\cal Q} \sigma_{xy} \rangle, 
\\
H_{X}(t) &=&
\langle [ e^{i {\cal QLQ} t} {\cal Q} X ] \, {\cal Q} \delta K \rangle,
\end{eqnarray}
\end{subequations}
in terms of the projection operator ${\cal Q}$
complementary to ${\cal P}$ defined in Eq.~(\ref{eq:P-tr-def}).
This feature 
is exactly the one sheared with the memory kernels
[see Eqs.~(\ref{eq:R-tr-def-1}),
(\ref{eq:GLE-tr-dum-32}), and 
(\ref{eq:M-tr-def})--(\ref{eq:N-prime-tr-def})].
Thus, no additional approximation than those introduced in 
Sec.~\ref{sec:MCT} is necessary to deal with 
$G_{X}(t)$ and $H_{X}(t)$. 
The only difference here is that,
since both $\sigma_{xy}$ and $\delta K$ are ``zero wave-vector'' quantities,
the second projection operator ${\cal P}_{2}$ given in Eq.~(\ref{eq:P2-def})
has to be replaced by ${\cal P}_{2}^{0}$ defined via
\begin{equation}
{\cal P}_{2}^{0} X \equiv
\sum_{{\bf k} > 0}
\langle X \rho_{\bf k} \rho_{\bf k}^{*} \rangle
\frac{1}{N^{2} S_{k}^{2}}
\rho_{\bf k} \rho_{\bf k}^{*}.
\label{eq:P2-0-tr-def}
\end{equation}
We thus obtain under the first mode-coupling approximation, 
in which the propagator $e^{i {\cal QLQ} t}$ is approximated
by the projected one ${\cal P}_{2}^{0} e^{i {\cal QLQ} t} {\cal P}_{2}^{0}$,
\begin{subequations}
\label{eq:G-H-dum-11}
\begin{eqnarray}
G_{X}(t) &=& 
\langle [ e^{i {\cal QLQ} t} {\cal P}_{2}^{0} {\cal Q} X ] \, 
{\cal P}_{2}^{0} \sigma_{xy} \rangle,
\\
H_{X}(t) &=&
\langle [ e^{i {\cal QLQ} t} {\cal P}_{2}^{0} {\cal Q} X ] \, 
{\cal P}_{2}^{0} \delta K \rangle.
\end{eqnarray}
\end{subequations}
Here we have noticed
${\cal Q} \sigma_{xy} = \sigma_{xy}$ and 
${\cal Q} \delta K = \delta K$
[see Eq.~(\ref{eq:Q-sigma-Q-deltaK})].

The evaluation of ${\cal P}_{2}^{0} \sigma_{xy}$
is presented in Appendix~\ref{appen:P2-sigma}
with the result
\begin{equation}
{\cal P}_{2}^{0} \sigma_{xy} =
- \frac{k_{\rm B}T}{N}
\sum_{{\bf k} > 0}
\frac{k_{x} k_{y}}{k}
\frac{S_{k}^{\prime}}{S_{k}^{2}} \,
\rho_{\bf k} \rho_{\bf k}^{*}.
\label{eq:P2-sigma}
\end{equation}
In view of Eq.~(\ref{eq:deltaK-def}),
one easily understands that
$\langle \delta K \rho_{\bf k} \rho_{\bf k}^{*} \rangle = 0$,
and hence,
\begin{equation}
{\cal P}_{2}^{0} \delta K = 0.
\label{eq:P2-deltaK}
\end{equation}
Thus, under the mode-coupling approximation, 
only those contributions abbreviated as $G_{X}(t)$ survive 
in the TTCF expressions for the steady-state properties. 

\subsection{Steady-state shear stress}

With the results in the previous subsection,
the TTCF expression (\ref{eq:ss-shear-TTCF})
for the steady-state shear stress 
under the mode-coupling approximation is given by
\begin{equation}
\sigma_{\rm ss} = \frac{\dot{\gamma}}{k_{\rm B}T V}
\int_{0}^{\infty} ds \,
\langle [ e^{i {\cal QLQ} s} {\cal P}_{2}^{0} \sigma_{xy} ] \, 
{\cal P}_{2}^{0} \sigma_{xy} \rangle.
\end{equation}
Substituting Eq.~(\ref{eq:P2-sigma}) into this expression yields
\begin{eqnarray}
\sigma_{\rm ss} &=&
\frac{k_{\rm B}T \dot{\gamma}}{V N^{2}}
\int_{0}^{\infty} ds \,
\sum_{{\bf k} > 0} \sum_{{\bf k}' > 0}
\frac{k_{x} k_{y}}{k}
\frac{S_{k}^{\prime}}{S_{k}^{2}} \,
\frac{k'_{x} k'_{y}}{k'}
\frac{S_{k'}^{\prime}}{S_{k'}^{2}} 
\nonumber \\
& & \qquad \qquad \qquad
\times \,
\langle
  [ e^{i {\cal QLQ} s} \rho_{\bf k} \rho_{\bf k}^{*} ] \,
  \rho_{{\bf k}'} \rho_{{\bf k}'}^{*}
\rangle.
\end{eqnarray}
Applying the factorization 
approximation (\ref{eq:factorization-app}),
one gets
\begin{eqnarray}
\langle
  [ e^{i {\cal QLQ} s} \rho_{\bf k} \rho_{\bf k}^{*} ] \,
  \rho_{{\bf k}'} \rho_{{\bf k}'}^{*}
\rangle &\approx&
\langle
  [ e^{i {\cal L} s} \rho_{\bf k} ] \, \rho_{{\bf k}^{\prime}}^{*}
\rangle \, 
\langle
  [ e^{i {\cal L} s} \rho_{\bf k}^{*} ] \, \rho_{{\bf k}^{\prime}}
\rangle 
\nonumber \\
&=&
\delta_{{\bf k}^{\prime}, {\bf k}(s)} N^{2} F_{\bf k}(s)^{2},
\end{eqnarray}
where in the final equality we have noticed that $F_{\bf k}(s)$ 
is a real function of time [see Eq.~(\ref{eq:reality-neq-tcf})].
This leads to the following MCT expression for the steady-state 
shear stress $\sigma_{\rm ss}$
in terms of the transient density correlators: 
\begin{equation}
\sigma_{\rm ss} =
\frac{k_{\rm B}T \dot{\gamma}}{2 (2 \pi)^{3}}
\int_{0}^{\infty} ds \,
\int d{\bf k} \,
\frac{k_{x}^{2} k_{y}k_y(s)}{k k(s)}
\frac{S_{k}^{\prime} S_{k(s)}^{\prime}}{S_{k}^{2} S_{k(s)}^{2}} \, F_{\bf k}(s)^{2}. 
\label{eq:ss-shear-MCT}
\end{equation}
The steady-state kinetic temperature $T_{\rm ss}$ can then be obtained via
Eq.~(\ref{eq:ss-temperature-TCF}).

\subsection{Steady-state density fluctuations}

With the remarks in Sec.~\ref{subsec:remarks-TTCF},
the TTCF expression (\ref{eq:ss-F-TTCF})
for the steady-state density correlator $F_{\bf q}^{\rm ss}(t)$
under the mode-coupling approximation is given by
\begin{widetext}
\begin{equation}
F_{\bf q}^{\rm ss}(t) = F_{\bf q}(t) - 
\frac{\dot{\gamma}}{N k_{\rm B}T}
\int_{0}^{\infty} ds \,
\langle 
  [ \, 
    e^{i {\cal QLQ} s} 
    {\cal Q} \{ \rho_{\bf q}(t) \rho_{{\bf q}(t)}^{*} \} \,
  ] \,
  {\cal P}_{2}^{0} \sigma_{xy}
\rangle.
\label{eq:ss-F-dum-11}
\end{equation}
Here we do not apply ${\cal P}_{2}^{0}$ to
${\cal Q} \{ \rho_{\bf q}(t) \rho_{{\bf q}(t)}^{*} \}$ since it already
has the form of the density product.
Let us notice that, 
since $\rho_{\bf k=0} = 0$ 
[see Eq.~(\ref{eq:rho-def})]
and 
$j_{\bf k=0}^{\mu} = (1/m) \sum_{i} p_{i}^{\mu} = 0$
[see the comment below Eq.~(\ref{eq:SLLOD-b})],
it follows from Eq.~(\ref{eq:shifted-tcf-11}) that
$\langle 
\rho_{\bf q}(t) \rho_{{\bf q}(t)}^{*} \, \rho_{\bf k}^{*}
\rangle = 
\delta_{{\bf k}, {\bf 0}} \,
\langle 
\rho_{\bf q}(t) \rho_{{\bf q}(t)}^{*} \, \rho_{\bf k=0}^{*}
\rangle = 0$ 
and 
$\langle 
\rho_{\bf q}(t) \rho_{{\bf q}(t)}^{*} \, j_{\bf k}^{\mu \, *}
\rangle = 
\delta_{{\bf k}, {\bf 0}} \,
\langle
\rho_{\bf q}(t) \rho_{{\bf q}(t)}^{*} \, j_{\bf k=0}^{\mu \, *}
\rangle = 0$.
One therefore obtains
${\cal P} \{ \rho_{\bf q}(t) \rho_{{\bf q}(t)}^{*} \} = 0$ 
[see Eq.~(\ref{eq:P-tr-def})], and hence,
${\cal Q} \{ \rho_{\bf q}(t) \rho_{{\bf q}(t)}^{*} \} = 
\rho_{\bf q}(t) \rho_{{\bf q}(t)}^{*}$.
Thus, we have for the integrand of 
Eq.~(\ref{eq:ss-F-dum-11})
\begin{equation}
\langle 
  [ \, 
    e^{i {\cal QLQ} s} 
     {\cal Q} \{ \rho_{\bf q}(t) \rho_{{\bf q}(t)}^{*} \} \,
  ] \,
  {\cal P}_{2}^{0} \sigma_{xy}
\rangle =
- \frac{k_{\rm B}T}{N}
\sum_{{\bf k} > 0}
\frac{k_{x} k_{y}}{k} \frac{S_{k}^{\prime}}{S_{k}^{2}}
\langle
  [ e^{i {\cal QLQ} s} \rho_{\bf q}(t) \rho_{{\bf q}(t)}^{*} ] \,
  \rho_{\bf k} \rho_{\bf k}^{*}
\rangle,
\label{eq:ss-F-dum-12}
\end{equation}
where we have used Eq.~(\ref{eq:P2-sigma}).
Here we apply the factorization approximation
[see Eq.~(\ref{eq:factorization-app})]
\begin{equation}
\langle
  [ e^{i {\cal QLQ} s} \rho_{\bf q}(t) \rho_{{\bf q}(t)}^{*} ] \,
  \rho_{\bf k} \rho_{\bf k}^{*}
\rangle \approx
\langle
  [ e^{i {\cal L} s} \rho_{\bf q}(t) ] \, \rho_{\bf k}^{*}
\rangle \,
\langle
  [ e^{i {\cal L} s} \rho_{{\bf q}(t)}^{*} ] \, \rho_{\bf k}
\rangle =
\delta_{{\bf k}, {\bf q}(t+s)} \,
N^{2} F_{\bf q}(t+s) F_{{\bf q}(t)}(s),
\label{eq:ss-F-dum-13}
\end{equation}
where in the final equality we have noticed that 
$F_{{\bf q}(t)}(s)$ 
is a real function of time [see Eq.~(\ref{eq:reality-neq-tcf})].
This yields the following MCT expression for the
steady-state density correlator $F_{\bf q}^{\rm ss}(t)$
in terms of the transient density correlators:
\begin{equation}
F_{\bf q}^{\rm ss}(t) = F_{\bf q}(t) +
\dot{\gamma} \int_{0}^{\infty} ds \,
\frac{q_{x} q_{y}(t+s)}{q(t+s)} 
\frac{S_{q(t+s)}^{\prime}}{S_{q(t+s)}^{2}}
F_{\bf q}(t+s) F_{{\bf q}(t)}(s).
\label{eq:ss-F-dum-14}
\end{equation}
As a corollary, we obtain for the steady-state structure factor 
$S_{\bf q}^{\rm ss} = F_{\bf q}^{\rm ss}(t=0)$
\begin{eqnarray}
S_{\bf q}^{\rm ss} = S_{q} +
\dot{\gamma} \int_{0}^{\infty} ds \,
\frac{q_{x} q_{y}(s)}{q(s)} 
\frac{S_{q(s)}^{\prime}}{S_{q(s)}^{2}} \, F_{\bf q}(s)^{2}.
\label{eq:ss-F-dum-15}
\end{eqnarray}
\end{widetext}

Let us see the connection between $\sigma_{\rm ss}$ and $S_{\bf q}^{\rm ss}$
under the mode-coupling approximation.
By comparing Eqs.~(\ref{eq:ss-shear-MCT}) and (\ref{eq:ss-F-dum-15}), one finds
\begin{equation}
\sigma_{\rm ss} =
\frac{k_{\rm B}T}{2 (2\pi)^{3}}
\int d{\bf k} \,
\frac{k_{x} k_{y}}{k} \frac{S_{k}^{\prime}}{S_{k}^{2}} \, S_{\bf k}^{\rm ss},
\label{eq:ss-shear-S-MCT}
\end{equation}
where we have noticed that the isotropic term in $S_{\bf k}^{\rm ss}$
does not contribute to the integral. 
Thus, $\sigma_{\rm ss}$ and $S_{\bf q}^{\rm ss}$ are 
handled on an equal footing naturally 
under the mode-coupling approximation.
Compared to Eq.~(\ref{eq:ss-shear-S-TTCF}), 
the kinetic part is missing here since only the
interaction part is dealt with under the mode-coupling approach.
In addition, since $S_{k}^{\prime} / S_{k}^{2} = \rho c_{k}^{\prime}$ 
[see Eq.~(\ref{eq:c-def})],
the bare potential in Eq.~(\ref{eq:ss-shear-S-TTCF}) is replaced by the 
``renormalized''~\cite{Hansen86} direct correlation function 
in Eq.~(\ref{eq:ss-shear-S-MCT}).
Finally, we notice that Eq.~(\ref{eq:ss-shear-S-MCT}) can directly be derived
from Eq.~(\ref{eq:P2-sigma}) by approximating
$\sigma_{\rm ss} \approx - \langle {\cal P}_{2}^{0} \sigma_{xy} \rangle_{\rm ss}/V$
and using the definition
$S_{\bf k}^{\rm ss} = (1/N) \langle \rho_{\bf k} \rho_{\bf k}^{*} \rangle_{\rm ss}$.

\section{Summary and discussion}
\label{sec:summary}

In this paper, we developed a nonequilibrium MCT for uniformly sheared systems
starting from microscopic, thermostatted SLLOD equations of motion.
Our theory aims at describing stationary-state properties 
including rheological ones, and this is accomplished via two steps.
Firstly, a set of self-consistent equations of motion is formulated 
for the transient density correlators $F_{\bf q}(t)$ 
based on the projection-operator formalism and on the mode-coupling approach,
which enables the calculation of $F_{\bf q}(t)$ provided the
static structure factor $S_{q}$ of the initial equilibrium state is given as input.
The transient time-correlation function formalism is then used
which, combined with the mode-coupling approximation, expresses
stationary-state properties in terms of $F_{\bf q}(t)$.
Thereby, steady-state quantities like 
the shear stress $\sigma_{\rm ss}$, temperature $T_{\rm ss}$,
density correlators $F_{\bf q}^{\rm ss}(t)$, and the structure factor $S_{\bf q}^{\rm ss}$ 
can all be calculated in terms of $S_{q}$. 
We also addressed how the steady-state temperature $T_{\rm ss}$ can 
be controlled using the constant-$\alpha$ model for the thermostat:
this can be done via a self-consistent treatment of the 
thermostatting multiplier $\alpha$
based on Eq.~(\ref{eq:ss-temperature-TCF}).
Our theory is able to treat $\sigma_{\rm ss}$ and $S_{\bf q}^{\rm ss}$ on an equal footing,
which is missing in the steady-state-fluctuations approach of 
Refs.~\cite{Miyazaki-sheared-MCT-all,Hayakawa08}.
In addition, we need not assume the validity of the fluctuation-dissipation
theorem in sheared states, which was necessary in Ref.~\cite{Miyazaki-sheared-MCT-all}.

The transient density correlators $F_{\bf q}(t)$ thus play a distinguished role
in our approach. 
Let us collect here all the relevant MCT equations for $F_{\bf q}(t)$ 
derived in Secs.~\ref{sec:GLE} and \ref{sec:MCT}
to highlight new features of our theory compared to the
equilibrium MCT~\cite{Goetze91b} and to the nonequilibrium MCT
developed by Fuchs and Cates (FC)
for sheared Brownian systems~\cite{Fuchs-Cates-sheared-MCT,Fuchs05}. 
The exact Zwanzig-Mori-type equations consist of
the continuity equation
\begin{widetext}
\begin{subequations}
\label{eq:summary-GLE-transient}
\begin{equation}
\Bigl[ \,
  \frac{\partial}{\partial t} -
  {\bf q} \cdot \mbox{\boldmath $\kappa$} \cdot
  \frac{\partial}{\partial {\bf q}} \,
\Bigr] F_{\bf q}(t) =
i {\bf q} \cdot {\bf H}_{\bf q}(t),
\label{eq:summary-GLE-transient-a}
\end{equation}
and the time-evolution equation for the 
transient density-current cross correlator $H_{\bf q}^{\lambda}(t)$ 
\begin{eqnarray}
\Bigl[ \,
  \frac{\partial}{\partial t} -
  {\bf q} \cdot \mbox{\boldmath $\kappa$} \cdot
  \frac{\partial}{\partial {\bf q}} \,
 \Bigr] H_{\bf q}^{\lambda}(t) &=&
 i q_{\lambda} \frac{v^{2}}{S_{q}} F_{\bf q}(t) -
 [\mbox{\boldmath $\kappa$} \cdot {\bf H}_{\bf q}(t)]^{\lambda} -
 \alpha H_{\bf q}^{\lambda}(t)
 \nonumber \\
 & &
- \,
\sum_{\mu}
\int_{0}^{t} ds \,
M_{\bf q}^{\lambda \mu}(s) \,
H_{{\bf q}(s)}^{\mu}(t-s) +
\dot{\gamma}
\int_{0}^{t} ds \,
i L_{\bf q}^{\lambda}(s) \,
F_{{\bf q}(s)}(t-s).
\label{eq:summary-GLE-transient-b}
\end{eqnarray}
\end{subequations}
In this equation, we already omitted those memory kernels which vanish under the
mode-coupling approximation (see Sec.~\ref{sec:MCT}). 
The MCT expressions for the memory kernels $M_{\bf q}^{\lambda \mu}(t)$ and 
$L_{\bf q}^{\lambda}(t)$ are given by
\begin{subequations}
\label{eq:summary-MCT-transient}
\begin{equation}
M_{\bf q}^{\lambda \mu}(t) =
\frac{\rho v^{2}}{2 (2 \pi)^{3}}
\int d{\bf k} \,
[ \, k_{\lambda} c_{k} + p_{\lambda} c_{p} \, ]
[ \, k_{\mu}(t) c_{k(t)} + p_{\mu}(t) c_{p(t)} \, ]
F_{\bf k}(t) F_{\bf p}(t),
\label{eq:summary-MCT-transient-a}
\end{equation}
\begin{equation}
L_{\bf q}^{\lambda}(t) = -
\frac{v^{2}}{2 (2 \pi)^{3}}
\int d{\bf k} \,
[ \, k_{\lambda} c_{k} + p_{\lambda} c_{p} \, ]
\Bigl[ \,
  \frac{k_{x} k_{y}(t)}{k(t)} \frac{S_{k(t)}^{\prime}}{S_{k(t)}} +
  \frac{p_{x} p_{y}(t)}{p(t)} \frac{S_{p(t)}^{\prime}}{S_{p(t)}} \,
\Bigr]
F_{\bf k}(t) F_{\bf p}(t).
\label{eq:summary-MCT-transient-b}
\end{equation}
\end{subequations}
\end{widetext}
Here ${\bf p} \equiv {\bf q} - {\bf k}$. 
Compared to the equilibrium MCT~\cite{Goetze91b}, new features entering here are	
(i) the replacement of $\partial/\partial t$ by
$[\partial/\partial t - 
{\bf q} \cdot \mbox{\boldmath $\kappa$} \cdot (\partial/\partial {\bf q})]$,
(ii) the presence of the second (shear)
and the third (thermostat) terms on the right-hand side of 
Eq.~(\ref{eq:summary-GLE-transient-b}),
(iii) the matrix structure of the memory kernel
$M_{\bf q}^{\lambda \mu}(t)$ describing the fluctuating-force correlations
which cannot be decomposed into the longitudinal and transversal parts,
and (iv) the presence of the additional memory kernel $L_{\bf q}^{\lambda}(t)$.
Furthermore, when compared with the FC theory~\cite{Fuchs-Cates-sheared-MCT,Fuchs05}, 
we see in addition to those rather trivial 
differences reflecting the Newtonian and Brownian short-time microscopic dynamics
(v) the memory kernel in the FC theory describing the 
fluctuating-force correlations, to be denoted as 
$M_{\bf q}^{\rm FC}(t, t')$, has a different mathematical structure in that
it depends on two times $t$ and $t'$ after the shearing force
is turned on, while only one time enters into our $M_{\bf q}^{\lambda \mu}(t)$, and
(vi) the memory kernel corresponding to $L_{\bf q}^{\lambda}(t)$
is absent also in the FC theory.
 
The first and second features just mentioned
arise from the shear part ($i {\cal L}_{\dot{\gamma}}$)
and the thermostat part ($i {\cal L}_{\alpha}$) in the $p$-Liouvillean
for the SLLOD equations [see Eqs.~(\ref{eq:iL-SLLOD})], which are absent
in the $p$-Liouvillean for quiescent systems.
The third feature reflects the anisotropic nature of the sheared system:
in the presence of shear, the longitudinal and transversal
current density fluctuations cannot be separately handled as in isotropic systems since
their cross correlators do not vanish. 
In this connection, we notice that the second term on the right-hand side of
Eq.~(\ref{eq:summary-GLE-transient-b}), which cannot be expressed in terms of the 
$\lambda$-component $H_{\bf q}^{\lambda}(t)$ alone,
also reflects the anisotropy of the sheared system.
Therefore, without introducing any further approximation (see below),
Eqs.~(\ref{eq:summary-GLE-transient-a}) and (\ref{eq:summary-GLE-transient-b})
cannot be combined to yield a single second-order
integro-differential equation for $F_{\bf q}(t)$ as in the equilibrium MCT~\cite{Goetze91b}. 
The fourth feature originates from the non-Hermitian nature of the 
$p$-Liouvillean describing nonequilibrium dynamics [see Eq.~(\ref{eq:Lp-inside})],
i.e., the presence of the additional memory kernel
$L_{\bf q}^{\lambda}(t)$ is expected on general grounds.

The fifth feature, when compared with the FC theory, 
is due to different strategies employed in deriving the Zwanzig-Mori-type equations 
for $F_{\bf q}(t)$:
the two-time structure in $M_{\bf q}^{\rm FC}(t,t')$ is an exact consequence 
of the Zwanzig-Mori-type equations (\ref{eq:FC-Zwanzig-Mori}) for $F_{\bf q}(t)$ 
upon which the FC theory is based (see Ref.~\cite{Fuchs05}),
while the one-time structure in our $M_{\bf q}^{\lambda \mu}(t)$ follows
from another exact equation~(\ref{eq:GLE-Hq-lambda-tr-1}) to which
the projection-operator formalism is applied (see Sec.~\ref{subsec:projection-operator-formalism}).
One therefore cannot judge which of the memory kernels is superior at the formal level:
we can only state that ours has a simpler mathematical structure.
Furthermore, both the memory kernels $M_{\bf q}^{\lambda \mu}(t)$ and
$M_{\bf q}^{\rm FC}(t,t')$ under the mode-coupling approximation describe 
essentially the same physics concerning the competition between the cage effect 
and the shear advection of density fluctuations (see also below). 

Thus, the sixth feature mentioned above, i.e., the presence/absence
of the memory kernel $L_{\bf q}^{\lambda}(t)$ is the major difference
between our and the FC theory. 
It is not likely that this difference originates from 
the different microscopic dynamics -- Newtonian or Brownian --
adopted in these theories since, as we stated above, 
the presence of such a memory kernel is expected on general grounds.

It is anticipated 
that the memory kernel $L_{\bf q}^{\lambda}(t)$ becomes relevant
only if significant anisotropy is developed in the density fluctuations.
This is because the shear stress $\sigma_{xy}$ entering into 
the defining equation (\ref{eq:L-tr-def}) of $L_{\bf q}^{\lambda}(t)$ 
is intrinsically an anisotropic quantity. 
We indeed confirmed from our preliminary numerical calculations
based on the MCT expression (\ref{eq:summary-MCT-transient-b}) that
the contribution from $L_{\bf q}^{\lambda}(t)$ is quite small
under the isotropic approximation for the density fluctuations 
to be discussed below.
It will be interesting to pursue in what circumstances 
this additional memory kernel becomes important whose presence is naturally
expected for nonequilibrium sheared systems. 

To further facilitate the comparison of our theory 
with the equilibrium MCT and with the FC theory,
the MCT equations
(\ref{eq:summary-GLE-transient}) and (\ref{eq:summary-MCT-transient})
shall be simplified using the isotropic approximation introduced in 
Appendix~\ref{appen:isotropic-app-1}.
Such a simplifying approximation is also useful in practical applications of our theory
to systems where anisotropy in the density fluctuations is small. 

The MCT equations (\ref{eq:iso-app-dum-53}), (\ref{eq:iso-app-dum-64}), and
(\ref{eq:iso-app-dum-73}) derived 
in Appendix~\ref{appen:isotropic-app-1}
under the isotropic approximation shall be rewritten
in the following form for the {\em normalized} transient density correlators
$\phi_{q}(t) \equiv F_{q}(t) / S_{q}$:
\begin{eqnarray}
& &
\ddot{\phi}_{q}(t) +
\Omega_{q}^{2} \phi_{q}(t) +
\alpha \dot{\phi}_{q}(t) 
\nonumber \\
& & \quad 
+ \,
\Omega_{q}^{2}
\int_{0}^{t} ds \, m_{q}^{\rm iso}(s) \,
\dot{\phi}_{\bar{q}(s)}(t-s) 
\nonumber \\
& & \qquad 
+ \, 
\dot{\gamma} \Omega_{q}^{2}
\int_{0}^{t} ds \,
l_{q}^{\rm iso}(s) \, \phi_{\bar{q}(s)}(t-s) = 0.
\label{eq:isotropic-GLE}
\end{eqnarray}
Here all the functions depend on the wave-vector modulus only; 
the dot denotes the partial time derivative; $\Omega_{q}^{2} \equiv
q^{2} v^{2} / S_{q}$ the square of the characteristic frequency
relevant for the short-time dynamics; 
and $\bar{q}(s) \equiv q [1 + (\dot{\gamma}s)^{2}/3]^{1/2}$ the
modulus of the advected wave vector under the isotropic approximation. 
The memory kernels, from which 
$\Omega_{q}^{2}$ is factored out 
following the convention in the equilibrium MCT~\cite{Goetze91b},
are given by
\begin{subequations}
\begin{eqnarray}
m_{q}^{\rm iso}(t) &=&
\int d{\bf k} \, V_{{\bf q}, {\bf k}, {\bf p}}^{(\dot{\gamma})}(t) \,
\phi_{k}(t) \phi_{p}(t),
\label{eq:isotropic-MCT-M}
\\
l_{q}^{\rm iso}(t) &=&
\int d{\bf k} \, V_{{\bf q}, {\bf k}, {\bf p}}^{(\dot{\gamma}) \, \prime}(t) \,
\phi_{k}(t) \phi_{p}(t),
\label{eq:isotropic-MCT-L}
\end{eqnarray}
with the time-dependent vertex functions
\begin{eqnarray}
V_{{\bf q}, {\bf k}, {\bf p}}^{(\dot{\gamma})}(t) &=&
\frac{\rho S_{q} S_{k} S_{p}}{2 (2 \pi)^{3} q^{4}} \,
[ {\bf q} \cdot {\bf k} c_{k} + {\bf q} \cdot {\bf p} c_{p} ] 
\nonumber \\
& & \qquad \qquad \quad
\times \,
[ {\bf q} \cdot {\bf k} c_{\bar{k}(t)} + {\bf q} \cdot {\bf p} c_{\bar{p}(t)} ],
\label{eq:isotropic-MCT-M-vertex}
\\
V_{{\bf q}, {\bf k}, {\bf p}}^{(\dot{\gamma}) \, \prime}(t) &=&
- \frac{ \dot{\gamma} t }{ 3 \sqrt{1+(\dot{\gamma}t)^{2}/3} }
\frac{S_{q} S_{k} S_{p}}{2 (2 \pi)^{3} q^{2}} 
[ {\bf q} \cdot {\bf k} c_{k} + {\bf q} \cdot {\bf p} c_{p} ] 
\nonumber \\
& & \qquad \qquad \quad
\times \,
\Bigl[ \, 
  k \frac{S_{\bar{k}(t)}^{\prime}}{S_{\bar{k}(t)}} +
  p \frac{S_{\bar{p}(t)}^{\prime}}{S_{\bar{p}(t)}} \,
\Bigr].
\label{eq:isotropic-MCT-L-vertex}
\end{eqnarray}
\end{subequations}
The resemblance of these equations to those in the equilibrium MCT~\cite{Goetze91b}
is apparent:
the major differences are the dephasing in the vertex function
$V_{qkp}^{(\dot{\gamma})}(t)$ for $m_{q}^{\rm iso}(t)$,
which enhances the relaxation of the density fluctuations, 
and the presence of the additional memory kernel $l_{q}^{\rm iso}(t)$.

Now, let us ``derive'' the MCT equations for sheared Brownian systems,
starting from Eq.~(\ref{eq:isotropic-GLE})
with the procedure adopted in Ref.~\cite{Franosch97}
for converting the microscopic dynamics from Newtonian to Brownian.
Assuming that the ``friction'' constant $\alpha$ is large, 
the inertia term in Eq.~(\ref{eq:isotropic-GLE}) shall be neglected.
As a result, the generalized oscillator
equation (\ref{eq:isotropic-GLE}) is specialized to generalized
relaxator equation
\begin{eqnarray}
& &
\dot{\phi}_{q}(t) +
\Gamma_{q} \phi_{q}(t) +
\Gamma_{q}
\int_{0}^{t} ds \, m_{q}^{\rm iso}(s) \,
\dot{\phi}_{\bar{q}(s)}(t-s) 
\nonumber \\
& &
\qquad \quad
+ \,
\dot{\gamma} \Gamma_{q} \int_{0}^{t} ds \,
l_{q}^{\rm iso}(s) \, \phi_{\bar{q}(s)}(t-s) = 0,
\label{eq:isotropic-GLE-Brownian}
\end{eqnarray}
where we have defined $\Gamma_{q} \equiv \Omega_{q}^{2}/\alpha$.
This equation, combined with 
Eqs.~(\ref{eq:isotropic-MCT-M}) and (\ref{eq:isotropic-MCT-M-vertex})
and neglecting $l_{q}^{\rm iso}(t)$ which is found to be small
from our preliminary calculations, 
is formally identical to the corresponding equation in the FC theory.
[See Eqs.~(4)--(6) of the second article cited in 
Ref.~\cite{Fuchs-Cates-sheared-MCT}.
There is a minor difference that 
$\dot{\phi}_{\bar{q}(s)}(t-s)$
at the advected wave number $\bar{q}(s)$ enters into
the third term in Eq.~(\ref{eq:isotropic-GLE-Brownian}), while
$\dot{\phi}_{q}(t-s)$ at the wave number $q$ 
appears in the corresponding FC equation.
Again, this reflects
the difference of the starting Zwanzig-Mori-type equations.] 
In this sense, our and the FC theory are equivalent.
But, it should be remembered that this holds only under the
isotropic approximation:
when anisotropy in the density fluctuations is significant, 
one has to go back to Eqs.~(\ref{eq:summary-GLE-transient}) and 
(\ref{eq:summary-MCT-transient}), and the presence/absence of the
memory kernel $L_{\bf q}^{\lambda}(t)$ 
may have significant consequences. 

Finally, we notice that the anisotropic nature of the steady-state 
quantities can still be 
captured within the mentioned isotropic approximation for the
transient density correlators.
This issue is discussed in Appendix~\ref{appen:isotropic-app-2}.
In particular, it is argued there why the steady-state shear stress $\sigma_{\rm ss}$
can be evaluated under the isotropic approximation,
as done in the application of the FC theory~\cite{Fuchs-Cates-sheared-MCT},  
although this quantity should vanish under isotropic density fluctuations.

\begin{acknowledgments}

S.-H.~C acknowledges financial support by 
Grant-in-Aids for scientific 
research from the 
Ministry of Education, Culture, Sports, Science and 
Technology of Japan (No.~20740245).
BK acknowledges financial support from 
Changwon National University Grant 2007. 

\end{acknowledgments}
 
\appendix

\section{Miscellaneous materials and details of some derivations}
\label{appen:derivation}

This appendix is devoted to a summary of miscellaneous materials
which are necessary in the main text, and to various technical manipulations 
in the derivations of some equations. 
In these derivations, we repeatedly use the relation
\begin{equation}
\langle ( i {\cal L}_{0} A ) \, B^{*} \rangle =
- \langle A \, ( i {\cal L}_{0} B )^{*} \rangle,
\label{eq:property-iL0}
\end{equation}
which holds for the unperturbed or
quiescent $p$-Liouvillean
$i {\cal L}_{0}$ given in Eq.~(\ref{eq:iL0}), and
\begin{equation}
\langle A F_{i}^{\lambda} \rangle =
- \Bigl\langle A \frac{\partial U}{\partial r_{i}^{\lambda}} 
\Bigr\rangle = - k_{\rm B}T 
\Bigl\langle \frac{\partial A}{\partial r_{i}^{\lambda}} \Bigr\rangle,
\label{eq:Yvon-theorem}
\end{equation}
where $F_{i}^{\lambda} = - \partial U / \partial r_{i}^{\lambda}$
denotes the $\lambda$ component of the
conservative force acting on the $i$th particle. 
These relations, well-known from equilibrium statistical
mechanics~\cite{Hansen86}, hold here 
since the averaging $\langle \cdots \rangle$
in this paper is defined with the canonical distribution
function [see Eq.~(\ref{eq:def-averaging})].
Also, terms involving odd number of momentum variables
vanish under such canonical averaging.

\subsection{Microscopic expression for stress tensor}
\label{appen:shear-stress}

Here we summarize the microscopic expression
for the stress tensor.
For simplicity, we deal with quiescent equilibrium system
for which the $p$-Liouvillean is given by
$i {\cal L} = i {\cal L}_{0}$
[see Eq.~(\ref{eq:iL0})].
In handling sheared systems, momenta appearing
in the following expressions should be understood as 
peculiar or SLLOD momenta~\cite{Evans90}.

The wave-vector dependent stress tensor
$\sigma_{\bf q}^{\lambda \mu}$ is introduced via the
continuity equation for the current density
fluctuation
$j_{\bf q}^{\lambda} =
\sum_{i} (p_{i}^{\lambda}/m) \exp( i {\bf q} \cdot {\bf r}_{i})$
\begin{equation}
i {\cal L}_{0} j_{\bf q}^{\lambda} =
\sum_{\mu} \frac{i q_{\mu}}{m} \,
\sigma_{\bf q}^{\lambda \mu},
\label{eq:continuity-for-j}
\end{equation}
and is given by~\cite{Hansen86}
\begin{equation}
\sigma_{\bf q}^{\lambda \mu} =
\sum_{i}
\Bigl[ \,
  p_{i}^{\lambda} p_{i}^{\mu} / m - \frac{1}{2} \sum_{j \ne i}
  \frac{r_{ij}^{\lambda} r_{ij}^{\mu}}{r_{ij}^{2}} 
  P_{\bf q}({\bf r}_{ij}) \,
\Bigr]
\exp(i {\bf q} \cdot {\bf r}_{i}).
\label{eq:q-denendent-sigma}
\end{equation}
Here ${\bf r}_{ij} = {\bf r}_{i} - {\bf r}_{j}$, 
$r_{ij} = | \, {\bf r}_{ij} \, |$, 
$r_{ij}^{\lambda} = r_{i}^{\lambda} - r_{j}^{\lambda}$, and
\begin{equation}
P_{\bf q}({\bf r}) =
r \phi'(r) 
\frac{1 - \exp(- i {\bf q} \cdot {\bf r})}{i {\bf q} \cdot {\bf r}},
\end{equation}
in which $\phi(r)$ denotes the pair-interaction potential.
Obviously, $\sigma_{\bf q}^{\lambda \mu}$ is a symmetric tensor.

The ``stress tensor'' referred to in the main text is 
the zero-wave-vector
limit of $\sigma_{\bf q}^{\lambda \mu}$:
\begin{equation}
\sigma_{\lambda \mu} \equiv
\sigma_{{\bf q} = {\bf 0}}^{\lambda \mu} =
\sum_{i}
\Bigl[ \,
  p_{i}^{\lambda} p_{i}^{\mu} / m - \frac{1}{2} \sum_{j \ne i}
  \frac{r_{ij}^{\lambda} r_{ij}^{\mu}}{r_{ij}} 
  \phi'(r_{ij}) \,
\Bigr].
\label{eq:appen-sigma-def}
\end{equation}
Exploiting the isotropy of the quiescent equilibrium system,
one can show that~\cite{Hansen86}
\begin{equation}
\langle \sigma_{\lambda \mu} \rangle = 0 \quad
(\lambda \ne \mu).
\label{eq:virial-theorem}
\end{equation}

The equivalence of the expression (\ref{eq:appen-sigma-def})
and Eq.~(\ref{eq:sigma-def}) in the main text can be demonstrated
as follows.
Since ${\bf F}_{i} = \sum_{j \ne i} {\bf F}_{ij}$ 
where ${\bf F}_{ij}$ denotes the force acting on the $i$th particle
from the $j$th particle 
and ${\bf F}_{ji} = - {\bf F}_{ij}$ due to Newton's third law, 
there holds
\begin{eqnarray}
\sum_{i} {\bf r}_{i} {\bf F}_{i} &=&
\frac{1}{2} 
\Bigl[ \, \sum_{i} {\bf r}_{i} \sum_{j \ne i} {\bf F}_{ij} + 
       \sum_{j} {\bf r}_{j} \sum_{i \ne j} {\bf F}_{ji} \,
\Bigr] 
\nonumber \\
&=&
\frac{1}{2} 
\sum_{i} \sum_{j \ne i}
{\bf r}_{ij} {\bf F}_{ij}.
\end{eqnarray}
Expressing the force ${\bf F}_{ij}$ in terms of the 
pair-interaction potential as
\begin{equation}
{\bf F}_{ij} = - \frac{\partial}{\partial {\bf r}_{i}}
\phi(r_{ij}) =
- \frac{{\bf r}_{ij}}{r_{ij}} \phi'(r_{ij}),
\end{equation}
one obtains
\begin{equation}
\sum_{i} r_{i}^{\lambda} F_{i}^{\mu} =
- \frac{1}{2} \sum_{i} \sum_{j \ne i}
  \frac{r_{ij}^{\lambda} r_{ij}^{\mu}}{r_{ij}} 
  \phi'(r_{ij}),
\end{equation}
indicating the equivalence of Eqs.~(\ref{eq:appen-sigma-def}) and (\ref{eq:sigma-def}). 

\subsection{Propagators under the global translation}
\label{appen:translational-invariance}

Here
we discuss how the $f$- and $p$-propagators
transform under the global translation 
${\bf \Gamma} \to {\bf \Gamma}^{\prime}$ 
defined by Eq.~(\ref{eq:shift-Gamma}).
The $p$-Liouvillean corresponding to 
the SLLOD equations 
is given by [see Eqs.~(\ref{eq:iL-SLLOD})]
\begin{equation}
i {\cal L}({\bf \Gamma}) =
\sum_{i}
\Bigl[ 
  ({\bf p}_{i}/m + \mbox{\boldmath $\kappa$} \cdot {\bf r}_{i})  
  \cdot \frac{\partial}{\partial {\bf r}_{i}} +
  ({\bf F}_{i} - \mbox{\boldmath $\kappa$} \cdot {\bf p}_{i} 
    - \alpha {\bf p}_{i})
  \cdot \frac{\partial}{\partial {\bf p}_{i}} 
\Bigr],
\label{eq:p-Liouvillean-SLLOD}
\end{equation}
and it transforms under ${\bf \Gamma} \to {\bf \Gamma}^{\prime}$ to
\begin{equation}
i {\cal L}({\bf \Gamma}^{\prime}) =
i {\cal L}({\bf \Gamma}) +
{\bf a} \cdot \mbox{\boldmath $\kappa$}^{\rm T} \cdot {\bf P} 
\,\,\, \mbox{with} \,\,\,
{\bf P} \equiv \sum_{i} \frac{\partial}{\partial {\bf r}_{i}} \, ,
\label{eq:iL-shift-dum-1}
\end{equation}
since ${\bf p}_{i}$ and 
${\bf F}_{i} = \sum_{j \ne i} {\bf F}_{ij}$ 
(where ${\bf F}_{ij}$ denotes 
the force acting on the $i$th particle by  the
$j$th particle and is a function of ${\bf r}_{ij}$ only) 
are not affected by ${\bf \Gamma} \to {\bf \Gamma}^{\prime}$.
Here $\mbox{\boldmath $\kappa$}^{\rm T}$ denotes the
transpose of $\mbox{\boldmath $\kappa$}$. 

Let us notice that, when $i {\cal L}({\bf \Gamma})$
acts on a phase variable $X({\bf \Gamma})$ that depends 
on momenta $\{ {\bf p}_{i} \}$ and particle separations
$\{ {\bf r}_{ij} \}$ only, there holds
${\bf P} X({\bf \Gamma}) = 0$.
Therefore, the only term in $i {\cal L}({\bf \Gamma})$ 
that does not commute with ${\bf P}$ is
the second term in
Eq.~(\ref{eq:p-Liouvillean-SLLOD}), for which we have
\begin{widetext}
\begin{equation}
P_{\nu} \sum_{i} \Bigl[ 
(\mbox{\boldmath $\kappa$} \cdot {\bf r}_{i}) \cdot
\frac{\partial}{\partial {\bf r}_{i}} \Bigr] =
\sum_{i}
\Bigl[ P_{\nu} (\mbox{\boldmath $\kappa$} \cdot {\bf r}_{i}) \Bigr]
\cdot \frac{\partial}{\partial {\bf r}_{i}} +
\sum_{i} \Bigl[ 
(\mbox{\boldmath $\kappa$} \cdot {\bf r}_{i}) \cdot
\frac{\partial}{\partial {\bf r}_{i}} \Bigr] P_{\nu}.
\end{equation}
We therefore obtain
\begin{eqnarray}
P_{\nu} i {\cal L}({\bf \Gamma}) - i {\cal L}({\bf \Gamma}) P_{\nu} =
\sum_{i}
\Bigl[ P_{\nu} (\mbox{\boldmath $\kappa$} \cdot {\bf r}_{i}) \Bigr]
\cdot \frac{\partial}{\partial {\bf r}_{i}} =
\sum_{i,j}
\Bigl[ 
  \frac{\partial}{\partial r_{j}^{\nu}}
  \Bigl( \sum_{\lambda, \mu} \kappa_{\lambda \mu} r_{i}^{\mu} \Bigr)
\Bigr]
\frac{\partial}{\partial r_{i}^{\lambda}} =
\sum_{i} \sum_{\lambda}
\kappa_{\lambda \nu} \frac{\partial}{\partial r_{i}^{\lambda}},
\end{eqnarray}
and hence, there holds
\begin{eqnarray}
( {\bf a} \cdot \mbox{\boldmath $\kappa$}^{\rm T} \cdot {\bf P} ) \,
i {\cal L}({\bf \Gamma}) - i {\cal L}({\bf \Gamma}) \,
( {\bf a} \cdot \mbox{\boldmath $\kappa$}^{\rm T} \cdot {\bf P} ) &=&
\sum_{\lambda, \mu}
a_{\lambda} \kappa_{\lambda \mu}^{\rm T} \, 
[ \, P_{\mu} i {\cal L}({\bf \Gamma}) - 
     i {\cal L}({\bf \Gamma}) P_{\mu} \, ]
\nonumber \\
&=&
\sum_{\lambda, \mu}
a_{\lambda} \kappa_{\mu \lambda} 
\sum_{i} \sum_{\lambda^{\prime}}
\kappa_{\lambda^{\prime} \mu} 
\frac{\partial}{\partial r_{i}^{\lambda^{\prime}}} 
\nonumber \\
&=&
\sum_{\lambda, \lambda^{\prime}}
a_{\lambda} \,
(\mbox{\boldmath $\kappa$} \cdot 
\mbox{\boldmath $\kappa$})_{\lambda^{\prime} \lambda} \,
P_{\lambda^{\prime}} = 0,
\label{eq:iL-shift-dum-11}
\end{eqnarray}
\end{widetext}
since the shear-rate tensor satisfies 
$\mbox{\boldmath $\kappa$} \cdot
\mbox{\boldmath $\kappa$} = 0$.
Thus, 
$i {\cal L}({\bf \Gamma})$ and 
${\bf a} \cdot \mbox{\boldmath $\kappa$}^{\rm T} \cdot {\bf P}$
commute.
This means that $f$-Liouvillean
$i {\cal L}^{\dagger}({\bf \Gamma})$ and
${\bf a} \cdot \mbox{\boldmath $\kappa$}^{\rm T} \cdot {\bf P}$
also commute since 
the difference between $f$- and $p$-Liouvilleans
for the SLLOD equations with the constant-$\alpha$
model for the thermostat is simply a constant
[see Eqs.~(\ref{eq:SLLOD-Lambda}) and 
(\ref{eq:relation-Liouville-operators})].

Using the Campbell-Baker-Hausdorff theorem 
which states that
$e^{{\cal A} + {\cal B}} = e^{\cal A} \, e^{\cal B}$
for commuting operators ${\cal A}$ and ${\cal B}$, 
one obtains from 
Eqs.~(\ref{eq:iL-shift-dum-1}) and (\ref{eq:iL-shift-dum-11})
\begin{equation}
e^{i {\cal L}({\bf \Gamma}^{\prime}) t} =
e^{i {\cal L}({\bf \Gamma}) t 
   + {\bf a} \cdot \mbox{\boldmath $\kappa$}^{\rm T} \cdot {\bf P} \, t} =
e^{i {\cal L}({\bf \Gamma}) t} \, 
e^{{\bf a} \cdot \mbox{\boldmath $\kappa$}^{\rm T} 
\cdot {\bf P} \, t}.
\label{eq:appen-shifted-p}
\end{equation}
Similarly, there holds for the $f$-propagator
\begin{equation}
e^{- i {\cal L}^{\dagger}({\bf \Gamma}^{\prime}) t} =
e^{- i {\cal L}^{\dagger}({\bf \Gamma}) t} \, 
e^{- {\bf a} \cdot \mbox{\boldmath $\kappa$}^{\rm T} 
\cdot {\bf P} \, t}. 
\label{eq:appen-shifted-f}
\end{equation}

\subsection{Derivation of Eq.~(\ref{eq:GLE-tr-dum-27})}
\label{appen:GLE-tr-dum-27}

Here we derive an expression for 
${\cal P} i {\cal L}_{0} j_{\bf q}^{\lambda}$.
To this end, one needs to evaluate the ensemble averages
$\langle [ i {\cal L}_{0} j_{\bf q}^{\lambda} ] \, 
\rho_{\bf k}^{*} \rangle$ and
$\langle [ i {\cal L}_{0} j_{\bf q}^{\lambda} ] \, 
j_{\bf k}^{\mu \, *} \rangle$
[see Eq.~(\ref{eq:P-tr-def})].
Using Eq.~(\ref{eq:property-iL0})
and the relation
$i {\cal L}_{0} \rho_{\bf q} = i {\bf q} \cdot {\bf j}_{\bf q}$, 
the former is given by
\begin{eqnarray}
& &
\langle [ i {\cal L}_{0} j_{\bf q}^{\lambda} ] \, 
\rho_{\bf k}^{*} \rangle =
- \delta_{{\bf q}, {\bf k}}
\langle j_{\bf q}^{\lambda} \, 
[ i {\cal L}_{0} \rho_{\bf q}]^{*} \rangle 
\nonumber \\
& & \qquad 
=
\delta_{{\bf q}, {\bf k}}
\langle j_{\bf q}^{\lambda} \, 
(i {\bf q} \cdot {\bf j}_{\bf q}^{*}) \rangle =
\delta_{{\bf q}, {\bf k}} \,
i q_{\lambda} N v^{2}. 
\label{eq:GLE-tr-dum-21-2}
\end{eqnarray}
For the latter, we use Eq.~(\ref{eq:continuity-for-j}) to obtain
\begin{equation}
\langle [ i {\cal L}_{0} j_{\bf q}^{\lambda} ] \, 
j_{\bf k}^{\mu \, *} \rangle = 
\delta_{{\bf q}, {\bf k}} 
\sum_{\nu}
\frac{i q_{\nu}}{m}
\langle \sigma_{\bf q}^{\lambda \nu} \, 
j_{\bf q}^{\mu \, *} \rangle = 0,
\label{eq:GLE-tr-dum-23}
\end{equation}
since only odd number of momentum variables are involved.
It thus follows from these results and Eq.~(\ref{eq:P-tr-def}) 
\begin{equation}
{\cal P} i {\cal L}_{0} j_{\bf q}^{\lambda} =
\sum_{\bf k}
\langle [ i {\cal L}_{0} j_{\bf q}^{\lambda} ] \, \rho_{\bf k}^{*} \rangle
\frac{1}{N S_{k}} \rho_{\bf k} =
i q_{\lambda} \frac{v^{2}}{S_{q}} \rho_{\bf q}.
\label{eq:appen-GLE-tr-dum-27}
\end{equation}

\subsection{Derivation of Eqs.~(\ref{eq:GLE-tr-dum-45}) 
and (\ref{eq:GLE-tr-dum-57})}
\label{appen:GLE-tr-dum-45-and-GLE-tr-dum-57}

Using Eq.~(\ref{eq:Lp-inside}), 
the ensemble averages in the integrands of
Eq.~(\ref{eq:GLE-tr-dum-33})
are given by
\begin{widetext}
\begin{eqnarray}
\langle 
[ i {\cal L} R_{\bf q}^{\lambda}(s) ] \, \rho_{{\bf q}(s)}^{*} 
\rangle &=&
- \langle 
R_{\bf q}^{\lambda}(s) \, [ i {\cal L} \rho_{{\bf q}(s)}]^{*} 
\rangle -
\frac{\dot{\gamma}}{k_{\rm B}T}
\langle 
R_{\bf q}^{\lambda}(s) \rho_{{\bf q}(s)}^{*} \sigma_{xy}
\rangle -
\frac{2 \alpha}{k_{\rm B}T}
\langle 
R_{\bf q}^{\lambda}(s) \rho_{{\bf q}(s)}^{*} \delta K
\rangle,
\label{eq:GLE-tr-dum-34}
\\
\langle 
[ i {\cal L} R_{\bf q}^{\lambda}(s) ] \, j_{{\bf q}(s)}^{\mu \, *}
\rangle &=&
- \langle 
R_{\bf q}^{\lambda}(s) \, [ i {\cal L} j_{{\bf q}(s)}^{\mu} ]^{*}
\rangle -
\frac{\dot{\gamma}}{k_{\rm B}T}
\langle 
R_{\bf q}^{\lambda}(s) j_{{\bf q}(s)}^{\mu \, *} \sigma_{xy}
\rangle -
\frac{2 \alpha}{k_{\rm B}T}
\langle 
R_{\bf q}^{\lambda}(s) j_{{\bf q}(s)}^{\mu \, *} \delta K
\rangle.
\label{eq:GLE-tr-dum-35}
\end{eqnarray}
Since ${\cal Q} R_{\bf q}^{\lambda}(s) = R_{\bf q}^{\lambda}(s)$
[see Eqs.~(\ref{eq:R-tr-def-1}) and (\ref{eq:R-tr-def-2})]
and the operator ${\cal Q}$ is idempotent and Hermitian, 
the above equations can be written as
\begin{eqnarray}
\langle 
[ i {\cal L} R_{\bf q}^{\lambda}(s) ] \, \rho_{{\bf q}(s)}^{*} 
\rangle &=&
- \langle 
R_{\bf q}^{\lambda}(s) \, [ {\cal Q} i {\cal L} \rho_{{\bf q}(s)}]^{*} 
\rangle -
\frac{\dot{\gamma}}{k_{\rm B}T}
\langle 
R_{\bf q}^{\lambda}(s) \, {\cal Q} [\rho_{{\bf q}(s)}^{*} \sigma_{xy}]
\rangle -
\frac{2 \alpha}{k_{\rm B}T}
\langle 
R_{\bf q}^{\lambda}(s) \, {\cal Q} [\rho_{{\bf q}(s)}^{*} \delta K]
\rangle,
\label{eq:GLE-tr-dum-34-2}
\\
\langle 
[ i {\cal L} R_{\bf q}^{\lambda}(s) ] \, j_{{\bf q}(s)}^{\mu \, *}
\rangle &=&
- \langle 
R_{\bf q}^{\lambda}(s) \, [ {\cal Q} i {\cal L} j_{{\bf q}(s)}^{\mu} ]^{*}
\rangle -
\frac{\dot{\gamma}}{k_{\rm B}T}
\langle 
R_{\bf q}^{\lambda}(s) \, {\cal Q} [j_{{\bf q}(s)}^{\mu \, *} \sigma_{xy}]
\rangle -
\frac{2 \alpha}{k_{\rm B}T}
\langle 
R_{\bf q}^{\lambda}(s) \, {\cal Q} [j_{{\bf q}(s)}^{\mu \, *} \delta K]
\rangle.
\label{eq:GLE-tr-dum-35-2}
\end{eqnarray}
In the following, we will show that
\begin{eqnarray}
& &
\langle 
R_{\bf q}^{\lambda}(s) \, [ {\cal Q} i {\cal L} \rho_{{\bf q}(s)}]^{*} 
\rangle =0,
\label{eq:GLE-tr-dum-34-3}
\\
& &
\langle 
R_{\bf q}^{\lambda}(s) \, [ {\cal Q} i {\cal L} j_{{\bf q}(s)}^{\mu} ]^{*}
\rangle =
\langle 
R_{\bf q}^{\lambda}(s) \, R_{{\bf q}(s)}^{\mu \, *}
\rangle.
\label{eq:GLE-tr-dum-35-3}
\end{eqnarray}
Substituting these results into 
Eqs.~(\ref{eq:GLE-tr-dum-34-2}) and (\ref{eq:GLE-tr-dum-35-2}) yields
Eqs.~(\ref{eq:GLE-tr-dum-45}) and (\ref{eq:GLE-tr-dum-57}),
respectively.

To derive Eq.~(\ref{eq:GLE-tr-dum-34-3}),
we first notice from Eq.~(\ref{eq:GLE-tr-dum-02})
\begin{equation}
i {\cal L} \rho_{{\bf q}(s)} =
i {\bf q}(s) \cdot {\bf j}_{{\bf q}(s)} +
\sum_{j}
i [ {\bf q}(s) \cdot \mbox{\boldmath $\kappa$} \cdot {\bf r}_{j} ] \,
e^{i {\bf q}(s) \cdot {\bf r}_{j}} =
i {\bf q}(s) \cdot {\bf j}_{{\bf q}(s)} +
{\bf q} \cdot \mbox{\boldmath $\kappa$} \cdot \frac{\partial}{\partial {\bf q}} \,
\rho_{{\bf q}(s)},
\label{eq:GLE-tr-dum-41}
\end{equation}
since $\mbox{\boldmath $\kappa$} \cdot \mbox{\boldmath $\kappa$} = 0$.
We therefore obtain,
since ${\cal Q} j_{{\bf q}(s)}^{\lambda} = 0$,
\begin{equation}
\langle 
R_{\bf q}^{\lambda}(s) \, [ {\cal Q} i {\cal L} \rho_{{\bf q}(s)}]^{*} 
\rangle =
\Bigl\langle 
R_{\bf q}^{\lambda}(s) \, 
\Bigl[
  {\bf q} \cdot \mbox{\boldmath $\kappa$} \cdot \frac{\partial}{\partial {\bf q}} \,
  \rho_{{\bf q}(s)}^{*}
\Bigr]
\Bigr\rangle.
\label{eq:GLE-tr-dum-42}
\end{equation}
On the other hand, 
it follows by taking a partial time derivative of the
first relation in Eq.~(\ref{eq:GLE-tr-dum-32}) that
\begin{eqnarray}
0 =
\frac{\partial}{\partial s}
\langle R_{\bf q}^{\lambda}(s) \rho_{{\bf q}(s)}^{*} \rangle 
&=&
\Bigl\langle
  \Bigl\{ \frac{\partial}{\partial s} 
    \Bigl[ e^{i {\cal QLQ} s} R_{\bf q}^{\lambda} \Bigr]
  \Bigr\}
  \rho_{{\bf q}(s)}^{*}
\Bigr\rangle +
\Bigl\langle
  R_{\bf q}^{\lambda}(s) 
  \Bigl[ \frac{\partial}{\partial s} \rho_{{\bf q}(s)}^{*} \Bigr]
\Bigr\rangle
\nonumber \\
&=&
\langle
  [ {\cal Q} i {\cal L} R_{\bf q}^{\lambda}(s) ] \,
  \rho_{{\bf q}(s)}^{*}
\rangle +
\Bigl\langle 
R_{\bf q}^{\lambda}(s) \, 
\Bigl[
  {\bf q} \cdot \mbox{\boldmath $\kappa$} \cdot \frac{\partial}{\partial {\bf q}} \,
  \rho_{{\bf q}(s)}^{*}
\Bigr]
\Bigr\rangle,
\label{eq:GLE-tr-dum-43}
\end{eqnarray}
where in the final equality 
Eq.~(\ref{eq:GLE-tr-new-dum-11}) has been used
for the second term. 
The first term in this expression is zero since ${\cal Q}$
is Hermitian and ${\cal Q} \rho_{{\bf q}(s)}^{*} = 0$.
We thus obtain
\begin{equation}
\Bigl\langle 
R_{\bf q}^{\lambda}(s) \, 
\Bigl[
  {\bf q} \cdot \mbox{\boldmath $\kappa$} \cdot \frac{\partial}{\partial {\bf q}} \,
  \rho_{{\bf q}(s)}^{*}
\Bigr]
\Bigr\rangle = 0.
\label{eq:GLE-tr-dum-44}
\end{equation}
Equation~(\ref{eq:GLE-tr-dum-34-3}) then follows from
Eqs.~(\ref{eq:GLE-tr-dum-42}) and (\ref{eq:GLE-tr-dum-44}).

We next derive Eq.~(\ref{eq:GLE-tr-dum-35-3}).
To this end, we notice from Eq.~(\ref{eq:GLE-tr-dum-12}) that,
since $\mbox{\boldmath $\kappa$} \cdot \mbox{\boldmath $\kappa$} = 0$,
\begin{equation}
i {\cal L} j_{{\bf q}(s)}^{\mu} = 
i {\cal L}_{0} j_{{\bf q}(s)}^{\mu} +
{\bf q} \cdot \mbox{\boldmath $\kappa$} \cdot \frac{\partial}{\partial {\bf q}} \,
j_{{\bf q}(s)}^{\mu} -
[ \mbox{\boldmath $\kappa$} \cdot {\bf j}_{{\bf q}(s)} ]^{\mu} -
\alpha j_{{\bf q}(s)}^{\mu}.
\label{eq:GLE-tr-dum-51}
\end{equation}
We therefore obtain, since
${\cal Q} j_{{\bf q}(s)}^{\lambda} = 0$, 
\begin{eqnarray}
\langle R_{\bf q}^{\lambda}(s) \,
{\cal Q} [ i {\cal L} j_{{\bf q}(s)}^{\mu} ]^{*} \rangle =
\langle R_{\bf q}^{\lambda}(s) \,
{\cal Q} [ i {\cal L}_{0} j_{{\bf q}(s)}^{\mu} ]^{*} \rangle +
\Bigl\langle 
  R_{\bf q}^{\lambda}(s) \,
  \Bigl[ 
    {\bf q} \cdot \mbox{\boldmath $\kappa$} \cdot \frac{\partial}{\partial {\bf q}} \,
    j_{{\bf q}(s)}^{\mu \, *}
  \Bigr]
\Bigr\rangle.
\label{eq:GLE-tr-dum-52}
\end{eqnarray}
Notice that the thermostat term $\alpha j_{{\bf q}(s)}^{\mu}$ does not contribute
here since we have adopted the constant-$\alpha$ model.
If, e.g., the Gaussian isokinetic thermostat is used,
the contribution 
$\langle R_{\bf q}^{\lambda}(s) \, 
{\cal Q}[ \alpha_{\rm G} j_{{\bf q}(s)}^{\mu \, *}] \rangle$
cannot be discarded. 

The vanishing of the second term in Eq.~(\ref{eq:GLE-tr-dum-52})
can be demonstrated as follows. 
Using the following equation
\begin{equation}
\frac{\partial}{\partial s}
j_{{\bf q}(s)}^{\mu \, *} =
{\bf q} \cdot \mbox{\boldmath $\kappa$} \cdot \frac{\partial}{\partial {\bf q}} \,
j_{{\bf q}(s)}^{\mu \, *},
\label{eq:GLE-tr-dum-53}
\end{equation}
a partial time derivative of the
second relation in Eq.~(\ref{eq:GLE-tr-dum-32}) is given by
\begin{eqnarray}
0 =
\frac{\partial}{\partial s}
\langle R_{\bf q}^{\lambda}(s) j_{{\bf q}(s)}^{\mu \, *} \rangle
&=&
\Bigl\langle
  \Bigl\{ \frac{\partial}{\partial s} 
    \Bigl[ e^{i {\cal QLQ} s} R_{\bf q}^{\lambda} \Bigr]
  \Bigr\}
  j_{{\bf q}(s)}^{\mu \, *} 
\Bigr\rangle +
\Bigl\langle
  R_{\bf q}^{\lambda}(s) 
  \Bigl[ \frac{\partial}{\partial s} j_{{\bf q}(s)}^{\mu \, *} \Bigr]
\Bigr\rangle
\nonumber \\
&=&
\langle
  [ {\cal Q} i {\cal L} R_{\bf q}^{\lambda}(s) ] \,
  j_{{\bf q}(s)}^{\mu \, *}
\rangle +
\Bigl\langle
  R_{\bf q}^{\lambda}(s) 
  \Bigl[
    {\bf q} \cdot \mbox{\boldmath $\kappa$} \cdot \frac{\partial}{\partial {\bf q}} \,
    j_{{\bf q}(s)}^{\mu \, *}
  \Bigr]
\Bigr\rangle.
\label{eq:GLE-tr-dum-54}
\end{eqnarray}
\end{widetext}
The first term is zero
since ${\cal Q} j_{{\bf q}(s)}^{\mu \, *} = 0$, 
and this leads to
\begin{equation}
\Bigl\langle
  R_{\bf q}^{\lambda}(s) 
  \Bigl[
    {\bf q} \cdot \mbox{\boldmath $\kappa$} \cdot \frac{\partial}{\partial {\bf q}} \,
    j_{{\bf q}(s)}^{\mu \, *}
  \Bigr]
\Bigr\rangle = 0.
\label{eq:GLE-tr-dum-55}
\end{equation}
One therefore obtains from 
Eqs.~(\ref{eq:GLE-tr-dum-52}) and (\ref{eq:GLE-tr-dum-55})
\begin{eqnarray}
\langle R_{\bf q}^{\lambda}(s) \,
{\cal Q} [ i {\cal L} j_{{\bf q}(s)}^{\mu} ]^{*} \rangle &=&
\langle R_{\bf q}^{\lambda}(s) \,
{\cal Q} [ i {\cal L}_{0} j_{{\bf q}(s)}^{\mu} ]^{*} \rangle 
\nonumber \\
&=&
\langle R_{\bf q}^{\lambda}(s) \,
R_{{\bf q}(s)}^{\mu \, *} \rangle,
\label{eq:GLE-tr-dum-56}
\end{eqnarray}
where in the final equality we have used
Eq.~(\ref{eq:R-tr-def-2}) for the
definition of the fluctuating force.
This completes the derivation of Eq.~(\ref{eq:GLE-tr-dum-35-3}).

\subsection{Derivation of Eq.~(\ref{eq:MCT-tr-dum-22})}
\label{appen:MCT-tr-dum-22}

Here we calculate the projected random force
${\cal P}_{2} R_{\bf q}^{\lambda}$:
\begin{equation}
{\cal P}_{2} R_{\bf q}^{\lambda} =
\sum_{{\bf k} > {\bf p}}
\langle R_{\bf q}^{\lambda} \rho_{\bf k}^{*} \rho_{\bf p}^{*}
\rangle
\frac{1}{N^{2} S_{k} S_{p}}
\rho_{\bf k} \rho_{\bf p}.
\label{eq:MCT-tr-dum-31}
\end{equation}
To this end, we need to evaluate
[see Eq.~(\ref{eq:R-tr-def-2})]
\begin{equation}
\langle 
R_{\bf q}^{\lambda} \rho_{\bf k}^{*} \rho_{\bf p}^{*} 
\rangle =
\langle
[ i {\cal L}_{0} j_{\bf q}^{\lambda} ] \, 
\rho_{\bf k}^{*} \rho_{\bf p}^{*} 
\rangle -
i q_{\lambda}
\frac{v^{2}}{S_{q}}
\langle
\rho_{\bf q} \rho_{\bf k}^{*} \rho_{\bf p}^{*} 
\rangle.
\label{eq:MCT-tr-dum-32}
\end{equation}
Using Eq.~(\ref{eq:property-iL0})
and the relation
$i {\cal L}_{0} \rho_{\bf q} = i {\bf q} \cdot {\bf j}_{\bf q}$, 
the first term is given by
\begin{eqnarray}
\langle
[ i {\cal L}_{0} j_{\bf q}^{\lambda} ] \,
\rho_{\bf k}^{*} \rho_{\bf p}^{*} 
\rangle &=&
- 
\langle
j_{\bf q}^{\lambda} \,
[ i {\cal L}_{0} \rho_{\bf k} ]^{*} \,
\rho_{\bf p}^{*} 
\rangle
- 
\langle
j_{\bf q}^{\lambda} \,
\rho_{\bf k}^{*} \,
[ i {\cal L}_{0} \rho_{\bf p} ]^{*} \,
\rangle
\nonumber \\
&=&
\langle
j_{\bf q}^{\lambda} \,
( i {\bf k} \cdot {\bf j}_{\bf k}^{*} ) \,
\rho_{\bf p}^{*} 
\rangle +
\langle
j_{\bf q}^{\lambda} \, 
\rho_{\bf k}^{*} \,
( i {\bf p} \cdot {\bf j}_{\bf p}^{*} )
\rangle 
\nonumber \\
&=&
\delta_{{\bf q}, {\bf k}+{\bf p}} \,
i N v^{2} [ k_{\lambda} S_{p} + p_{\lambda} S_{k} ].
\label{eq:MCT-tr-dum-33}
\end{eqnarray}
For the second term in Eq.~(\ref{eq:MCT-tr-dum-32}), we use 
the convolution approximation (\ref{eq:convolution-app}):
\begin{equation}
i q_{\lambda}
\frac{v^{2}}{S_{q}}
\langle
\rho_{\bf q} \rho_{\bf k}^{*} \rho_{\bf p}^{*} 
\rangle \approx
\delta_{{\bf q}, {\bf k} + {\bf p}} \,
i Nv^{2} \, q_{\lambda} \, S_{k} S_{p}.
\label{eq:MCT-tr-dum-34}
\end{equation}
One thus obtains from 
Eqs.~(\ref{eq:MCT-tr-dum-32})--(\ref{eq:MCT-tr-dum-34})
\begin{eqnarray}
\langle
R_{\bf q}^{\lambda} \rho_{\bf k}^{*} \rho_{\bf p}^{*}
\rangle =
- 
\delta_{{\bf q}, {\bf k} + {\bf p}} \,
i N \rho v^{2} S_{k} S_{p}
[ k_{\lambda} c_{k} + p_{\lambda} c_{p} ],
\end{eqnarray}
in terms of the direct correlation function [see Eq.~(\ref{eq:c-def})].
Substituting this result into Eq.~(\ref{eq:MCT-tr-dum-31}) 
finally yields
\begin{equation}
{\cal P}_{2} R_{\bf q}^{\lambda} =
-i \frac{\rho v^{2}}{N}
\sum_{{\bf k} > {\bf p}}
\delta_{{\bf q}, {\bf k}+{\bf p}}
[ k_{\lambda} c_{k} + p_{\lambda} c_{p} ]
\rho_{\bf k} \rho_{\bf p}.
\label{eq:appen-MCT-tr-dum-22}
\end{equation}

\subsection{Derivation of Eq.~(\ref{eq:MCT-tr-dum-41})}
\label{appen:MCT-tr-dum-41}

Here we derive the MCT expression for the memory kernel
$L_{\bf q}^{\lambda}(t)$ defined in Eq.~(\ref{eq:L-tr-def}).
Under the first mode-coupling approximation
$e^{i {\cal QLQ} t} \approx {\cal P}_{2} e^{i {\cal QLQ} t} {\cal P}_{2}$
(see Sec.~\ref{sec:MCT}),
one obtains
\begin{equation}
L_{\bf q}^{\lambda}(t) \approx
i \frac{1}{N k_{\rm B} T S_{q(t)}}
\langle 
[ e^{i {\cal QLQ} t} {\cal P}_{2} R_{\bf q}^{\lambda} ] \, 
{\cal P}_{2} {\cal Q} [ \rho_{{\bf q}(t)}^{*} \sigma_{xy} ] \rangle.
\label{eq:MCT-tr-dum-42}
\end{equation}
Since ${\cal P}_{2} R_{\bf q}^{\lambda}$ is already given in 
Eq.~(\ref{eq:appen-MCT-tr-dum-22}), 
we only need to consider
\begin{equation}
{\cal P}_{2} {\cal Q} [ \rho_{{\bf q}(t)} \sigma_{xy} ] =
\sum_{{\bf k} > {\bf p}}
\langle 
\{ {\cal Q} [ \rho_{{\bf q}(t)} \sigma_{xy} ] \} \,
\rho_{\bf k}^{*} \rho_{\bf p}^{*}
\rangle
\frac{1}{N^{2} S_{k} S_{p}}
\rho_{\bf k} \rho_{\bf p}.
\label{eq:MCT-tr-dum-43}
\end{equation}

Let us start from 
${\cal Q} [ \rho_{{\bf q}(t)} \sigma_{xy} ]$,
for which we need to know the averages
$\langle [ \rho_{{\bf q}(t)} \sigma_{xy} ] \rho_{\bf k}^{*} \rangle$
and 
$\langle [ \rho_{{\bf q}(t)} \sigma_{xy} ] 
j_{\bf k}^{\mu \, *} \rangle$
[see Eq.~(\ref{eq:P-tr-def})].
The latter is zero,
$\langle [ \rho_{{\bf q}(t)} \sigma_{xy} ] 
j_{\bf k}^{\mu \, *} \rangle = 0$,
since this term involves odd number of momentum variables only.
For the former, one obtains using Eq.~(\ref{eq:sigma-def})
\begin{eqnarray}
& &
\langle
[\rho_{{\bf q}(t)} \sigma_{xy} ] \, \rho_{\bf k}^{*}
\rangle =
\delta_{{\bf k}, {\bf q}(t)}
\langle
\rho_{{\bf q}(t)} \sigma_{xy} \rho_{{\bf q}(t)}^{*}
\rangle 
\nonumber \\
& & \qquad
= 
- \delta_{{\bf k}, {\bf q}(t)}
\Bigl\langle
\sum_{i,j,l}
x_{j} \frac{\partial U}{\partial y_{j}}
e^{i {\bf q}(t) \cdot ({\bf r}_{i} - {\bf r}_{\ell})}
\Bigr\rangle,
\label{eq:MCT-tr-dum-51}
\end{eqnarray}
since the kinetic-part contribution from $\sigma_{xy}$ vanishes.
Applying Eq.~(\ref{eq:Yvon-theorem}) to this equation yields
\begin{widetext}
\begin{eqnarray}
\langle
[ \rho_{{\bf q}(t)} \sigma_{xy} ] \, \rho_{\bf k}^{*}
\rangle &=&
- \delta_{{\bf k}, {\bf q}(t)} \,
k_{\rm B}T 
\Bigl\langle
\sum_{i,j,\ell}
x_{j}
\frac{\partial}{\partial y_{j}}
e^{i {\bf q}(t) \cdot ({\bf r}_{i} - {\bf r}_{\ell})}
\Bigr\rangle
\nonumber \\
&=&
- \delta_{{\bf k}, {\bf q}(t)} \,
k_{\rm B}T \, 
q_{y}(t) \,
\langle
\sum_{i,\ell}
i
(x_{i} - x_{\ell})
e^{i {\bf q}(t) \cdot ({\bf r}_{i} - {\bf r}_{\ell})}
\rangle
\nonumber \\
&=&
- \delta_{{\bf k}, {\bf q}(t)} \,
N k_{\rm B}T 
q_{y}(t)
\frac{\partial S_{q(t)}}{\partial q_{x}(t)} =
- \delta_{{\bf k}, {\bf q}(t)} \,
N k_{\rm B}T \frac{q_{x} q_{y}(t)}{q} S_{q(t)}^{\prime},
\label{eq:MCT-tr-dum-52}
\end{eqnarray}
since $q_{x}(t) = q_{x}$ [see Eq.~(\ref{eq:advected-q})] and
$\partial S_{q(t)} / \partial q_{x} = (q_{x}/q) S_{q(t)}^{\prime}$. 
We therefore obtain from Eq.~(\ref{eq:P-tr-def})
\begin{equation}
{\cal P}
[ \rho_{{\bf q}(t)} \sigma_{xy} ] =
\sum_{\bf k}
\langle
[ \rho_{{\bf q}(t)} \sigma_{xy} ] \, \rho_{\bf k}^{*}
\rangle \frac{1}{N S_{k}} \rho_{\bf k} =
- k_{\rm B} T 
\frac{q_{x} q_{y}(t)}{q} \frac{S_{q(t)}^{\prime}}{S_{q(t)}}
\rho_{{\bf q}(t)},
\label{eq:MCT-tr-dum-54}
\end{equation}
and hence
\begin{equation}
{\cal Q}
[ \rho_{{\bf q}(t)} \sigma_{xy} ] =
\rho_{{\bf q}(t)} \sigma_{xy} +
k_{\rm B} T 
\frac{q_{x} q_{y}(t)}{q} \frac{S_{q(t)}^{\prime}}{S_{q(t)}} 
\rho_{{\bf q}(t)}.
\label{eq:MCT-tr-dum-55}
\end{equation}

Now let us calculate
\begin{equation}
\langle
[ {\cal Q} \{ \rho_{{\bf q}(t)} \sigma_{xy} \} ] \,
\rho_{\bf k}^{*} \rho_{\bf p}^{*}
\rangle =
\langle
[\rho_{{\bf q}(t)} \sigma_{xy} ] \,
\rho_{\bf k}^{*} \rho_{\bf p}^{*}
\rangle +
k_{\rm B}T 
\frac{q_{x} q_{y}(t)}{q} \frac{S_{q(t)}^{\prime}}{S_{q(t)}} 
\langle
\rho_{{\bf q}(t)}
\rho_{\bf k}^{*} \rho_{\bf p}^{*}
\rangle.
\label{eq:MCT-tr-dum-61}
\end{equation}
Using Eq.~(\ref{eq:sigma-def}), the first term is given by
\begin{eqnarray}
\langle
[\rho_{{\bf q}(t)} \sigma_{xy} ] \,
\rho_{\bf k}^{*} \rho_{\bf p}^{*}
\rangle &=&
- \delta_{{\bf q}(t),{\bf k}+{\bf p}}
\Bigl\langle
\sum_{i} e^{i {\bf q}(t) \cdot {\bf r}_{i}}
\sum_{j} x_{j} \frac{\partial U}{\partial y_{j}}
\sum_{\ell} e^{- i {\bf k} \cdot {\bf r}_{\ell}}
\sum_{m} e^{- i {\bf p} \cdot {\bf r}_{m}}
\Bigr\rangle 
\nonumber \\
&=&
- \delta_{{\bf q}(t),{\bf k}+{\bf p}} \, 
k_{\rm B} T
\Bigl\langle
\sum_{i,j,\ell,m}
x_{j}
\frac{\partial}{\partial y_{j}}
\Bigl(
  e^{i {\bf q}(t) \cdot {\bf r}_{i}}
  e^{- i {\bf k} \cdot {\bf r}_{\ell}}
  e^{- i {\bf p} \cdot {\bf r}_{m}}
\Bigr)
\Bigr\rangle,
\end{eqnarray}
where we have employed Eq.~(\ref{eq:Yvon-theorem}) in the second equality.
The calculation of this term can be continued in the same manner
as in Eq.~(\ref{eq:MCT-tr-dum-52}) with the result
\begin{equation}
\langle
[ \rho_{{\bf q}(t)} \sigma_{xy} ] \,
\rho_{\bf k}^{*} \rho_{\bf p}^{*}
\rangle =
- \delta_{{\bf q}(t),{\bf k}+{\bf p}}
k_{\rm B} T
\Bigl(
  k_{y} \frac{\partial}{\partial k_{x}} +
  p_{y} \frac{\partial}{\partial p_{x}}
\Bigr)
\langle
\rho_{{\bf q}(t)}
\rho_{\bf k}^{*}
\rho_{\bf p}^{*}
\rangle.
\label{eq:MCT-tr-dum-63}
\end{equation}
Substituting this into Eq.~(\ref{eq:MCT-tr-dum-61}) yields
\begin{eqnarray}
\langle
[ {\cal Q} \{ \rho_{{\bf q}(t)} \sigma_{xy} \} ] \, 
\rho_{\bf k}^{*} \rho_{\bf p}^{*}
\rangle =
- \delta_{{\bf q}(t),{\bf k}+{\bf p}}
k_{\rm B} T
\Bigl\{
  k_{y} \frac{\partial}{\partial k_{x}} +
  p_{y} \frac{\partial}{\partial p_{x}} -
  \frac{q_{x} q_{y}(t)}{q} \frac{S_{q(t)}^{\prime}}{S_{q(t)}} 
\Bigr\}
\langle 
\rho_{{\bf q}(t)}
\rho_{\bf k}^{*}
\rho_{\bf p}^{*}
\rangle.
\label{eq:MCT-tr-dum-64}
\end{eqnarray}
This expression can further be simplified under
the convolution approximation (\ref{eq:convolution-app}),
\begin{equation}
\langle 
\rho_{{\bf q}(t)}
\rho_{\bf k}^{*}
\rho_{\bf p}^{*}
\rangle \approx
\delta_{{\bf q}(t),{\bf k}+{\bf p}} \,
N S_{q(t)} S_{k} S_{p}.
\label{eq:MCT-tr-dum-65}
\end{equation}
Let us notice that, when ${\bf q}(t) = {\bf k} + {\bf p}$, 
there holds
\begin{equation}
\frac{\partial}{\partial k_{x}}
[S_{q(t)} S_{k} S_{p}] =
\frac{q_{x}}{q(t)} S_{q(t)}^{\prime} S_{k} S_{p} +
\frac{k_{x}}{k} S_{q(t)} S_{k}^{\prime} S_{p}.
\label{eq:MCT-tr-dum-66}
\end{equation}
Similarly, we have
\begin{equation}
\frac{\partial}{\partial p_{x}}
[S_{q(t)} S_{k} S_{p}] =
\frac{q_{x}}{q(t)} S_{q(t)}^{\prime} S_{k} S_{p} +
\frac{p_{x}}{p} S_{q(t)} S_{k} S_{p}^{\prime}.
\label{eq:MCT-tr-dum-68}
\end{equation}
It then follows from 
Eqs.~(\ref{eq:MCT-tr-dum-64})--(\ref{eq:MCT-tr-dum-68}) that
\begin{eqnarray}
\langle
[ {\cal Q} \{ \rho_{{\bf q}(t)} \sigma_{xy} \} ] \,
\rho_{\bf k}^{*} \rho_{\bf p}^{*}
\rangle =
- \delta_{{\bf q}(t),{\bf k}+{\bf p}}
N k_{\rm B} T
S_{q(t)} S_{k} S_{p} 
\Bigl\{
  \frac{k_{x} k_{y}}{k}
  \frac{S_{k}^{\prime}}{S_{k}} +
  \frac{p_{x} p_{y}}{p}
  \frac{S_{p}^{\prime}}{S_{p}}
\Bigr\},
\label{eq:MCT-tr-dum-69}
\end{eqnarray}
and substituting this into Eq.~(\ref{eq:MCT-tr-dum-43}) yields
\begin{equation}
{\cal P}_{2} {\cal Q} [ \rho_{{\bf q}(t)} \sigma_{xy} ] =
- \frac{k_{\rm B}T}{N}
S_{q(t)}
\sum_{{\bf k} > {\bf p}}
\delta_{{\bf q}(t), {\bf k}+{\bf p}}
\Bigl\{
  \frac{k_{x} k_{y}}{k}
  \frac{S_{k}^{\prime}}{S_{k}} +
  \frac{p_{x} p_{y}}{p}
  \frac{S_{p}^{\prime}}{S_{p}}
\Bigr\}
\rho_{\bf k} \rho_{\bf p}.
\label{eq:MCT-tr-dum-70}
\end{equation}

Substituting 
Eqs.~(\ref{eq:appen-MCT-tr-dum-22}) and (\ref{eq:MCT-tr-dum-70})
into Eq.~(\ref{eq:MCT-tr-dum-42})
and then using the factorization 
approximation (\ref{eq:factorization-app}),
we finally obtain with ${\bf p} \equiv {\bf q} - {\bf k}$
\begin{equation}
L_{\bf q}^{\lambda}(t) =
- \frac{v^{2}}{2 (2 \pi)^{3}}
\int d{\bf k} \,
[ k_{\lambda} c_{k} + p_{\lambda} c_{p} ] \, 
\Bigl[
  \frac{k_{x} k_{y}(t)}{k(t)} \frac{S_{k(t)}^{\prime}}{S_{k(t)}} +
  \frac{p_{x} p_{y}(t)}{p(t)} \frac{S_{p(t)}^{\prime}}{S_{p(t)}}
\Bigr] \,
F_{\bf k}(t) F_{\bf p}(t).
\label{eq:appen-MCT-tr-dum-41}
\end{equation}
\end{widetext}

\subsection{Derivation of Eq.~(\ref{eq:MCT-tr-dum-81})}
\label{appen:MCT-tr-dum-81}

Here we show that the memory kernel
$L_{\bf q}^{\prime \, \lambda \mu}(t)$ 
defined in Eq.~(\ref{eq:L-prime-tr-def}) vanishes 
under the mode-coupling approximation formulated with ${\cal P}_{2}$. 
We start from
\begin{equation}
L_{\bf q}^{\prime \, \lambda \mu}(t) \approx
\frac{m}{N (k_{\rm B}T)^{2}}
\langle 
[ e^{i {\cal QLQ} t} {\cal P}_{2} R_{\bf q}^{\lambda}(t) ] \, 
{\cal P}_{2} {\cal Q} [ j_{{\bf q}(t)}^{\mu \, *} \sigma_{xy} ] \rangle,
\label{eq:MCT-tr-dum-82}
\end{equation}
under the first mode-coupling approximation
$e^{i {\cal QLQ} t} \approx {\cal P}_{2} e^{i {\cal QLQ} t} {\cal P}_{2}$
(see Sec.~\ref{sec:MCT}).
In the following, we demonstrate
$\langle
\{ {\cal Q} [ j_{{\bf q}(t)}^{\mu} \sigma_{xy} ] \} \,
\rho_{\bf k}^{*} \rho_{\bf p}^{*} \rangle = 0$, i.e.,
${\cal P}_{2} {\cal Q} [ j_{{\bf q}(t)}^{\mu} \sigma_{xy} ] = 0$, 
which completes the derivation of 
$L_{\bf q}^{\prime \, \lambda \mu}(t) = 0$. 

Let us start from 
${\cal Q} [ j_{{\bf q}(t)}^{\mu} \sigma_{xy} ]$,
for which we need to know the averages
$\langle [ j_{{\bf q}(t)}^{\mu} \sigma_{xy} ] 
\rho_{\bf k}^{*} \rangle$
and 
$\langle [ j_{{\bf q}(t)}^{\mu} \sigma_{xy} ] 
j_{\bf k}^{\nu \, *} \rangle$
[see Eq.~(\ref{eq:P-tr-def})].
The former is zero,
$\langle [ j_{{\bf q}(t)}^{\mu} \sigma_{xy} ] 
\rho_{\bf k}^{*} \rangle = 0$,
since this term involves odd number of momentum variables only.
Using Eq.~(\ref{eq:sigma-def}), the latter reads
\begin{widetext}
\begin{equation}
\langle 
[ j_{{\bf q}(t)}^{\mu} \sigma_{xy} ] \, 
j_{\bf k}^{\nu \, *} 
\rangle =
\delta_{{\bf k}, {\bf q}(t)}
\langle
j_{{\bf q}(t)}^{\mu} \sigma_{xy} 
j_{{\bf q}(t)}^{\nu \, *}
\rangle =
\delta_{{\bf k}, {\bf q}(t)}
\Bigl\langle
\sum_{i}
\frac{p_{i}^{\mu}}{m} e^{i {\bf q}(t) \cdot {\bf r}_{i}}
\sum_{j}
\Bigl(
  \frac{p_{j}^{x} p_{j}^{y}}{m} - x_{j} \frac{\partial U}{\partial y_{j}}
\Bigr)
\sum_{\ell}
\frac{p_{\ell}^{\nu}}{m} e^{- i {\bf q}(t) \cdot {\bf r}_{\ell}}
\Bigr\rangle.
\label{eq:MCT-tr-dum-84}
\end{equation}
\end{widetext}
In this equation,
the kinetic-part contribution survives only when
(i) $i=j=\ell$, $\mu = x$, $\nu = y$, and
(ii) $i=j=\ell$, $\mu = y$, $\nu = x$,
and the potential-term contribution survives only
when $i=\ell$, $\mu = \nu$.
This leads to
\begin{eqnarray}
\langle 
[ j_{{\bf q}(t)}^{\mu} \sigma_{xy} ] \,
j_{\bf k}^{\nu \, *} 
\rangle =
\delta_{{\bf k}, {\bf q}(t)}
N mv^{4}
( \delta_{\mu x} \delta_{\nu y} + \delta_{\mu y} \delta_{\nu x} ),
\label{eq:MCT-tr-dum-85}
\end{eqnarray}
since the potential-term contribution vanishes
after applying Eq.~(\ref{eq:Yvon-theorem}). 
We therefore obtain from Eq.~(\ref{eq:P-tr-def})
\begin{eqnarray}
{\cal P}
[ j_{{\bf q}(t)}^{\mu} \sigma_{xy} ] &=&
\sum_{\bf k} \sum_{\nu}
\langle 
[ j_{{\bf q}(t)}^{\mu} \sigma_{xy} ] \,
j_{\bf k}^{\nu \, *} 
\rangle 
\frac{1}{Nv^{2}} j_{\bf k}^{\nu} 
\nonumber \\
&=&
mv^{2}
[ \, \delta_{\mu x} j_{{\bf q}(t)}^{y} +
     \delta_{\mu y} j_{{\bf q}(t)}^{x}   \, ],
\end{eqnarray}
and hence
\begin{equation}
{\cal Q}
[ j_{{\bf q}(t)}^{\mu} \sigma_{xy} ] =
j_{{\bf q}(t)}^{\mu} \sigma_{xy} -
mv^{2}
[ \delta_{\mu x} j_{{\bf q}(t)}^{y} +
  \delta_{\mu y} j_{{\bf q}(t)}^{x}   ].
\end{equation}
Since the right-hand side of this equation involves
odd number of momentum variables only, there holds
\begin{equation}
\langle
\{ {\cal Q} [ j_{{\bf q}(t)}^{\mu} \sigma_{xy} ] \} \,
\rho_{\bf k}^{*} \rho_{\bf p}^{*} \rangle =0.
\end{equation}

\subsection{Derivation of Eq.~(\ref{eq:MCT-tr-dum-91})}
\label{appen:MCT-tr-dum-91}

Here we show that the memory kernels
$N_{\bf q}^{\lambda}(t)$ and 
$N_{\bf q}^{\prime \, \lambda \mu}(t)$ 
defined in 
Eqs.~(\ref{eq:N-tr-def}) and (\ref{eq:N-prime-tr-def}) vanish 
under the mode-coupling approximation formulated with ${\cal P}_{2}$. 
We start from the following expressions 
under the approximation
$e^{i {\cal QLQ} t} \approx {\cal P}_{2} e^{i {\cal QLQ} t} {\cal P}_{2}$
(see Sec.~\ref{sec:MCT}):
\begin{equation}
N_{\bf q}^{\lambda}(t) \approx
i \frac{2}{N k_{\rm B}T S_{q(t)}}
\langle 
[ e^{i {\cal QLQ} t} {\cal P}_{2} R_{\bf q}^{\lambda}(t) ] \, 
{\cal P}_{2} {\cal Q} [ \rho_{{\bf q}(t)}^{*} \delta K ] \rangle,
\label{eq:MCT-tr-dum-92-1}
\end{equation}
\begin{equation}
N_{\bf q}^{\prime \, \lambda \mu}(t) \approx
\frac{2m}{N (k_{\rm B}T)^{2}}
\langle 
[ e^{i {\cal QLQ} t} {\cal P}_{2} R_{\bf q}^{\lambda}(t) ] \, 
{\cal P}_{2} {\cal Q} [ j_{{\bf q}(t)}^{\mu \, *} \delta K ] \rangle.
\label{eq:MCT-tr-dum-92-2}
\end{equation}
In the following, we demonstrate
$\langle
\{ {\cal Q} [ \rho_{{\bf q}(t)} \delta K ] \} \,
\rho_{\bf k}^{*} \rho_{\bf p}^{*} \rangle = 0$ and
$\langle
\{ {\cal Q} [ j_{{\bf q}(t)}^{\mu} \delta K ] \} \,
\rho_{\bf k}^{*} \rho_{\bf p}^{*} \rangle = 0$. 
This means
${\cal P}_{2} {\cal Q} [ \rho_{{\bf q}(t)} \delta K ] = 0$ and
${\cal P}_{2} {\cal Q} [ j_{{\bf q}(t)}^{\mu} \delta K ] = 0$,
and hence, completes the derivation of
$N_{\bf q}^{\lambda}(t) = 0$ and
$N_{\bf q}^{\prime \, \lambda \mu}(t) = 0$.

Let us start from 
${\cal Q} [ \rho_{{\bf q}(t)} \delta K ]$,
for which we need to know the averages
$\langle [ \rho_{{\bf q}(t)} \delta K ] \rho_{\bf k}^{*} \rangle$
and 
$\langle [ \rho_{{\bf q}(t)} \delta K ] 
j_{\bf k}^{\mu \, *} \rangle$
[see Eq.~(\ref{eq:P-tr-def})].
In view of Eq.~(\ref{eq:deltaK-def}),
one easily obtains
\begin{equation}
\langle [ \rho_{{\bf q}(t)} \delta K ] \rho_{\bf k}^{*} \rangle = 0,
\quad 
\langle [ \rho_{{\bf q}(t)} \delta K ] 
j_{\bf k}^{\mu \, *} \rangle = 0.
\end{equation}
Thus, 
${\cal P} [ \rho_{{\bf q}(t)} \delta K ] = 0$, and hence,
\begin{equation}
{\cal Q} [ \rho_{{\bf q}(t)} \delta K ] = 
\rho_{{\bf q}(t)} \delta K.
\end{equation}
Likewise, one obtains
\begin{equation}
{\cal Q} [ j_{{\bf q}(t)}^{\mu} \delta K ] = 
j_{{\bf q}(t)}^{\mu} \delta K.
\end{equation}
It is then obvious that
\begin{equation}
\langle
\{ {\cal Q} [ \rho_{{\bf q}(t)} \delta K ] \} \,
\rho_{\bf k}^{*} \rho_{\bf p}^{*} \rangle = 0, \,\,\,
\langle
\{ {\cal Q} [ j_{{\bf q}(t)}^{\mu} \delta K ] \} \,
\rho_{\bf k}^{*} \rho_{\bf p}^{*} \rangle = 0.
\end{equation}

\subsection{Derivation of Eqs.~(\ref{eq:G-H-dum-2})}
\label{appen:remark-TTCF}

Here we show that the functions $G_{X}(t)$ and $H_{X}(t)$ defined
in Eq.~(\ref{eq:G-H-dum-1}) evolve in time within the
subspace orthogonal to $\{ \rho_{\bf k}, j_{\bf k}^{\mu} \}$,
i.e., there hold
\begin{eqnarray}
G_{X}(t) =
\langle [ e^{i {\cal L} t} X ] \sigma_{xy} \rangle =
\langle [ e^{i {\cal QLQ} t} {\cal Q} X ] {\cal Q} \sigma_{xy} \rangle,
\label{eq:remark-TTCF-dum-01}
\\
H_{X}(t) =
\langle [ e^{i {\cal L} t} X ] \delta K \rangle =
\langle [ e^{i {\cal QLQ} t} {\cal Q} X ] {\cal Q} \delta K \rangle,
\label{eq:remark-TTCF-dum-02}
\end{eqnarray}
in terms of the projection operator ${\cal Q}$
complementary to ${\cal P}$ defined in Eq.~(\ref{eq:P-tr-def}).
Before embarking on the derivation, 
let us notice
\begin{equation}
{\cal Q} \sigma_{xy} = \sigma_{xy}
\,\,\, \mbox{ and } \,\,\,
{\cal Q} \delta K = \delta K.
\label{eq:Q-sigma-Q-deltaK}
\end{equation}
The first relation follows from
$\langle \sigma_{xy} \rho_{\bf k} \rangle =
\delta_{{\bf k}, {\bf 0}}
\langle \sigma_{xy} \rho_{\bf k=0} \rangle = 0$
[see Eq.~(\ref{eq:rho-def})]
and $\langle \sigma_{xy} j_{\bf k}^{\mu \, *} \rangle = 0$, and
the second relation can be derived in a similar manner.
Thus, the presence of the operator ${\cal Q}$
in front of $\sigma_{xy}$ and $\delta K$ in 
Eqs.~(\ref{eq:remark-TTCF-dum-01}) and
(\ref{eq:remark-TTCF-dum-02})
is irrelevant. 

In the following, we shall deal with the function
$G_{X}(t)$ only, since $H_{X}(t)$ can be 
handled in a similar manner.
Applying the identity
\begin{equation}
e^{i {\cal L} t} =
e^{i {\cal LQ} t} + \int_{0}^{t} ds \,
e^{i {\cal L} (t-s)} \, i {\cal LP} \,
e^{i {\cal LQ} s},
\label{eq:remark-TTCF-dum-11}
\end{equation}
[notice the difference in the order of operators
compared to the identity (\ref{eq:operator-identity-1})],
one finds
\begin{widetext}
\begin{eqnarray}
& &
e^{i {\cal L} t} X =
e^{i {\cal LQ} t} X +
\int_{0}^{t} ds \,
e^{i {\cal L} (t-s)} \, i {\cal LP} \,
e^{i {\cal LQ} s} X
\nonumber \\
& & 
=
e^{i {\cal LQ} t} X +
\sum_{\bf k} \frac{1}{NS_{k}}
\int_{0}^{t} ds 
\langle [ e^{i {\cal LQ} s} X ] \rho_{\bf k}^{*} \rangle 
e^{i {\cal L} (t-s)} i {\cal L} \rho_{\bf k} +
\sum_{\bf k} \sum_{\mu}
\frac{1}{Nv^{2}}
\int_{0}^{t} ds 
\langle [ e^{i {\cal LQ} s} X ] j_{\bf k}^{\mu \, *} \rangle
e^{i {\cal L} (t-s)} i {\cal L} j_{\bf k}^{\mu},
\label{eq:remark-TTCF-dum-12}
\end{eqnarray}
\end{widetext}
where we have used the definition (\ref{eq:P-tr-def})
of the operator ${\cal P}$ and noticed that the
ensemble averaged terms are independent of the phase
and are unaffected by the Liouvillean and the propagator.
Let us notice here that
\begin{equation}
i {\cal L} \rho_{\bf k} \to 0 
\,\,\, \mbox{ and } \,\,\,
i {\cal L} j_{\bf k}^{\mu} \to 0 
\,\,\, \mbox{ for } \,\,\,
{\bf k} \to 0.
\label{eq:remark-TTCF-dum-13}
\end{equation}
The former is obvious in view of Eq.~(\ref{eq:GLE-tr-dum-02}),
while the latter can be derived
on the basis of Eq.~(\ref{eq:GLE-tr-dum-12})
by noticing 
$i {\cal L}_{0} j_{\bf k}^{\lambda} = 
(1/m) \sum_{\mu} i k_{\mu} \sigma_{\bf k}^{\lambda \mu}$
[see Eq.~(\ref{eq:continuity-for-j})]
and 
$j_{\bf k=0}^{\lambda} = (1/m) \sum_{i} p_{i}^{\lambda} = 0$
[see the comment below Eq.~(\ref{eq:SLLOD-b})].
Equation~(\ref{eq:remark-TTCF-dum-13}) simply 
expresses the fact that the density and the current density,
the latter being defined for sheared systems 
in terms of the peculiar momenta,
are conserved variables. 

Now let us consider the transient correlator
$\langle [ e^{i {\cal L} t} X ] \, \sigma_{xy} \rangle$
formed with the ``zero wave-vector'' quantity $\sigma_{xy}$. 
The translational invariance implies that
\begin{equation}
\langle [ e^{i {\cal L} (t-s)} A_{\bf k} ] \, \sigma_{xy} \rangle =
\delta_{{\bf k}, {\bf 0}} \,
\langle [ e^{i {\cal L} (t-s)} A_{\bf k=0} ] \, \sigma_{xy} \rangle 
= 0,
\label{eq:remark-TTCF-dum-21}
\end{equation}
for $A_{\bf k} = i {\cal L} \rho_{\bf k}$ and
$i {\cal L} j_{\bf k}^{\mu}$ 
because of Eq.~(\ref{eq:remark-TTCF-dum-13}).
Thus, there is no contribution to 
$\langle [ e^{i {\cal L} t} X ] \, \sigma_{xy} \rangle$
from the second and third terms on the right-hand side of
Eq.~(\ref{eq:remark-TTCF-dum-12}), and we obtain
\begin{equation}
G_{X}(t) =
\langle [ e^{i {\cal L} t} X ] \, \sigma_{xy} \rangle =
\langle [ e^{i {\cal LQ} t} X ] \, \sigma_{xy} \rangle. 
\label{eq:remark-TTCF-dum-22}
\end{equation}
Since the operator ${\cal Q}$ is idempotent and Hermitian,
one finds using Eq.~(\ref{eq:Q-sigma-Q-deltaK})
\begin{eqnarray}
G_{X}(t) &=&
\langle [ {\cal Q} e^{i {\cal LQ} t} X ] \, 
{\cal Q} \sigma_{xy} \rangle 
\nonumber \\
&=&
\langle [ {\cal Q} e^{i {\cal QLQ} t} {\cal Q} X ] \, 
{\cal Q} \sigma_{xy} \rangle 
\nonumber \\
&=&
\langle [ e^{i {\cal QLQ} t} {\cal Q} X ] \, 
{\cal Q} \sigma_{xy} \rangle,
\label{eq:remark-TTCF-dum-23}
\end{eqnarray}
where in the second equality we have noticed
\begin{equation}
{\cal Q} e^{i {\cal LQ} t} =
{\cal Q} e^{i {\cal QLQ} t} {\cal Q}.
\label{eq:remark-TTCF-dum-24}
\end{equation}
This completes the derivation of 
Eq.~(\ref{eq:remark-TTCF-dum-01}), and
Eq.~(\ref{eq:remark-TTCF-dum-02}) can be derived 
in a similar manner.

\subsection{Derivation of Eq.~(\ref{eq:P2-sigma})}
\label{appen:P2-sigma}

Here we derive the expression for 
${\cal P}_{2}^{0} \sigma_{xy}$.
For this purpose, we need to know
the average $\langle \sigma_{xy} \rho_{\bf k} \rho_{\bf k}^{*} \rangle$.
Using Eq.~(\ref{eq:sigma-def}), this average can be
written as 
\begin{eqnarray}
\langle \sigma_{xy} \, \rho_{\bf k} \rho_{\bf k}^{*} \rangle &=&
- \sum_{i} 
\Bigl\langle
  x_{i} \frac{\partial U}{\partial y_{i}} \, 
  \rho_{\bf k} \rho_{\bf k}^{*}
\Bigr\rangle 
\nonumber \\
&=&
- k_{\rm B}T \sum_{i} 
\Bigl\langle
  x_{i} \frac{\partial}{\partial y_{i}} \,
  [ \rho_{\bf k} \rho_{\bf k}^{*} ]
\Bigr\rangle,
\end{eqnarray}
where we have used Eq.~(\ref{eq:Yvon-theorem}).
Since 
$\partial \rho_{\bf k} /\partial y_{i} =
i k_{y} e^{i {\bf k} \cdot {\bf r}_{i}}$ and
$\partial \rho_{\bf k} / \partial k_{x} =
i \sum_{i} x_{i} e^{i {\bf k} \cdot {\bf r}_{i}}$, 
we obtain
\begin{widetext}
\begin{eqnarray}
\langle \sigma_{xy} \, \rho_{\bf k} \rho_{\bf k}^{*} \rangle &=&
- k_{\rm B}T 
\Bigl\{
  i k_{y} 
  \Bigl\langle \Bigl( 
  \sum_{i} x_{i} e^{i {\bf k} \cdot {\bf r}_{i}} \Bigr)
  \rho_{\bf k}^{*} \Bigr\rangle -
  i k_{y} 
  \Bigl\langle \Bigl( 
  \sum_{i} x_{i} e^{- i {\bf k} \cdot {\bf r}_{i}} \Bigr)
  \rho_{\bf k} \Bigr\rangle
\Bigr\}
\nonumber \\
&=&
- k_{\rm B}T \, k_{y}
\Bigl\{
  \Bigl\langle \Bigl( 
  \frac{\partial}{\partial k_{x}} \rho_{\bf k} \Bigr)
  \rho_{\bf k}^{*} \Bigr\rangle +
  \Bigl\langle \rho_{\bf k} \Bigl( 
  \frac{\partial}{\partial k_{x}} \rho_{\bf k}^{*} \Bigr)
  \Bigr\rangle
\Bigr\} =
- N k_{\rm B} T \,
\frac{k_{x} k_{y}}{k} \, S_{k}^{\prime}.
\end{eqnarray}
\end{widetext}
It then follows from Eq.~(\ref{eq:P2-0-tr-def}) that
\begin{eqnarray}
{\cal P}_{2}^{0} \sigma_{xy} &=&
\sum_{{\bf k} > 0}
\langle \sigma_{xy} \, \rho_{\bf k} \rho_{\bf k}^{*} \rangle
\frac{1}{N^{2} S_{k}^{2}}
\rho_{\bf k} \rho_{\bf k}^{*} 
\nonumber \\
&=&
- \frac{k_{\rm B}T}{N}
\sum_{{\bf k} > 0}
\frac{k_{x} k_{y}}{k}
\frac{S_{k}^{\prime}}{S_{k}^{2}} \,
\rho_{\bf k} \rho_{\bf k}^{*}.
\end{eqnarray}

\section{Isotropic approximation}
\label{appen:isotropic-app}

In this appendix, we shall introduce the isotropic approximation 
which considerably simplifies the wave-vector-dependent MCT equations 
(\ref{eq:summary-GLE-transient}) and (\ref{eq:summary-MCT-transient})
for the transient density correlators.
Such a simplifying approximation is useful in practical applications of
our theory to systems where anisotropy in the density fluctuations is small.
We also argue that the anisotropic nature of steady-state quantities
like the shear stress can nevertheless be captured within such an approximation.

\subsection{MCT equations for the transient correlators}
\label{appen:isotropic-app-1}

The isotropic approximation consists of the following three assumptions.
First, it is assumed that $F_{\bf q}(t)$ depends only on the modulus 
$q = |{\bf q}|$, i.e.,
\begin{equation}
F_{\bf q}(t) \approx F_{q}(t).
\label{eq:iso-app-dum-21}
\end{equation}
Second, we introduce a corresponding approximation for the transient
cross correlator $H_{\bf q}^{\lambda}(t)$ formed with current
density fluctuations. 
Since $H_{\bf q}^{\lambda}(t)$ is a vector correlator whose
orientational dependence comes also from the dependence
on $\lambda$, one cannot introduce such a simple
approximation like 
$H_{\bf q}^{\lambda}(t) \approx H_{q}^{\lambda}(t)$.
Instead, we assume that the following relation,
valid for isotropic quiescent systems, to hold:
\begin{equation}
H_{\bf q}^{\lambda}(t) \approx
- i \frac{q_{\lambda}}{q^{2}} \, \frac{\partial}{\partial t} F_{q}(t).
\label{eq:iso-app-dum-31}
\end{equation}
The third assumption concerns the modulus of the advected
wave vector ${\bf q}(t)$ [see Eq.~(\ref{eq:advected-q})]:
\begin{equation}
q(t)^{2} = q^{2} + 2 (\dot{\gamma}t) \, q_{x} q_{y} +
(\dot{\gamma}t)^{2} q_{x}^{2}.
\label{eq:iso-app-dum-42}
\end{equation}
We assume that $q(t)^{2}$ can be approximated by its orientational
average.
This is equivalent to neglecting the anisotropic term $q_{x} q_{y}$ and
approximating $q_{x}^{2}$ by $q^{2}/3$ in Eq.~(\ref{eq:iso-app-dum-42}),
leading to
\begin{equation}
q(t) \approx 
q \sqrt{ 1 + (\dot{\gamma}t)^{2} /3 } \equiv \bar{q}(t).
\label{eq:iso-app-dum-43}
\end{equation}
In the following, we shall see consequences of these assumptions.

It follows from the approximation (\ref{eq:iso-app-dum-31}) 
\begin{equation}
\frac{\partial}{\partial t} F_{q}(t) = i {\bf q} \cdot {\bf H}_{\bf q}(t),
\label{eq:iso-app-dum-101}
\end{equation}
implying that ${\bf q} \cdot {\bf H}_{\bf q}(t)$ also becomes an isotropic quantity.
Equation~(\ref{eq:iso-app-dum-101}) is consistent with Eq.~(\ref{eq:summary-GLE-transient-a})
under the isotropic approximation.
To see this, we first rewrite Eq.~(\ref{eq:summary-GLE-transient-a}) as 
\begin{equation}
\frac{\partial}{\partial t} F_{\bf q}(t) =
\dot{\gamma} q_{x} \frac{\partial}{\partial q_{y}} \,
F_{\bf q}(t) + i {\bf q} \cdot {\bf H}_{\bf q}(t),
\label{eq:iso-app-dum-15}
\end{equation}
where the specific form 
$\kappa_{\lambda \mu} = 
\dot{\gamma} \delta_{\lambda x} \delta_{\mu y}$
for the shear-rate tensor has been used. 
The application of the approximation (\ref{eq:iso-app-dum-21}) then yields
\begin{equation}
\frac{\partial}{\partial t} F_{q}(t) =
\dot{\gamma} \frac{q_{x} q_{y}}{q} \frac{\partial}{\partial q} F_{q}(t) +
i {\bf q} \cdot {\bf H}_{\bf q}(t),
\label{eq:iso-app-dum-22}
\end{equation}
because
$\partial F_{q}(t)/\partial q_{y} = (q_{y}/q) \partial F_{q}(t)/\partial q$. 
Since now the left-hand side depends only on the modulus $q$,
the orientational averaging of this expression gives Eq.~(\ref{eq:iso-app-dum-101}). 

From a partial time derivative of Eq.~(\ref{eq:iso-app-dum-101}), we obtain
\begin{equation}
\frac{\partial^{2}}{\partial t^{2}} F_{q}(t) =
i {\bf q} \cdot \frac{\partial}{\partial t} {\bf H}_{\bf q}(t).
\label{eq:iso-app-dum-11}
\end{equation}
Substituting Eq.~(\ref{eq:summary-GLE-transient-b}) into the right-hand side yields
\begin{widetext}
\begin{eqnarray}
\frac{\partial^{2}}{\partial t^{2}} F_{q}(t) &=&
\sum_{\lambda}
i q_{\lambda} 
\Bigl[ {\bf q} \cdot \mbox{\boldmath $\kappa$} \cdot
  \frac{\partial}{\partial {\bf q}} \Bigr] H_{\bf q}^{\lambda}(t) 
- q^{2} \frac{v^{2}}{S_{q}} F_{\bf q}(t) -
\sum_{\lambda} i q_{\lambda}
[\mbox{\boldmath $\kappa$} \cdot {\bf H}_{\bf q}(t)]^{\lambda} -
\alpha [ i {\bf q} \cdot {\bf H}_{\bf q}(t) ]
\nonumber \\
& &
- \,
\sum_{\lambda,\mu}
\int_{0}^{t} ds \,
i q_{\lambda} M_{\bf q}^{\lambda \mu}(s) \,
H_{{\bf q}(s)}^{\mu}(t-s) -
\dot{\gamma}
\sum_{\lambda}
\int_{0}^{t} ds \,
q_{\lambda} L_{\bf q}^{\lambda}(s) \,
F_{{\bf q}(s)}(t-s).
\label{eq:iso-app-dum-12}
\end{eqnarray}
Applying the approximation~(\ref{eq:iso-app-dum-21}) to the second and sixth terms,
(\ref{eq:iso-app-dum-31}) to the fifth term, and
(\ref{eq:iso-app-dum-101}) to the fourth term, we obtain
\begin{eqnarray}
\ddot{F}_{q}(t) &=&
\sum_{\lambda}
i q_{\lambda} 
\Bigl[ {\bf q} \cdot \mbox{\boldmath $\kappa$} \cdot
  \frac{\partial}{\partial {\bf q}} \Bigr] H_{\bf q}^{\lambda}(t) 
- q^{2} \frac{v^{2}}{S_{q}} F_{q}(t) -
\sum_{\lambda} i q_{\lambda}
[\mbox{\boldmath $\kappa$} \cdot {\bf H}_{\bf q}(t)]^{\lambda} -
\alpha \dot{F}_{q}(t)
\nonumber \\
& & 
- \,
\int_{0}^{t} ds \,
\Bigl[
  \sum_{\lambda,\mu}
  q_{\lambda} M_{\bf q}^{\lambda \mu}(s) q_{\mu}(s) / q(s)^{2}
\Bigr] 
\dot{F}_{q(s)}(t-s) -
\dot{\gamma}  \int_{0}^{t} ds \,
\Bigl[ \sum_{\lambda} q_{\lambda} L_{\bf q}^{\lambda}(s) \Bigr] 
F_{q(s)}(t-s),
\label{eq:iso-app-dum-13}
\end{eqnarray}
where the dot denotes the partial time derivative. 
The first and third terms on the right-hand side of this equation
can be manipulated as 
\begin{eqnarray}
\sum_{\lambda}
i q_{\lambda} 
\Bigl[ {\bf q} \cdot \mbox{\boldmath $\kappa$} \cdot
  \frac{\partial}{\partial {\bf q}} \Bigr] H_{\bf q}^{\lambda}(t) &=&
\sum_{\lambda}
\Bigl[ {\bf q} \cdot \mbox{\boldmath $\kappa$} \cdot
  \frac{\partial}{\partial {\bf q}} \Bigr] 
[ i q_{\lambda} H_{\bf q}^{\lambda}(t) ] -
\sum_{\lambda}
\Bigl[ {\bf q} \cdot \mbox{\boldmath $\kappa$} \cdot
  \frac{\partial}{\partial {\bf q}} i q_{\lambda} \Bigr] 
H_{\bf q}^{\lambda}(t)
\nonumber \\
&=& 
\dot{\gamma} q_{x} \frac{\partial}{\partial q_{y}} \, 
[ i {\bf q} \cdot {\bf H}_{q}(t) ] -
\sum_{\lambda}
\Bigl[ \dot{\gamma} q_{x} \frac{\partial}{\partial q_{y}} i q_{\lambda} \Bigr]
H_{\bf q}^{\lambda}(t)
\nonumber \\
&= &
\dot{\gamma} q_{x} \frac{\partial}{\partial q_{y}} \, \dot{F}_{q}(t) -
\dot{\gamma} \, [iq_{x} H_{\bf q}^{y}(t)] =
\dot{\gamma} \frac{q_{x} q_{y}}{q} 
\Bigl[ \frac{\partial}{\partial q} - \frac{1}{q} \Bigr] \dot{F}_{q}(t),
\label{eq:iso-app-dum-17}
\end{eqnarray}
\begin{equation}
\sum_{\lambda} i q_{\lambda}
[\mbox{\boldmath $\kappa$} \cdot {\bf H}_{\bf q}(t)]^{\lambda} =
\sum_{\lambda} i q_{\lambda}
\Bigl[ \sum_{\mu} \dot{\gamma} \delta_{\lambda x} \delta_{\mu y} 
H_{\bf q}^{\mu}(t)  \Bigr] =
\dot{\gamma} \, [iq_{x} H_{\bf q}^{y}(t)] =
\dot{\gamma} \frac{q_{x} q_{y}}{q^{2}} 
\dot{F}_{q}(t),
\label{eq:iso-app-dum-18}
\end{equation}
in deriving which 
we have used the approximations~(\ref{eq:iso-app-dum-31}) and (\ref{eq:iso-app-dum-101}).
Both of these terms are anisotropic, and vanish after taking the orientational average. 
[Remember that the left-hand side of Eq.~(\ref{eq:iso-app-dum-13}) depends only on the
modulus $q$.]
We therefore obtain 
\begin{equation}
\ddot{F}_{q}(t) +
q^{2} \frac{v^{2}}{S_{q}} F_{q}(t) +
\alpha \dot{F}_{q}(t) +
\int_{0}^{t} ds \, M_{\bf q}^{\rm iso}(s) \,
\dot{F}_{\bar{q}(s)}(t-s) +
\dot{\gamma}
\int_{0}^{t} ds \,
L_{\bf q}^{\rm iso}(s) \, F_{\bar{q}(s)}(t-s) = 0,
\label{eq:iso-app-dum-53}
\end{equation}
where we have employed the approximation~(\ref{eq:iso-app-dum-43}) 
for the modulus of the advected wave number in the fourth and fifth terms, and
introduced
\begin{eqnarray}
M_{\bf q}^{\rm iso}(t) &\equiv&
\sum_{\lambda, \mu} 
q_{\lambda} M_{\bf q}^{\lambda \mu}(t) q_{\mu}(t) / q(t)^{2},
\label{eq:iso-app-dum-54}
\\
L_{\bf q}^{\rm iso}(t) &\equiv&
\sum_{\lambda} q_{\lambda} L_{\bf q}^{\lambda}(t).
\label{eq:iso-app-dum-55}
\end{eqnarray}

Now, we are left with the kernels
$M_{\bf q}^{\rm iso}(t)$ and $L_{\bf q}^{\rm iso}(t)$ which still
depend on the wave vector.
From Eqs.~(\ref{eq:iso-app-dum-54}) and (\ref{eq:summary-MCT-transient-a}),
one gets for $M_{\bf q}^{\rm iso}(t)$
under the approximations (\ref{eq:iso-app-dum-21}) and (\ref{eq:iso-app-dum-43})
\begin{equation}
M_{\bf q}^{\rm iso}(t) =
\frac{\rho v^{2}}{2 (2 \pi)^{3}}
\frac{1}{q^{2} [ 1 + (\dot{\gamma}t)^{2} / 3 ]}
\int d{\bf k} \,
[ {\bf q} \cdot {\bf k} c_{k} + {\bf q} \cdot {\bf p} c_{p} ] \, 
[ {\bf q}(t) \cdot {\bf k}(t) c_{\bar{k}(t)} + {\bf q}(t) \cdot {\bf p}(t) c_{\bar{p}(t)} ] \,
F_{k}(t) F_{p}(t).
\label{eq:iso-app-dum-61}
\end{equation}
It is clear from this expression that the wave-vector dependence of the
this memory kernel stems from the terms 
${\bf q}(t) \cdot {\bf k}(t)$ and ${\bf q}(t) \cdot {\bf p}(t)$.
Using Eq.~(\ref{eq:advected-q}),
the explicit expression for the former reads
\begin{equation}
{\bf q}(t) \cdot {\bf k}(t) =
{\bf q} \cdot {\bf k} + (\dot{\gamma}t) (q_{x} k_{y} + q_{y} k_{x}) +
(\dot{\gamma} t)^{2} q_{x} k_{x},
\label{eq:iso-app-dum-62}
\end{equation}
and a similar expression holds for ${\bf q}(t) \cdot {\bf p}(t)$.
With the same spirit as in the approximation~(\ref{eq:iso-app-dum-43}),
the anisotropic terms $q_{x} k_{y}$ and $q_{y} k_{x}$ shall be neglected,
and $q_{x} k_{x}$ approximated by
${\bf q} \cdot {\bf k} / 3$, i.e., 
\begin{equation}
{\bf q}(t) \cdot {\bf k}(t) \approx
{\bf q} \cdot {\bf k} \, [ 1 + (\dot{\gamma}t)^{2} / 3 ].
\label{eq:iso-app-dum-63}
\end{equation}
Substituting this and a similar approximation for ${\bf q}(t) \cdot {\bf p}(t)$ into
Eq.~(\ref{eq:iso-app-dum-61}) yields the following expression 
which now depends only on the modulus $q$:
\begin{eqnarray}
M_{q}^{\rm iso}(t) =
\frac{\rho v^{2}}{2 (2 \pi)^{3} q^{2}}
\int d{\bf k} \,
[ {\bf q} \cdot {\bf k} c_{k} + {\bf q} \cdot {\bf p} c_{p} ] \, 
[ {\bf q} \cdot {\bf k} c_{\bar{k}(t)} + {\bf q} \cdot {\bf p} c_{\bar{p}(t)} ] 
F_{k}(t) F_{p}(t).
\label{eq:iso-app-dum-64}
\end{eqnarray}

Concerning $L_{\bf q}^{\rm iso}(t)$, one gets from 
Eqs.~(\ref{eq:iso-app-dum-55}) and (\ref{eq:summary-MCT-transient-b})
under the approximation (\ref{eq:iso-app-dum-21}) 
\begin{equation}
L_{\bf q}^{\rm iso}(t) = -
\frac{v^{2}}{2 (2 \pi)^{3}}
\int d{\bf k} \,
[ {\bf q} \cdot {\bf k} c_{k} + {\bf q} \cdot {\bf p} c_{p} ]
\Bigl[ \,
  \frac{k_{x} k_{y}(t)}{k(t)} \frac{S_{k(t)}^{\prime}}{S_{k(t)}} +
  \frac{p_{x} p_{y}(t)}{p(t)} \frac{S_{p(t)}^{\prime}}{S_{p(t)}} \,
\Bigr]
F_{k}(t) F_{p}(t).
\label{eq:iso-app-dum-71}
\end{equation}
In this expression, the wave vector dependence comes from
\begin{equation}
k_{x} k_{y}(t) = k_{x} k_{y} + (\dot{\gamma}t) k_{x}^{2},
\label{eq:iso-app-dum-72}
\end{equation}
and $p_{x} p_{y}(t)$. 
Here again, the anisotropic term $k_{x} k_{y}$ shall be neglected,
and $k_{x}^{2}$ approximated by $k^{2}/3$, i.e., 
\begin{equation}
k_{x} k_{y}(t) \approx (\dot{\gamma}t) k^{2} / 3,
\label{eq:iso-app-dum-72-2}
\end{equation}
and $p_{x} p_{y}(t)$ shall be approximated similarly.
Along with the approximation~(\ref{eq:iso-app-dum-43}), one then obtains
the following expression which now depends on the modulus $q$ only:
\begin{equation}
L_{q}^{\rm iso}(t) = -
\frac{v^{2}}{2 (2 \pi)^{3}} \,
\frac{ \dot{\gamma} t }{ 3 \sqrt{1+(\dot{\gamma}t)^{2}/3} } 
\int d{\bf k} \,
[ {\bf q} \cdot {\bf k} c_{k} + {\bf q} \cdot {\bf p} c_{p} ]
\Bigl[ \, 
  k \frac{S_{\bar{k}(t)}^{\prime}}{S_{\bar{k}(t)}} +
  p \frac{S_{\bar{p}(t)}^{\prime}}{S_{\bar{p}(t)}} \,
\Bigr]
F_{k}(t) F_{p}(t).
\label{eq:iso-app-dum-73}
\end{equation}

\subsection{Steady-state quantities}
\label{appen:isotropic-app-2}

Under the isotropic approximation~(\ref{eq:iso-app-dum-21}) 
for the transient density correlators 
and (\ref{eq:iso-app-dum-43}) for the modulus of the advected wave vector,
the MCT expressions (\ref{eq:ss-F-dum-14}) and (\ref{eq:ss-F-dum-15})
for the steady-state density correlator and structure factor are given by
\begin{eqnarray}
& &
F_{\bf q}^{\rm ss}(t) = 
F_{q}(t) +
\dot{\gamma} \int_{0}^{\infty} ds \,
\frac{[ q_{x} q_{y} + \dot{\gamma}(t+s) q_{x}^{2}]}{\bar{q}(t+s)}
\frac{S_{\bar{q}(t+s)}^{\prime}}{S_{\bar{q}(t+s)}^{2}}
F_{q}(t+s) F_{\bar{q}(t)}(s),
\label{eq:iso-app-ss-11}
\\
& &
S_{\bf q}^{\rm ss} =
S_{q} +
\dot{\gamma} \int_{0}^{\infty} ds \,
\frac{( q_{x} q_{y} + \dot{\gamma}s \, q_{x}^{2})}{\bar{q}(s)} 
\frac{S_{\bar{q}(s)}^{\prime}}{S_{\bar{q}(s)}^{2}} \, F_{q}(s)^{2}.
\label{eq:iso-app-ss-12}
\end{eqnarray}
From a numerical point of view, it is not necessary to further simplify these expressions 
since these steady-state quantities are the final output of the theory rather than
the ones involved in the self-consistent calculations.
Of course, it is instructive to consider their averages over the orientation
$\hat{\bf q} \equiv {\bf q}/q$
\begin{eqnarray}
& &
F_{q}^{\rm ss}(t) \equiv \frac{1}{4 \pi} \int d\hat{\bf q} \, 
F_{\bf q}^{\rm ss}(t) = 
F_{q}(t) +
\int_{0}^{\infty} ds \,
\frac{(\dot{\gamma})^{2} (t+s)}{3 \sqrt{1 + [\dot{\gamma}(t+s)]^{2}/3}} \, 
\frac{q S_{\bar{q}(t+s)}^{\prime}}{S_{\bar{q}(t+s)}^{2}}
F_{q}(t+s) F_{\bar{q}(t)}(s),
\label{eq:iso-app-ss-21}
\\
& &
S_{q}^{\rm ss} \equiv \frac{1}{4 \pi} \int d\hat{\bf q} \, 
S_{\bf q}^{\rm ss} =
S_{q} +
\int_{0}^{\infty} ds \,
\frac{(\dot{\gamma})^{2} s}{3 \sqrt{1+(\dot{\gamma}s)^{2}/3}} \, 
\frac{q S_{\bar{q}(s)}^{\prime}}{S_{\bar{q}(s)}^{2}} \, F_{q}(s)^{2},
\label{eq:iso-app-ss-22}
\end{eqnarray}
to which only those terms in 
Eqs.~(\ref{eq:iso-app-ss-11}) and (\ref{eq:iso-app-ss-12}) proportional to $q_{x}^{2}$
contribute. 
However, it is more informative to regard 
Eqs.~(\ref{eq:iso-app-ss-11}) and (\ref{eq:iso-app-ss-12})
as the approximate expressions in which the anisotropic nature of the steady-state density fluctuations 
is retained to the lowest order: such anisotropy arises from 
the terms in Eqs.~(\ref{eq:iso-app-ss-11}) and (\ref{eq:iso-app-ss-12})
proportional to $q_{x} q_{y}$.
Indeed, such a viewpoint is necessary 
to correctly understand ``the isotropic approximation for the steady-state shear stress'',
adopted in Ref.~\cite{Fuchs-Cates-sheared-MCT}, 
which sounds contradictory since the shear stress is intrinsically an anisotropic quantity
and vanishes under isotropic density fluctuations.
We shall come back to this point in a moment. 

Under the isotropic approximations~(\ref{eq:iso-app-dum-21}) and 
(\ref{eq:iso-app-dum-43}), one obtains from the MCT expression (\ref{eq:ss-shear-MCT})
for the steady-state shear stress $\sigma_{\rm ss}$
\begin{eqnarray}
\sigma_{\rm ss} =
\frac{k_{\rm B}T \dot{\gamma}}{2 (2 \pi)^{3}}
\int_{0}^{\infty} ds \,
\int d{\bf k} \,
\frac{k_{x}^{2} k_{y} (k_{y} + \dot{\gamma} s \, k_{x})}{k \bar{k}(s)}
\frac{S_{k}^{\prime} S_{\bar{k}(s)}^{\prime}}{S_{k}^{2} S_{\bar{k}(s)}^{2}} \,
F_{k}(s)^{2}. 
\label{eq:iso-app-ss-31}
\end{eqnarray}
One easily understands that only the term proportional to $k_{x}^{2} k_{y}^{2}$
survives after the integration over the orientation $\hat{\bf k} \equiv {\bf k}/k$, yielding
the following expression for $\sigma_{\rm ss}$ under the isotropic approximation:
\begin{eqnarray}
\sigma_{\rm ss} =
\frac{k_{\rm B}T \dot{\gamma}}{60 \pi^{2}}
\int_{0}^{\infty} ds \,
\frac{1}{\sqrt{1+(\dot{\gamma}s)^{2}/3}} 
\int_{0}^{\infty} dk \, k^{4} 
\frac{S_{k}^{\prime} S_{\bar{k}(s)}^{\prime}}{S_{k}^{2} S_{\bar{k}(s)}^{2}} \,
F_{k}(s)^{2}. 
\label{eq:iso-app-ss-32}
\end{eqnarray}
This is essentially the same expression as adopted in Ref.~\cite{Fuchs-Cates-sheared-MCT}.

To connect such an isotropic expression for $\sigma_{\rm ss}$ with anisotropic
density fluctuations, we rewrite the term in Eq.~(\ref{eq:iso-app-ss-31}) which survives 
after the integration over $\hat{\bf k}$ in the following form
\begin{eqnarray}
\sigma_{\rm ss} =
\frac{k_{\rm B}T}{2 (2 \pi)^{3}}
\int d{\bf k} \,
\frac{k_{x} k_{y}}{k}
\frac{S_{k}^{\prime}}{S_{k}^{2}} \,
\Bigl\{
  \dot{\gamma}
  \int_{0}^{\infty} ds \,
  \frac{k_{x} k_{y}}{\bar{k}(s)}
\frac{S_{\bar{k}(s)}^{\prime}}{S_{\bar{k}(s)}^{2}} \, F_{k}(s)^{2}
\Bigr\} .
\label{eq:iso-app-ss-41}
\end{eqnarray}
\end{widetext}
The quantity in the curly brackets is exactly the aforementioned anisotropic term in
Eq.~(\ref{eq:iso-app-ss-12}). 
Thus, the steady-state shear stress $\sigma_{\rm ss}$ can be handled
within the isotropic approximation
since its MCT expression takes a form of the product of two anisotropic terms,
one from $k_{x} k_{y}$ and the other from the anisotropic part
of the density fluctuations, which altogether behaves as an isotropic term
inside the integral.

\end{document}